\documentclass{aastex631}

\shorttitle{Repeating Fast Radio Bursts}
\shortauthors{Hu \& Huang}
\graphicspath{{./}{figures/}}
\usepackage{CJK}
\usepackage{amsmath}
\usepackage{longtable}
\usepackage{booktabs}
\usepackage{hyperref}
\usepackage[all]{hypcap}
\usepackage{CJK}
\makeatletter

\newcommand{\Rmnum}[1]{\expandafter\@slowromancap\romannumeral #1@}
\makeatother

\begin{document}
\begin{CJK*}{UTF8}{gbsn}
\title{A Comprehensive Analysis on Repeating Fast Radio Bursts}
\correspondingauthor{Yong-Feng Huang}
\email{hyf@nju.edu.cn}
\author[0000-0002-5238-8997]{Chen-Ran Hu (胡宸然)}
\affiliation{School of Astronomy and Space Science, Nanjing
University\\ Nanjing 210023, People's Republic of China}

\author[0000-0001-7199-2906]{Yong-Feng Huang (黄永锋)}
\affiliation{School of Astronomy and Space Science, Nanjing
University\\ Nanjing 210023, People's Republic of China}
\affiliation{Key Laboratory of Modern Astronomy and Astrophysics
(Nanjing University)\\ Ministry of Education, People's Republic of
China}

\begin{abstract}

Nearly 700 fast radio burst (FRB) sources have been detected till
now, among which 29 FRBs are found to burst out repeatedly.
Although a firm connection between at least some FRBs and
magnetars has been established, the trigger mechanism and
radiation process in these enigma phenomena are still highly
controversial. In this study, we build a sample of 16 repeating
FRBs from which at least five bursts have been detected, including
the most active four repeaters of FRBs 20121102A, 20180916B,
20190520B and 20201124A. Various key parameters of their bursts
are collected from the literature, which include
the arrival time, pulse width, dispersion
measure (DM), Faraday rotation measure (RM), bandwidth, waiting
time, peak flux and fluence. The distribution and time evolution
of these parameters are investigated. Potential correlations
between various parameter pairs are also extensively explored.
The behaviors of different repeaters are then compared.
It is found that the DM of FRB
20121102A seems to increase continuously on a long timescale.
While the DM of most repeaters varies in a narrow range of $\pm
3$ cm$^{-3}$ pc, FRB 20190520B is found to have a large variation
range of $\pm 12$ cm$^{-3}$ pc. RM evolves with time
in a much more chaos behavior in different repeaters.
A linear correlation is found between the absolute mean RM
and $\rm {DM}_{Host}$, which may provide a method to estimate the
redshift of FRBs. Generally, the waiting time shows a similar bimodal
distribution for the active repeating sources.
The implications of these features to the underlying physics are
discussed.


\end{abstract}



\section{Introduction}
\label{sec1:introduction}

FRB 010724 was reported by \citet{1} in 2007 as the first detected fast
radio burst (FRB). It was discovered during an analysis of
archived pulsar survey data, recorded simultaneously in three
adjacent beams by the Parkes radio telescope
of Australia. It has a high Galactic latitude, but with a
relatively low signal-noise ratio. The second FRB, FRB 010621, was
discovered by \citet{2}, which has a lower galactic latitude
$(b=-4.003^\circ)$. Subsequently, \citet{3} reported a sample of
four new FRBs. Since then, the study of FRBs has become an
important new field in high energy astrophysics.

The observation of FRBs has entered a new era since 2017. Nearly
700 FRB sources have been detected so far. However, due to the
limited field of view of radio telescopes, the observed event rate
is far less than the intrinsic FRB rate in the whole sky. FRBs can
be divided into two main groups, i.e. repeating FRBs and one-off
(non-repeating) FRBs. So far, the number of observed one-off FRB
sources is much more than those of repeating FRBs. With the
accumulation of observational data, it is getting more and more
clear that FRBs should be connected with compact objects,
especially highly magnetized neutron stars. Viable models include:
giant flares from magnetars \citep{4,5,56}, giant pulses of young
pulsars \citep{6,7,8}, star quakes of young magnetars \citep{9},
the interaction between magnetar flares and their surroundings
\citep{57,58,59,60}, lightning of pulsar magnetospheres
\citep{10}, planets around pulsars \citep{11}, collapse of massive
neutron stars to black holes \citep{12,13,14}, neutron star-white
dwarf/Be star binaries \citep{15,16}, binary neutron stars mergers
\citep{17,18}, collisions between neutron stars and asteroids
\citep{19,20}. Other novel models even include: cosmic string
collisions \citep{21}, stellar flares \citep{22}, binary white
dwarf mergers \citep{23}, evaporation of primary black holes
\citep{24}, mergers of charged binary black holes
\citep{25,26,27}, collapses of the crust of strange stars
\citep{28}, light sails of extragalactic civilizations \citep{29},
collisions of relativistic field structures \citep{30}, and
interactions of strong plasma stream with pulsar magnetospheres
\citep{31}. In this new field of high energy astrophysics, further
exploration of FRBs will help us better understand binary neutron
star mergers \citep{16,32}, supernova remnants
\citep{33,34,35,36}, cosmic baryon content \citep{37,38}, compact
faint objects \citep{39,40}, new cosmic ray acceleration
mechanisms \citep{41}. It will also help examining basic physics
\citep{42,43,44,45,46,47} and constraining cosmology models and
dark energy content \citep{48,49,50,51,52,53,54}.

It is interesting to note that FRBs could be connected with some
kinds of X-ray bursts (XRBs) from magnetars. In history,
magnetars were initially proposed to explain soft $\gamma$-ray repeaters
(SGRs) and anomalous X-ray pulsars \citep{55,189}.
They are a special type of pulsars
characterized by very strong magnetic fields, being born as the
result of the evolution of massive stars towards the end of their
lives or the binary neutron star mergers. Magnetars continuously
emit high energy electromagnetic radiation, mainly in X-rays and
$\gamma$-rays, through the decay of its magnetic field.
As slowly rotating magnetars, SGRs enter an active phase every few
years or decades, generating a series of X-ray and soft $\gamma$-ray
bursts. Recent observations have revealed that at least some FRBs
are connected with XRBs from magnetars, but the detailed mechanisms
are still unknown.

To be specific, some repeating FRBs show a quasi-periodical behavior in
their activities \citep{61,62,63}. As a valuable SGR found to be closely
related to FRBs, SGR J1935+2154 was first detected by the Swift
Burst Alert Telescope (BAT) in 2014 \citep{68}. FRB 20200428A was
observed to be originated from this active SGR by the Hard X-ray
Modulation Telescope (HXMT) and the Canadian Hydrogen Intensity
Mapping Experiment telescope (CHIME) \citep{69,70}, in association
with one of a series of XRBs \citep{61,71,72,73,74}. Such an
association explicitly supports the idea that at least some FRBs
might be connected with magnetars. Inspired by FRB 20200428A,
researchers have extensively investigated the relation between
periodically repeating FRBs and SGRs \citep{64,65,66,67}.

However, since FRB 20200428A is the only sample observed to be
associated with an XRB so far, the origin and trigger mechanism of
FRBs are still highly debated. In order to better reveal the
nature of FRBs, here we present a comprehensive study on repeating
FRBs. A large sample of repeating FRBs observed by 11 different
telescopes are collected. The data are analyzed and compared for
each FRB source, aiming to provide useful guidance for future
observation strategies and modeling of repeating FRBs.


\section{Data and Methods}
\label{sec2:data and methods}

Till now, nearly 700 FRB sources have been detected, 29 of
which are repeating ones (see e.g.,
\href{https://www.chime-frb.ca}{chime-frb-web},
\href{https://www.herta-experiment.org/frbstats}{HeRTA: FRBSTATS},
\href{https://www.wis-tns.org}{Transient Name Server},
\href{https://www.frbcat.org}{FRB Catalogue} and
\href{https://ecommons.cornell.edu/handle/1813/67195}{Fast Radio
Burst Community Newsletter}). It is not clear whether different
repeating FRBs have the same physical origin, but it is
interesting to see whether the observed parameters of repeating
FRBs are similar or not. We have collected the observational data
of repeating FRBs from the literature, and will present a
statistical analysis on them. Since new observational data on
repeating FRBs are continuously coming out, we have set a deadline
for the data used in this study, i.e. as of June 26, 2022.
Four FRBs, i.e. FRB
20121102A, 20180916B, 20190520B and 20201124A, are the
most active repeating sources, which will be paid special attention
to in this study.

FRB 20121102A, as the first detected repeating FRB, has the
longest observation history. It was first discovered by the 305-m
William E. Gordon Telescope at the Arecibo Observatory (Arecibo)
\citep{178} and was later followed by a large number of
observations \citep{118,119,120,121}. More than 2000 FRBs have
been detected from this source. Especially, between August 29 and
October 29, 2019, 1652 bursts were detected by the
Five-hundred-meter Aperture Spherical radio Telescope (FAST)
\citep{111}. The average dispersion measure (DM) is 565 $\rm
cm^{-3}~pc$, and the average Faraday rotation measure (RM) is
78153 $\rm rad~m^{-2}$. The latest analysis suggests that there is
no periodicity in its activities \citep{75}. It locates in an
irregular dwarf galaxy with a low metallicity, at a redshift of
$z=0.193$. The source is in a star-forming region, possibly associated with
a luminous persistent radio source. The properties of the
persistent radio source are consistent with a low luminosity,
accreting massive black hole \citep{120,150,151}.

FRB 20180916B was first discovered by CHIME \citep{108}, with
about 90 bursts being observed. Between April and
December 2020, 54 bursts were further observed by the Westerbork
Synthesis Radio Telescope (WSRT), which consists of twelve 25-m
dishes in a new system called the Aperture Tile in Focus (Apertif)
\citep{125}. The total number of bursts detected from FRB
20180916B exceeds 150 to date. The average DM is 349\ $\rm\ cm^{-3}~pc$
and the average RM is -108\ $\rm rad~m^{-2}$. The activities of
this FRB source show a period of 16.35 days \citep{81,82}. It locates
in a nearby massive spiral galaxy at a redshift of $z=0.0337$. The
source is in a star-forming region, but is not associated with any
persistent radio counterpart \citep{152}.

FRB 20190520B was first discovered by FAST, with four bursts being
reported initially. Subsequently, 75 bursts were further observed
by FAST between April 25 and September 19, 2020 \citep{114}.
Between June 2021 and January 2022, Parkes detected 113 additional
bursts \citep{127}. The total number of bursts being observed from
this source is close to 200. The average DM is 1207\ $\rm\
cm^{-3}~pc$, and the average RM is 3603\ $\rm rad~m^{-2}$. This
FRB has a very distinct DM excess and no periodicity is
found in its activity \citep{114}. It locates in a dwarf galaxy
with a high specific star formation rate (SFR), at a redshift of
$z=0.241$. The source is in an extreme magneto-ionic and dense
environment, in association with a compact persistent radio
source. Such an environment may hint a distinctive origin or an
earlier evolutionary stage for this FRB source \citep{114}.

FRB 20201124A, as a newly detected active repeating FRB, was first
discovered by CHIME. CHIME detected more than 33 bursts from it
(\href{https://www.chime-frb.ca}{chime-frb-web}). From April 1 to
June 11, 2021, FAST also detected 1863 bursts from this source
\citep{115}. The total number of bursts observed from FRB
20201124A thus exceeds 2000. The average DM is 413\ $\rm\
cm^{-3}~pc$, and the average RM is -586\ $\rm rad~m^{-2}$. The RM
of this FRB was found to vary significantly over time \citep{164}.
No periodicity is found in its activity \citep{115}. It locates in
a Milky-Way-sized barred spiral galaxy with a high metallicity, at
a redshift of $z=0.09795$. The source resides in an inter-arm
region at an intermediate distance from the center of its host,
with a low stellar density. Such an environment does not favor a
young magnetar origin, which requires an extreme explosion of a
massive star \citep{115,147,153}.

For these active repeating FRBs, statistical studies have been
conducted extensively on individual FRBs. For instance, studies on
the periodicity \citep{75}, spectral features \citep{76,77}, radio
variability \citep{78}, and parameter statistics \citep{79,80} are
carried out for FRB 20121102A. For FRB 20180916B, periodicity
\citep{81,82} and multi-band alignment \citep{83} are
investigated. Also, RM variation of FRB 20190520B \citep{127,149}
has been explored,  and polarization variation \citep{84} and
parameter statistics \citep{85,86} are conducted for FRB
20201124A. Meanwhile, some researchers have studied the
classification of FRB populations through various features
connected to pulse profile, brightness temperature \citep{87}, DM
 \citep{88}, location within host \citep{89}, SFR
\citep{90}, source density \citep{91}, and parameter synthesis
\citep{92,93}. Notably, there is a dominant view that repeating
FRBs and one-off FRBs may be two distinct populations
\citep{94,95,96,97}. With the accumulation of observational data,
new statistical researches on various parameters are updated
\citep{98,99,100,101,102,103,104,105}, constantly promoting our
understanding of these interesting sources. Note that there are
also some statistical studies that concentrate specifically on
one-off FRBs \citep{106,107} or repeating FRBs \citep{108,109}.

\subsection{Data Collection}
\label{data collection}

A recent study by \citet{110} suggests that there are some
statistical similarities between one-off and repeating FRBs.
However, note that it is actually a very difficult issue to
compare them directly. Repeating sources are
usually constantly monitored at high sensitivity, which can lead
to the detection of a large number of weak FRBs. As a result, the
distribution of parameters may be systematically different for
repeating FRBs and one-off FRBs. In this aspect, it has been
suggested that the difference between one-off FRBs and the first
detected burst of repeating sources should be somewhat comparable
and could present meaningful information on their origin \citep{116}.
Additionally, it is still controversial as to whether the one-off
FRBs actually repeat or not. Considering these issues, here we will
limit our researches to repeating FRBs only. We have collected the
observational data from 11 different telescopes, including FAST of
China, CHIME in Canada, Arecibo in USA, the 76-m Lovell telescope
(Lovell) in UK, the Effelsberg 100-m Radio Telescope (Effelsberg)
in Germany, the Apertif in Netherlands, the LOw Frequency ARrray
(LOFAR) in Europe, the upgraded Giant Metrewave Radio Telescope
(uGMRT) in India, the Parkes in Australia, the Robert C. Byrd
Green Bank Telescope (GBT) and the KarlG. Jansky Very Large Array
(VLA) in USA. The data correspond to various repeating FRB
sources, including multiple frequency observations at different
times. An overall view of the data sources are shown in Table
\ref{table1}.

For all the burst events in Table \ref{table1}, we have
collected various key parameters from the literature (marked as
references in the table), including the arrival time expressed in modified Julian date (MJD),
pulse width, DM, RM, bandwidth, peak flux and fluence. For
example, the physical parameters of 1652 bursts from FRB 20121102A
detected by FAST are obtained from Reference (1) \citep{111}.
The arrival time (MJD) corresponds to the time at the solar system
barycenter.
For each FRB, the waiting time between two adjacent bursts is calculated
but this parameter is applicable only for those events that
are detected in a continuous monitoring campaign.
DM is usually measured by maximizing the band-integrated structure
of the bursts, which provides a smaller range than optimizing the
signal-noise ratio. Equivalent/Boxcar pulse width is often defined
as the width of an assumptive rectangular burst that has the same
fluence as the observed burst. The burst bandwidth is defined as
the frequency range of the burst detected by the telescope,
which however could be limited by the passband of the receiver itself.
Note that the parameters in different data sets sometimes are
measured with different standards and thus could be troubled
by systematic errors. To overcome this problem, we have tried
to carry out our analysis on each data set separately as far
as possible.

\subsection{Statistical Methods}
\label{statistical methods}

Various statistical analyses have been conducted on these repeating sources
in our study. Unbiased estimation under normal or lognormal distribution is
used to extract and standardize the statistical characteristics of each
physical parameter for each repeater, which could help reduce the
systematical difference between various telescopes.

The significance of correlation between various parameter pairs is
assessed through the Kendall's tau method \citep{196}, which gives an index
generally called the tau-b coefficient. Here, for the sake of clarity, we
designate the coefficient as $Cor$. As a non-parametric statistics,
the Kendall's tau method applies to all kinds of monotonic correlations,
including those described by a simple linear function \citep{196}. It has
also been widely used in astrophysics \citep{199,200}. The correlation
coefficient in Kendall's tau statistics ($Cor$) is calculated as \citep{201}
\begin{eqnarray}
\label{formula1}
Cor = \frac{n_{\rm C}-n_{\rm D}}{\sqrt{\left( n_{\rm C} + n_{\rm D} +
n_{\rm x} \right) \left( n_{\rm C} + n_{\rm D} + n_{\rm y} \right) } },
\end{eqnarray}
where $n_{\rm C}$ is the number of concordant pairs (the pair of data points
with a positive slope), $n_{\rm D}$ is the number of discordant pairs (the
pair of data points with a negative slope), $n_{\rm x}$ and $n_{\rm y}$
are the numbers of the so called ties. Here a tie refers to a pair of
data points that have the same X parameter or the same Y parameter. In other
words, the slope between them is either zero or infinity. Here, $n_{\rm x}$
is the number of ties in ranking $X$, and $n_{\rm y}$ is the number of
ties in ranking $Y$. Comparing with the Pearson correlation method, an obvious
advantage of the Kendall's tau statistics is that the ties can be
fully taken into account. On the other hand, the existence of ties usually
leads to the failure of the Pearson method. Note that ties are not so rare
in repeating FRBs, especially for those active repeaters.

Generally speaking, a positive (negative) $Cor$ indicates a positive
(negative) correlation, and its absolute value (the maximum value is 1)
characterizes the significance of the correlation, which can be classified as
very strong ($0.8 < |Cor| \le 1$), strong ($0.6 < |Cor| \le 0.8$),
moderate ($0.4 < |Cor| \le 0.6$), weak ($0.2 < |Cor| \le 0.4$) and very
weak ($0 < |Cor| \le 0.2$). The robustness of the correlation is also examined
by using the so called p-statistics, which gives an index of $P$.
Here, $P$ represents the probability of obtaining the derived
$Cor$ value simply by chance when no intrinsic correlation exists between
the two parameters \citep{202}. The smaller the $P$ value is, the more
likely a correlation would exist. Usually, people take $p = 0.05$ as
a criterion value. However, note that sometimes a slightly larger $P$ value
could be due to the small sample size rather than the absence of correlation.
Therefore, for those cases with an impressive $Cor$ value ($|Cor| \ge 0.4$)
but with a relatively high $P$ value ($0.050 < P \le 0.333$), some further
examination is needed.


In our study, single-parameter analyses (Section \ref{sec3:single-parameter
analysis}) and double-parameter analyses (Section \ref{sec4:double-parameter
analysis}) are carried out to explore intrinsic features of repeating FRBs.
Since the observational data points are generally quite scattered and the
correlation between parameter pairs is usually very weak, we mainly use linear
functions to fit the data points for simplicity. Other kinds of functions such
as the exponential function are applied only in those cases when the data
points show a clear tendency. For the histogram plots illustrating the
single-parameter distribution of FRBs, lognormal functions are usually adopted.
Based on these investigations, conclusions are presented for each source in
Section \ref{sec5:conclusions and discussion}.

\begin{deluxetable}{cccccccccccccccccccc}[htbp]
\tabletypesize{\scriptsize} \tabcolsep=2pt \tablecaption{An
overview of the main data used in this study. The data were
collected as of June 26, 2022. In the top half of the table, the
number of bursts detected by different telescopes is presented for
each repeating FRB source. Note that FRBs 20121102A, 20180916B,
20190520B and 20201124A are the most active sources. There are
still some other inactive repeating FRBs with less than 10 bursts
being observed (mainly by CHIME). They are not listed in
this table but are used for DM analysis below. The bottom half of
the table indicates the parameters available from each dataset.
} \label{table1} \tablehead{
    \colhead{$\rm FRB name$}   &
    \multicolumn{4}{c}{$\rm FAST$}   &
    \colhead{$\rm CHIME^{\star}$}   &
    \colhead{$\rm Arecibo$}   &
    \colhead{$\rm Lovell$}   &
    \multicolumn{3}{c}{$\rm Effelsberg$}   &
    \colhead{$\rm Apertif$}   &
    \colhead{$\rm LOFAR$}   &
    \colhead{$\rm uGMRT$}   &
    \multicolumn{2}{c}{$\rm Parkes$}   &
    \multicolumn{2}{c}{$\rm GBT^{\dotplus}$}   &
    \colhead{$\rm VLA$}   &
    \colhead{$\rm Total$}
}
\startdata
 20121102A & 1652  &       &       &       &       & 518   & 32    & 4     &       &       &       &       &       &       &       & 93    &       & 2     & 2301 \\
20201124A &       &       &       & 1863  & 34    &       &       &       &       & 20    &       &       & 48    &       &       &       &       &       & 1965 \\
20190520B &       &       & 79    &       &       &       &       &       &       &       &       &       &       &       & 113   &       & 3     &       & 195 \\
20180916B &       &       &       &       & 91    &       &       &       &       &       & 54    & 9     &       &       &       &       &       &       & 154 \\
20200120E &       &       &       &       & 8     &       &       &       & 60    &       &       &       &       &       &       &       &       &       & 68 \\
20190303A &       &       &       &       & 27    &       &       &       &       &       &       &       &       &       &       &       &       &       & 27 \\
20180814A &       &       &       &       & 22    &       &       &       &       &       &       &       &       &       &       &       &       &       & 22 \\
20180301A &       & 20    &       &       &       &       &       &       &       &       &       &       &       & 1     &       &       &       &       & 21 \\
20181119A &       &       &       &       & 13    &       &       &       &       &       &       &       &       &       &       &       &       &       & 13 \\
20190417A &       &       &       &       & 13    &       &       &       &       &       &       &       &       &       &       &       &       &       & 13 \\
20190208A &       &       &       &       & 11    &       &       &       &       &       &       &       &       &       &       &       &       &       & 11 \\
20190212A &       &       &       &       & 11    &       &       &       &       &       &       &       &       &       &       &       &       &       & 11 \\
\hline
$\rm MJD^{\divideontimes}$ & √     & √     & √     & √     & √     & √     & √     & √     & √     & √     & √     & √     & √     & √     & √     & √     & √     & √     &  \\
DM & √     &√& √     & √     & √     & √& √     & √&       &       & √     & √     &       & √     & √     & √     & √     & √     &  \\
RM &       &√&       & √     & Part  & √&       & √&       & √     &       &       &       & √     & √     & √& √     & √     &  \\
Width & √     & √     & √     & √     & Part  & √     & √     & √     & √     & √     &       &       & √     & √     & √     & √     &       & √     &  \\
Bandwidth & √     &       &       & √     & Part  & √ &       & √& √     &       &       &       &       & √     &       & √&       &       &  \\
Peak Flux & √     & √     &       & √     & Part  & √&       & √     &       & √     &       &       & √     & √     & √     & √&       & √     &  \\
Fluence & √     &√& √     & √     & Part  & √     & √     & √& √     & √     & √     & √     & √     & √     &       & √     &       & √     &  \\
Reference & (1) & (2,3) & (4) & (5) & (6,7) & (8,9,10,11) & (12)  & (11,13) & (14) & (15) & (16) & (16) & (17)  & (18) & (19) & (20,21) & (22) & (11) &  \\
\enddata
\tablecomments{
References: (1) \citet{111}; (2) \citet{112}; (3) \citet{113}; (4) \citet{114}; (5) \citet{115}; (6) \citet{116}; (7) \citet{117}; (8) \citet{118}; (9) \citet{119}; (10) \citet{120}; (11) \citet{121}; (12) \citet{62}; (13) \citet{122}; (14) \citet{123}; (15) \citet{124}; (16) \citet{125}; (17) \citet{85}; (18) \citet{126}; (19) \citet{127}; (20) \citet{128}; (21) \citet{129}; (22) \citet{130}.  \\
${^\star}$ CHIME provides MJD and DM data for each burst, but many other physical parameters are available only for part of these bursts.  \\
${^\dotplus}$ FRB 20121102A was observed by GBT, but note that the parameters reported in these two references are slightly different. \\
${^\divideontimes}$ Modified Julian date (MJD) is useful in calculating the waiting time.
}
\end{deluxetable}


\section{Single-Parameter Analysis}
\label{sec3:single-parameter analysis}

The emission of FRBs is mainly in radio wavebands ($\sim$ GHz),
lasting typically for a few milliseconds with the flux density
being of the order of Jansky. FRBs have significant
extragalactic DM excess. Such an excess is mainly caused by the
intergalactic medium (IGM), suggesting that FRBs should be of
extragalactic or even cosmological origin
\citep{135,136,137,138,139}. Direct spectroscopic observations of the
optical counterpart of FRB 20121102A show that the redshift is
$z=0.19273 \pm 0.00008$, confirming its extragalactic
origin \citep{131}. A typical FRB can release an isotropic energy
up to $\rm 10^{41}~erg$ during the burst. The short duration of
FRBs implies that the source should be small and may be connected
with compact stars. The intensity distribution function can also
provide useful clues on the origin of FRBs \citep{132}. It has
been noted that the galactic latitudes of FRBs are usually
high (ranging from $20^\circ$ to $70^\circ$). By analyzing the
dipole moments ($\cos\theta$) and quadrupole moments
($\sin^2b-1/3$) of FRB's distribution on the sky, it is found that
they are quite isotropic \citep{134}, which is further confirmed
by \citet{190} recently.

Considering the intense emission at radio frequencies, the
brightness temperatures of FRBs are generally as high as  $\rm
\sim {10}^{35}~K$, i.e.
\begin{eqnarray}
\label{formula2}
T_{\rm B} \sim\ F_{\rm \nu} D^2/{2}\pi
k\left(vT\right)^2\approx1.1\times{10}^{35}K{\ \
\left(\frac{F_{\rm \nu}}{\rm Jy}\right)\left(\frac{\nu}{\rm
GHz}\right)}^{-2} \left(\frac{T}{\rm ms}\right)^{-2}
\left(\frac{D}{\rm Gpc}\right)^{-2},
\end{eqnarray}
where, $I_{\rm \nu}=F_{\rm \nu}/\pi\theta^2$ is the radiation intensity at
frequency $\nu$, $\theta$ represents the angular radius of the
source, $k$ is the Boltzmann constant, and $\lambda$ is the
wavelength. $\theta$ can be approximated as $c T/D$, where $c$ is
the speed of light, $T$ is the time scale of the FRB, and $D$ is
the distance of the source. The high brightness temperature of
FRBs indicates that non-thermal radiation mechanisms should be
involved \citep{181,182}. The
light curves of FRBs mainly show a single pulse structure, but
\citet{140} found a bimodal pulse profile in one burst from the
source of FRB 20121002A. Multi-peak bursts are also found
from FRBs 20190520B \citep{114} and 20201124A \citep{191} recently.
More interestingly, \citet{141} reported that the profile of FRB 20201020A has a
sub-millisecond quasi-periodic structure. FRBs are well characterized by
several key parameters, such as $\rm DM$, $\rm RM$, bandwidth, pulse width, waiting time,
peak flux and fluence. In this section, we present a detailed analysis on these
parameters.

\subsection{Dispersion Measure}
\label{dispersion measure}

When the electromagnetic wave propagates in plasma, it interacts
with free electrons so that its group velocity becomes
frequency-dependent. This is the so called dispersion. As a result,
high frequency emission arrives earlier than low frequency waves.
The dispersion measure is defined as the integral of electron
number density ($n_{\rm e}$) along the line of sight, which can be
expressed as
\begin{eqnarray}
\label{formula3}
{\rm DM}=\int_{0}^{D}{\frac{n_{\rm e}\left(l\right)}{1+z\left(l\right)}dl_{\rm \parallel}}=\left\langle
n_{\rm e}\right\rangle D .
\end{eqnarray}

Obviously, $\rm DM$ can be divided into several parts \citep{183},
\begin{eqnarray}
\label{formula4}
 {\rm DM} = {\rm DM}_{\rm MW}+{\rm DM}_{\rm IGM}+\frac{{\rm DM}_{\rm Host}}{1+z} .
\end{eqnarray}
Here, ${\rm DM}_{\rm MW}$ is the contribution from the interstellar medium
(ISM) within the Milky Way (MW), ${\rm DM}_{\rm IGM}$ is the contribution
of the intergalactic medium (IGM), and ${\rm DM}_{\rm Host}$ corresponds
to the contribution of ISM within the host galaxy and the local
circum-burst environment near the source.

The arrival time of the electromagnetic waves is frequency
dependent, and the delay time at frequency $\nu$ is related to
$\rm DM$ as \citep{184}
\begin{eqnarray}
\label{formula5} {\Delta}t_{\rm \nu}\left(s\right)=\frac{e^2}{2\pi
m_{\rm e}c}\frac{\rm DM\left({\rm pc{\cdot}cm^{-3}}\right)}{\nu^2},
\end{eqnarray}
where $e$ is the electron charge and $m_{\rm e}$ is the electron mass.
Using this equation, we can estimate the $\rm DM$ of FRBs through
multi-wavelength radio observations,
\begin{eqnarray}
\label{formula6} {\rm DM}=\left({\Delta}t_{\rm l}-{\Delta}t_{\rm h}\right)
\frac{2\pi m_{\rm e}c}{e^2} \frac{v_{\rm h}^2v_{\rm l}^2}{v_{\rm h}^2-v_{\rm l}^2},
\end{eqnarray}
where $\mathrm{\Delta}t_{\rm l}$ represents the time delay of the pulse
at a lower frequency $\nu_{\rm l}$, and $\mathrm{\Delta}t_{\rm h}$ represents
the delay at a higher frequency $\nu_{\rm h}$.
Sometimes, multipath propagation caused by scattering makes the
plasma act like a lens, and the $\rm DM$ can be used to establish
plasma lens models.

$\rm DM$ is an important parameter of FRBs and is usually used as a
distance indicator. Firstly, we can estimate ${\rm DM}_{\rm MW}$ by using the
widely used MW free electron models such as NE2001 \citep{142} or
YMW16 \citep{143}. With the help of the electron distribution derived
from cosmological simulations on the
epoch after reionization \citep{144}, we can also calculate
${\rm DM}_{\rm IGM}$ at redshift $z$ \citep{145,146}. Combining
${\rm DM}_{\rm Host}$ estimated by assuming a particular galaxy type
\citep{133}, we can finally infer the redshift and luminosity
distance of the source by considering Equation (\ref{formula4}). Moreover,
taking into account the visibility of various observing epochs can
also help estimate the distance of the FRB source \citep{147}.
On the other hand, it is also possible for us to estimate
${\rm DM}_{\rm IGM}$ through the Macquart relation (i.e. the ${\rm DM}_{\rm IGM}-z$
relation) when the redshift is available for a repeating FRB source
\citep{148}.

\begin{figure}[htbp]
\gridline{\fig{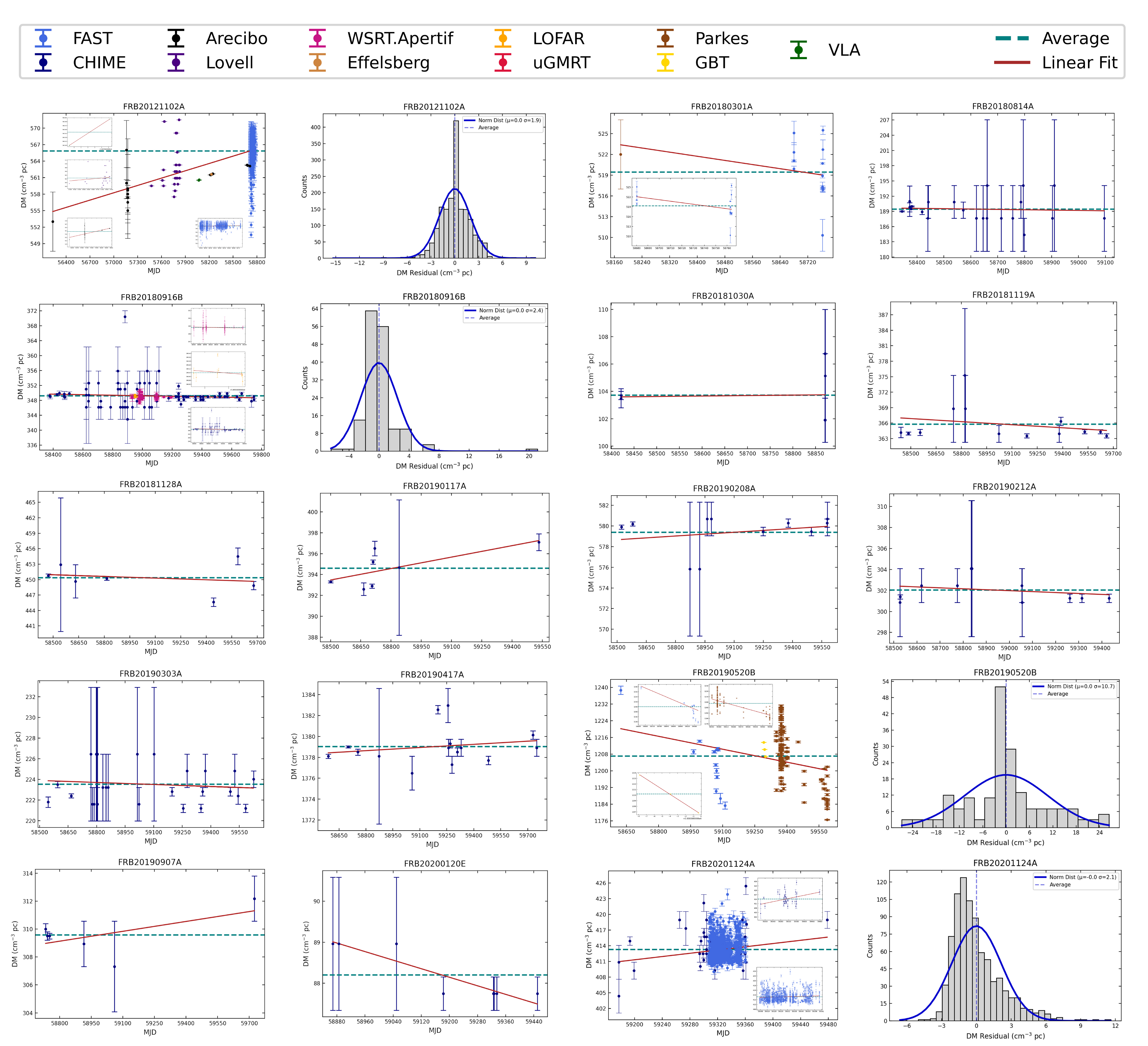}{1\textwidth}{}} \caption{$\rm DM$ plotted
versus time for each repeating FRB source. The horizontal dashed
line corresponds to the DM average of each FRB, and the solid line
is the linear fit to the observational data points. The inset
shows the observational data points from different telescopes. For
the four most active FRBs, the residuals of data points with
respect to the linear fit are also illustrated via histogram. }
\label{figure1}
\end{figure}

Figure \ref{figure1} plots the $\rm DM$ of the repeating FRBs as a function of time.
We see that in most cases, $\rm DM$ does not have an obvious evolution
trend over time. However, there are still several special cases.
The $\rm DM$ of FRB 20190520B shows a rapid declining trend
($Cor$ = -0.216, $P$ = 8.759 $\times 10^{-6}$),
while the $\rm DM$ of FRB 20121102A
($Cor$ = 0.104, $P$ = 2.258 $\times 10^{-10}$) shows an
increasing trend (see the insets in Figure \ref{figure1}).
For the four most active FRBs, we have fitted their $\rm DM$ evolution
tendency by using a linear function, and then pay special attention to the
$\rm DM$ residuals with respect to the linear function. The residual
(dispersion of $\rm DM$) reflects the random variation range of $\rm DM$.
The residuals are also plotted in Figure \ref{figure1}. We see
that the DM residuals generally follow a normal distribution for
all the four sources. It is interesting to note that the $\rm DM$
variation is mainly in a range of $\pm 3$ cm$^{-3}$ pc. However,
FRB 20190520B seems to be special since its variation is as large
as $\pm 12$ cm$^{-3}$ pc. It is still unclear whether such a
variation is due to the change of the local circum-burst
environment or not.

\begin{figure}[htbp]
\gridline{\fig{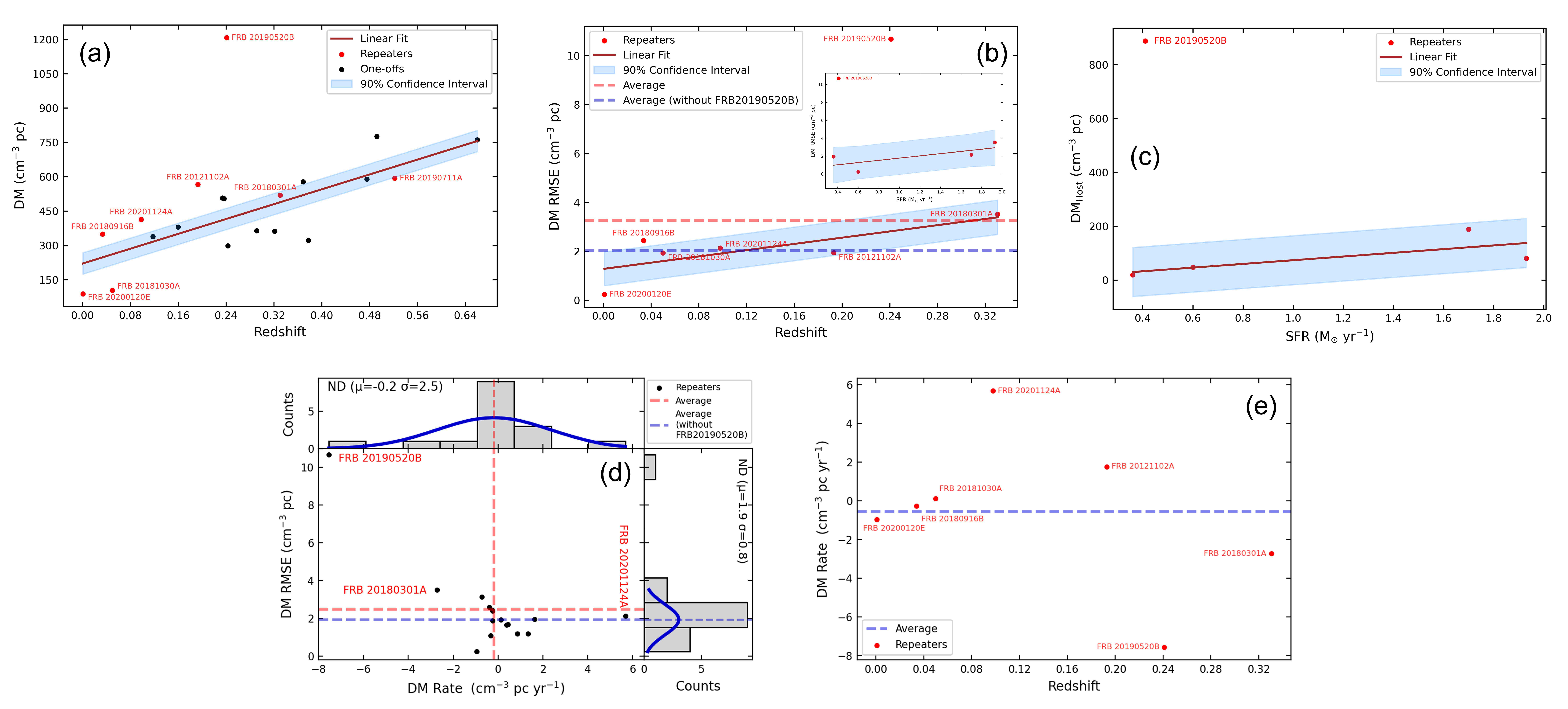}{1\textwidth}{}}
 \caption{Correlations
between the redshift, SFR and DM for the repeating FRBs. Panel (a):
the relation between DM and redshift. Panel (b): the relation
between DM RMSE and redshift.
The inset shows the relation between DM RMSE and SFR.
Panel (c): the relation between $\rm {DM}_{Host}$
and SFR. Panel (d): the relation between DM RMSE and DM
time-varying rate. Panel (e): the relation between DM time-varying
rate and redshift.}
\label{figure2}
\end{figure}

In Figure \ref{figure2}, possible correlations between various
parameter pairs involving the redshift, SFR and $\rm DM$ are examined
for repeating FRBs. In Panel (a), we see
that $\rm DM$ is linearly correlated with the redshift
($Cor$ = 0.544, $P$ = 0.001). The correlation can be best fitted by a linear function as
\begin{eqnarray}
\label{formula7}
{\rm DM}=811.409z+220.365\ {\rm pc\ cm^{-3}}.
\end{eqnarray}
It is consistent
with the results of \citet{148}, which reflects the major role
played by ${\rm DM}_{\rm IGM}$ in $\rm DM$. The root mean square error (RMSE)
is used to characterize the dispersion level of $\rm DM$ with respect to
the linear fit for each repeating FRB.
In Figure \ref{figure2}(b), we see that the RMSE of $\rm DM$ is also positively
correlated with redshift ($Cor$ = 0.467, $P$ = 0.272).
\citet{203} have argued that the $\rm DM$ variations caused by
the fluctuations in large-scale structures, HII regions and the cosmological
expansion are all very small. The variation induced by the fluctuations of gas density
in the Milky Way is also negligible \citep{203}. Therefore the RMSE of DM should be
mainly contributed by the local environment of the FRB sources. Consequently,
the increase of the RMSE of DM with redshift as shown in Figure \ref{figure2}(b)
may reflect the cosmological evolution of the FRB sources. It is interesting to
note that the SFR also increases with redshift when $z \le 1$ \citep{204}, which
means that the number of younger magnetars should increase with redshift.
Young magnetars are generally in an active phase which can frequently give birth
to various bursts due to instabilities connected with their strong magnetic fields \citep{189}.
It naturally leads to some fluctuations in the medium around them, resulting in
a relatively large variation of DM in their local environment \citep{203}. So,
the increase of the RMSE of DM with redshift essentially supports the connection
between FRBs and young magnetars. Such an explanation is further supported by
the positive correlation between the RMSE of DM and SFR ($Cor$ = 0.667, $P$ = 0.333)
shown in the inset of Figure \ref{figure2}(b).
Note that the correlation between RMSE and the redshift still remains tight
even if FRB 20190520B, which is an obvious outlier, is excluded.

Figure \ref{figure2}(c) shows that ${\rm DM}_{\rm Host}$ increases
with the SFR of the host galaxy ($Cor$ = 0.667, $P$ = 0.333).
It should be caused by the increase of ${\rm DM}_{\rm Host}$
due to star formation activities in the host galaxy. However,
differences in local activities may make the correlation somewhat dispersive.
In Figure \ref{figure2}(d), we see that the RMSE and the time-varying rate of
$\rm DM$ are both normally distributed. There is also no correlation between
these two parameters.
Some Detailed values concerning DM are presented in Table \ref{table2}.
From this table and Figure \ref{figure2}(d), we notice that the repeaters
can be roughly divided into three classes according to their time-varying rate of DM.
The first class are characterized by a rapid decrease rate of DM, which
include FRB 20190520B (with a time-varying rate of -7.55 $\rm\ cm^{-3}~pc~yr^{-1}$)
and FRB 20180301A (with a time-varying rate of -2.73 $\rm\ cm^{-3}~pc~yr^{-1}$).
The second class are featured by a rapid DM increase, which include
FRB 20201124A (with a time-varying rate of 5.67 $\rm\ cm^{-3}~pc~yr^{-1}$).
Other repeaters have a slow varying rate of DM (with the absolute rate
less than $2 \rm\ cm^{-3}~pc~yr^{-1}$), which form the third class.
The evolution rate of DM may reflect the triggering mechanism or local
environment of FRBs. For example, \citet{203} argued that shortly after a
supernova explosion, the SNR will initially be in a high speed free
expansion stage, during which the temperature will decrease quickly and
the ionization ratio will also rapidly decrease due to adiabatic cooling.
It will lead to a rapid decrease of DM. Later, when the SNR enters the
Sedov-Taylor phase and the snowplow deceleration phase, the interaction
of the non-relativistic shocks with the circumambient medium will
ionize the neutral medium and give birth to some additional free electrons.
The DM will then slowly increase.
On the other hand, a rapid increase of DM can be driven by the strong wind
of young pulsars/magnetars. Relativistic electron-positron pairs streaming
out from these rapidly-rotating compact objects will significantly increase
the number of free electrons in the circum-stellar medium and boost the DM.
Additionally, note that the plasma lensing effect can also affect the DM,
but it generally leads to some small amplitude perturbations. Further more,
the binary motion of the central engine and the interaction with its companion
may lead to some periodical evolution of DM. In realistic cases, all the above
mechanisms may take effect jointly. More detailed observations on the evolution
and variation of DM in the future may help diagnose the trigger mechanism and
local environment of FRBs.


Figure \ref{figure2}(e) shows
that the time-varying rate of $\rm DM$ seems to be uncorrelated with the redshift.
However, as mentioned in the above paragraph, there might be three
classes of repeaters when the time-varying rate of DM is considered, i.e.
those with DM decreases or increases rapidly, or with a slow evolution of DM.
In the future, when the sample size of repeating FRBs becomes large enough,
it is interesting to further examine whether there is any correlation between
the time-varying rate of DM and $z$ for each class of FRB sources.
In all the five panels of Figure \ref{figure2}, we notice that FRB 20190520B
is very different. It is always an outlier with respect to other FRBs. The DM
abnormality in FRB 20190520B has also been pointed out by \citet{114}.
It could be caused by some special conditions in the local environment or in
its host galaxy, or by strong turbulence on the line of sight \citep{149}.


\begin{deluxetable}{ccccccccccc}[htbp]
\tabletypesize{\scriptsize}
 \tablecaption{Statistical characteristics of $\rm DM$ for all FRBs with an observed host galaxy.}
 \label{table2}
 \tablehead{
    \colhead{$\rm Type$}   &
    \colhead{$\rm FRB name$}   &
    \colhead{$\rm Redshift$}   &
    \colhead{$\rm References^{\star}$}   &
    \colhead{$\rm \overline{\rm DM}$}   &
    \colhead{$\rm DM~Rate$}   &
    \colhead{$\rm DM~RMSE$}   &
    \colhead{$\rm DM_{\rm Exc}^{\dotplus}$}   &
    \colhead{$\rm DM_{\rm Host}^{\divideontimes}$}   &
    \colhead{$\rm SFR$}   &
    \colhead{$\rm \overline{\rm RM}$}  \\
    \colhead{}  &
    \colhead{}  &
    \colhead{}  &
    \colhead{}  &
    \colhead{($\rm\ cm^{-3}~pc$)}  &
    \colhead{($\rm\ cm^{-3}~pc~yr^{-1}$)}  &
    \colhead{($\rm\ cm^{-3}~pc$)}  &
    \colhead{($\rm\ cm^{-3}~pc$)}  &
    \colhead{($\rm\ cm^{-3}~pc$)}  &
    \colhead{($\rm\ M_{\rm \odot}~yr^{-1}$)}  &
    \colhead{($\rm rad~m^{-2}$)}
}
\startdata
Repeating & 20121102A & 0.1930  & (1,2,3) & 565.83  & 1.76  & 1.95  & 347.80  & 180.17  &       & 78152.95  \\
& 20180916B & 0.0337  & (4) & 349.20  & -0.26  & 2.44  & 150.40  & 121.13  &       & -107.83  \\
& 20190520B & 0.2410  & (5) & 1207.04  & -7.55  & 10.68  & 1097.04  & 887.72  & 0.41  & 3603.18  \\
& 20201124A & 0.0980  & (6,7,8) & 413.27  & 5.67  & 2.14  & 273.27  & 188.20  & 1.70  & -586.59  \\
& 20180301A & 0.3304  & (9) & 519.44  & -2.73  & 3.52  & 367.44  & 80.48  & 1.93  & 81.96  \\
& 20181030A & 0.0500  & (10) & 103.70  & 0.12  & 1.93  & 62.30  & 18.87  & 0.36  &  \\
& 20200120E & 0.0008  & (11,12) & 88.20  & -0.95  & 0.24  & 48.20  & 47.51  & 0.40 - 0.80 &  \\
& 20190711A & 0.5220  & (13) & 593.10  &       &       &       &       &       &  \\
\hline
One-off & 20150418A & 0.4920  & (14) & 776.20  &       &       &       &       &       &  \\
& 20180924B & 0.3212  & (15) & 361.42  &       &       &       &       &       &  \\
& 20181112A & 0.4755  & (16) & 589.27  &       &       &       &       &       &  \\
& 20190102C & 0.2912  & (9) & 363.60  &       &       &       &       &       &  \\
& 20190523A & 0.6600  & (17) & 760.80  &       &       &       &       &       &  \\
& 20190608B & 0.1178  & (18) & 338.70  &       &       &       &       &       &  \\
& 20190611B & 0.3778  & (9) & 321.40  &       &       &       &       &       &  \\
& 20190714A & 0.2365  & (9) & 504.00  &       &       &       &       &       &  \\
& 20191001A & 0.2340  & (19) & 506.92  &       &       &       &       &       &  \\
& 20191228A & 0.2430  & (9) & 297.90  &       &       &       &       &       &  \\
& 20200430A & 0.1600  & (9) & 380.10  &       &       &       &       &       &  \\
& 20200906A & 0.3688  & (9) & 577.80  &       &       &       &       &       &  \\
\enddata
\tablecomments{
References: (1) \citet{120}; (2) \citet{150}; (3) \citet{151}; (4) \citet{152}; (5) \citet{114}; (6) \citet{115}; (7) \citet{147}; (8) \citet{153}; (9) \citet{105}; (10) \citet{154}; (11) \citet{155}; (12) \citet{156}; (13) \citet{157}; (14) \citet{158}; (15) \citet{159}; (16) \citet{160}; (17) \citet{161}; (18) \citet{162}; (19) \citet{163}.\\
${^\star}$ Redshift and SFR data are taken from these references. \\
${^\dotplus}$ The extragalactic DM is calculated by using the NE2001 model \citep{142}.\\
${^\divideontimes}$ DM component contributed by the local and
host galaxy is calculated by using the Macquart relation \citep{148}.\\
}
\end{deluxetable}

\subsection{Faraday Rotation Measure}
\label{faraday rotation measure}

$\rm RM$ is another key parameter of FRBs and is often used to infer
the magnetic field on the line of sight. Due to the Faraday
effect, the polarization angle ${\Psi}_{\rm obs}\left(\lambda\right)$
at wavelength $\lambda$ can be expressed as \citep{183}
\begin{eqnarray}
\label{formula8}
{\Psi}_{\rm obs}(\lambda)={\Psi}_{\rm 0}+{\Delta\Psi}={\Psi}_{\rm 0} +
\frac{e^3\lambda^2}{2\pi m_{\rm e}^2c^4}
\int_{0}^{D}{\frac{n_{\rm e}\left(l\right)
B_{\rm \parallel}(l)}{\left[1+z\left(l\right)\right]^2}dl},
\end{eqnarray}
where ${\Psi}_{\rm 0}$ is the initial polarization angle,
$\mathrm{\Delta\Psi}$ is the variation of the polarization angle
induced by the Faraday effect, and ${B}_{\rm \parallel}$ represents the
magnetic field parallel to the line of sight. $\rm RM$ is defined as
\begin{eqnarray}
\label{formula9} {\rm RM} = \frac{e^3}{2\pi m_{\rm e}^2c^4}
\int_{0}^{D}{\frac{n_{\rm e}\left(l\right)B_{\rm \parallel}(l)}{\left[1+z\left(l\right)\right]^2}
dl}.
\end{eqnarray}
Then ${\Psi}_{\rm obs}\left(\lambda\right)$ can be simplified as:
\begin{eqnarray}
\label{formula10} {\Psi}_{\rm obs}\left(\lambda\right) =
{\Psi}_{\rm 0}+\lambda^2{\rm RM} = 0.5\arctan\frac{U}{Q}.
\end{eqnarray}
${\Psi}_{\rm obs}\left(\lambda\right)$ can be calculated from the
observed Stokes parameters $U$ and $Q$, so that $\rm RM$ can also be
derived through observations \citep{185,186,187}. Furthermore, combining $\rm DM$ and
$\rm RM$, we can estimate the parallel magnetic field as \citep{188}
\begin{eqnarray}
\label{formula11} \left\langle B_{\rm \parallel}\right\rangle =
\frac{\rm RM}{0.812{\rm DM}} \approx 1.232 \left(\frac{\rm RM}{\rm rad\ m^{-2}}
\right) \left(\frac{\rm DM}{\rm pc\ cm^{-3}} \right)^{-1}{\rm \mu G}.
\end{eqnarray}
Note that the linear polarization of FRBs will be significantly
affected during the propagation. A recent study by \citet{109}
shows that FRBs depolarize at lower frequencies, i.e. the degree
of linear polarization decreases with frequency.

\begin{figure}[htbp]
\gridline{\fig{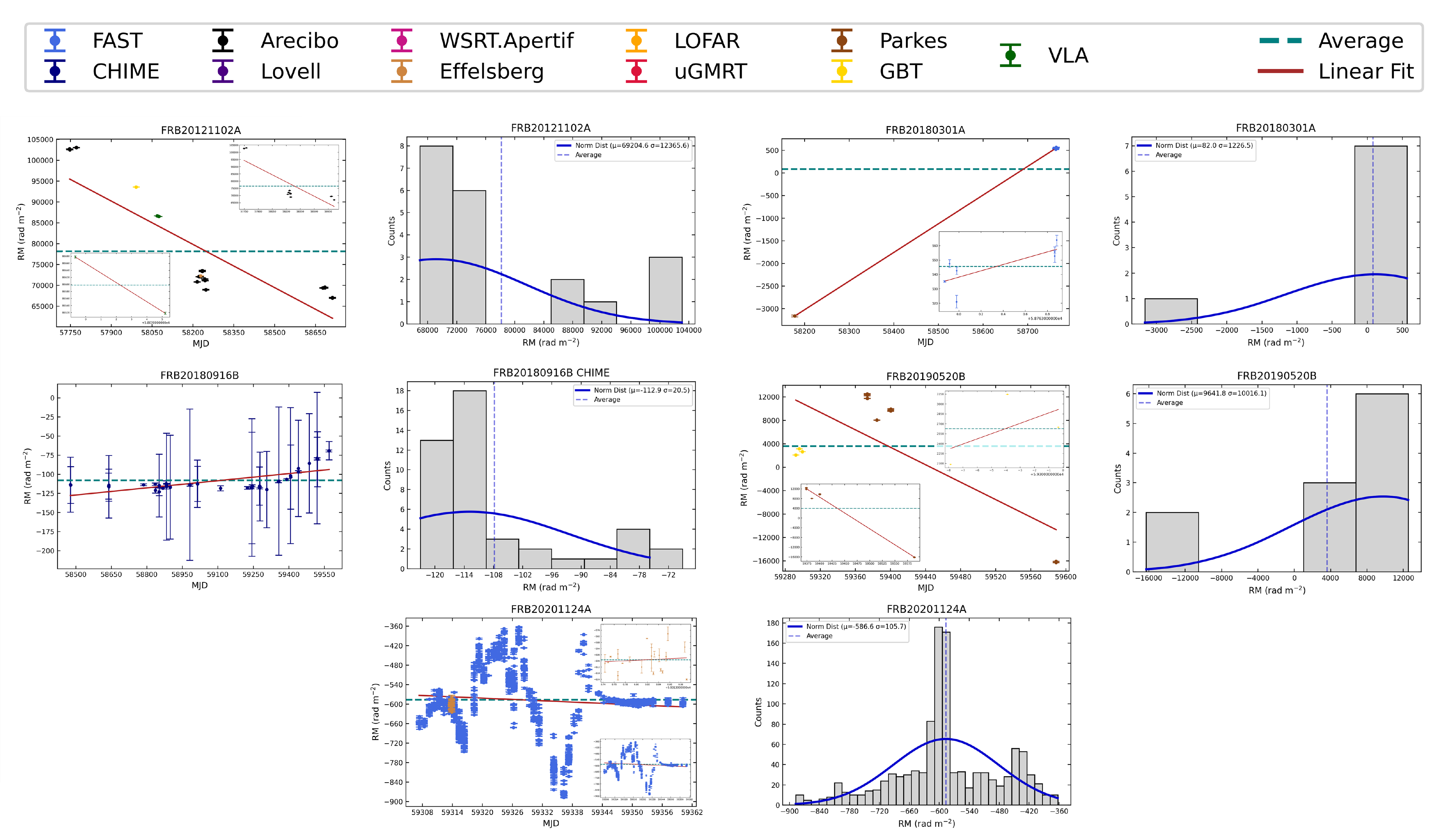}{1\textwidth}{}}
 \caption{Evolution of $\rm RM$ for repeating FRBs. The horizontal dashed line
indicates the average value of $\rm RM$ for each FRB. The solid line
is the best linear fit to the data points. For each FRB, the
distribution of $\rm RM$ is illustrated via histogram. The inset shows
the observed data points of each telescopes. }
\label{figure3}
\end{figure}

Figure \ref{figure3} plots $\rm RM$ versus time for all the repeating
FRBs with $\rm RM$ data available. We see that $\rm RM$ generally has a
strong evolution over time \citep{164,127}. This can be seen in
all the five FRBs shown in Figure \ref{figure3}, i.e. FRBs
20121102A, 20180301A, 20190520B, 20201124A, and 20201124A.
However, the evolution trend is different for different sources. In
some FRBs, $\rm RM$ increases with time, but in others, $\rm RM$ decreases
with time. In the case of FRB 20201124A, which has the most
prolific $\rm RM$ data, the evolution is very complicated. Note
that the $\rm RM$ of FRB 20180301A and FRB 20190520B experienced
variations between positive and negative values.
Useful information could be extracted from these observations,
which may help constrain the trigger mechanism of FRBs.
For example, the marked variation of RM could be
caused by the inversion of magnetic field in the emission
region, but it is also possible that FRBs could happen in more
than one region near the source. To some extent, the former case
does not support the young magnetar models unless there is a
special magnetization mechanism; while orbital motions may
be involved in the latter case. Also, considering the above
time scale and variation amplitude of $\rm RM$ evolution,
the triggering position of FRBs may constantly
vary between the equator and the polar regions.

Some interesting models have been proposed to explain the
various phenomena associated with the variation of RM.
\citet{164} argued that repeating FRBs could be produced by a
magnetar interacting with the decretion disk of a Be star. The orbital
motion of the magnetar within the disk can then naturally lead to
the special RM evolution pattern as observed in FRB 20201124A.
\citet{127} suggested that the RM evolution could arise from a turbulent
magnetized plasma screen in the local environment of the source. Such
turbulence could be excited by the stellar wind of the binary companion
of the FRB source.
Finally, note that multipolar magnetic field of magnetars may be
involved in the central engine of FRBs \citep{206}. Significant
$\rm RM$ evolution would be naturally expected in the framework
of such a complex magnetic field.
Orbital motions or binary interactions are involved in many
of the above explanations, which may lead to some obvious periodicity
in the variation of RM. However, whether there is any signature of
such periodicity in observations still needs to be examined with more
observational data in the future.

The histogram plots of Figure \ref{figure3} show
the variation amplitude of $\rm RM$ in each repeating source. In the case
of FRB 20201124A, the histogram can be fitted by a normal
distribution, but with a high peak at the center.
Detailed statistical analyses are shown in Table \ref{table3}.

\begin{figure}[htbp]
\gridline{\fig{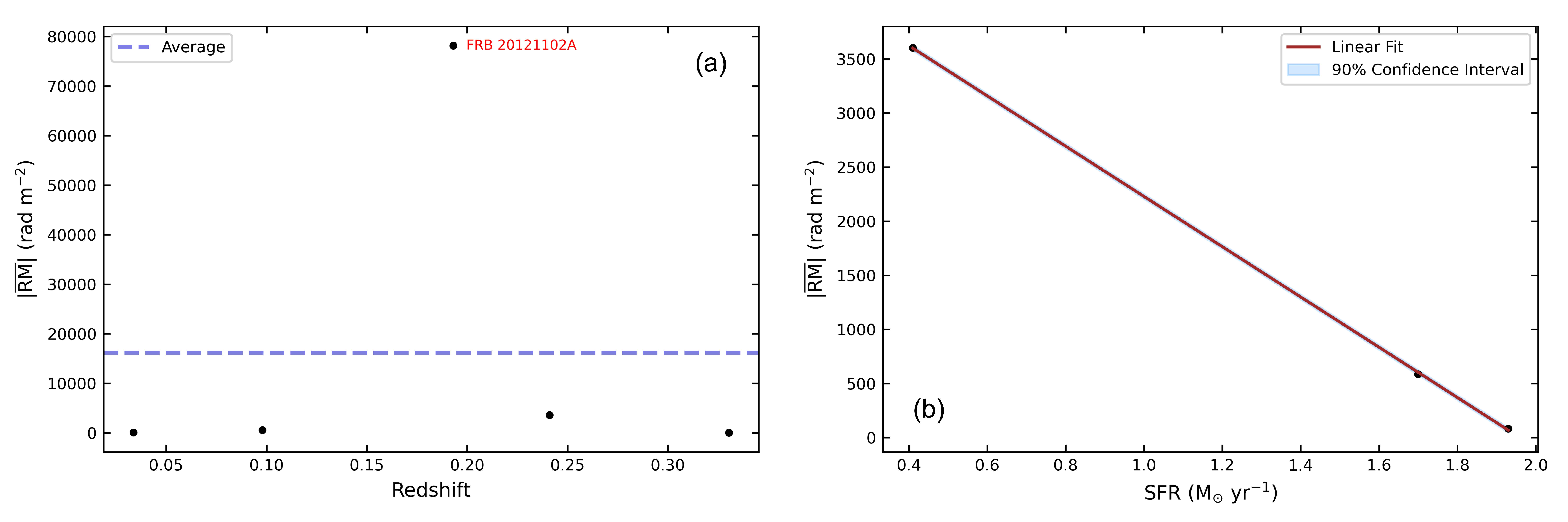}{1\textwidth}{}}
 \caption{Correlations
between the redshift, SFR and $\overline{\rm RM}$ for repeating FRBs.
Panel (a): the relation between $|\overline{\rm RM}|$ and redshift.
Panel (b): the relation between $|\overline{\rm RM}|$ and SFR. Note
that the linear fit in this panel is very tight so that the $90\%$
confidence interval is too narrow to be illustrated.} \label{figure4}
\end{figure}

$\overline{\rm RM}$, the mean value of $\rm RM$, can be regarded as an indication
of the magnetic field strength. Here we use $|\overline{\rm RM}|$ to
refer to the absolute value of $\overline{\rm RM}$. Figure \ref{figure4} shows the
relations between the redshift, SFR and $|\overline{\rm RM}|$ for
repeating FRBs. We see that $|\overline{\rm RM}|$ has no correlation
with the redshift, but it correlates negatively with SFR
($Cor$ = -1.000, $P$ = 0.333).
It is surprising to notice that the negative correlation between
$|\overline{\rm RM}|$ and SFR in Figure \ref{figure4}(b) is very tight,
with the observational data points almost in a straight line.
However, note that there are only three data points here, thus more
observations are needed to test whether such a correlation is true or
not in the future. Theoretically, FRBs are likely powered by
magnetars \citep{193,192,194}. Whether the negative correlation between
$|\overline{\rm RM}|$ and SFR can give any hints on the central engines
also needs further study.

\begin{figure}[htbp]
\gridline{\fig{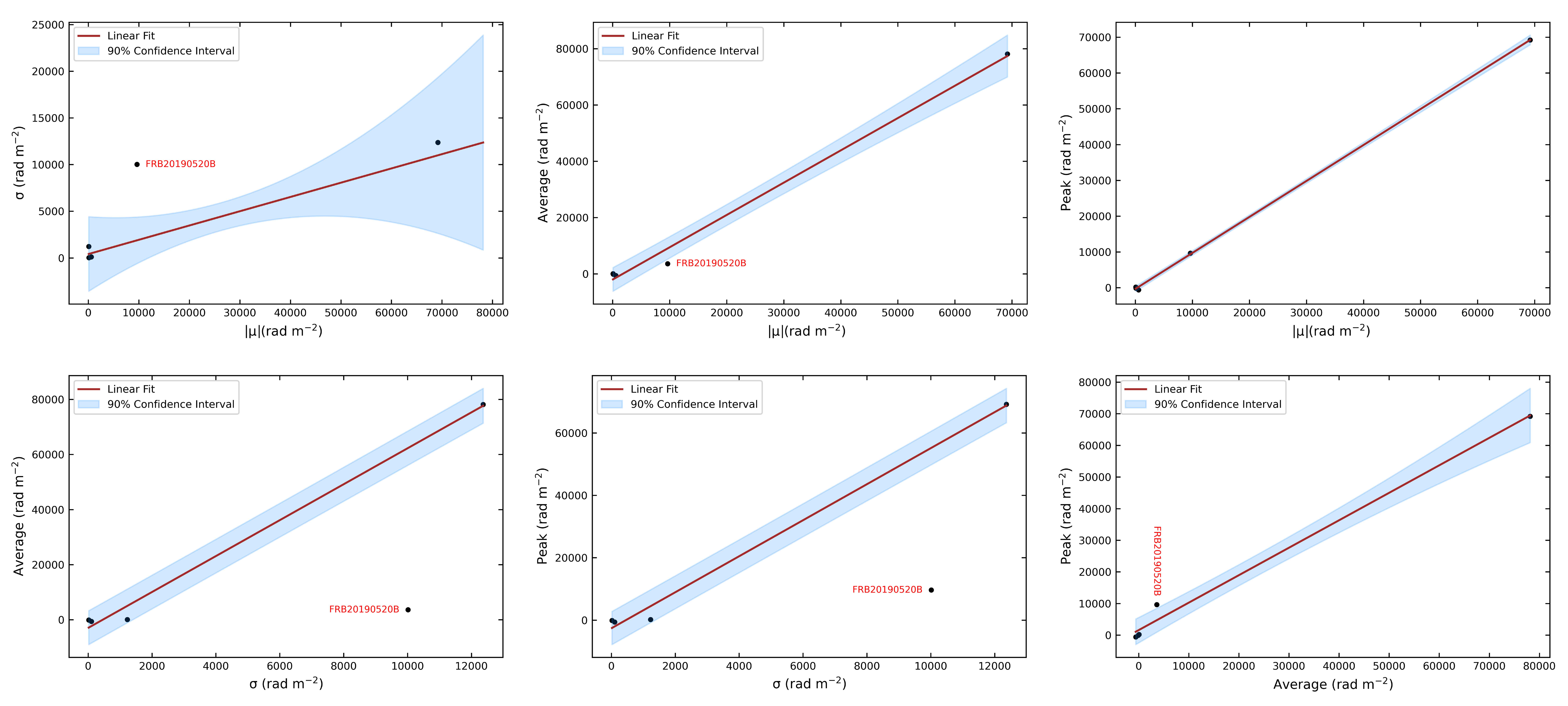}{1\textwidth}{}}
 \caption{Correlation between the statistical parameters of $\rm RM$ for repeating FRBs.
 $\mu$ and $\sigma$ represent the two parameters of normal distribution of $\rm RM$ for
 each FRB, characterizing the typical value and the dispersion range, respectively.
 $Average$ stands for the mean value, which is slightly different from $\mu$ here because
 $\mu$ is derived by a normal distribution fitting. $Peak$ represents the $\rm RM$ value
 that is recorded most frequently in observations.}
\label{figure5}
\end{figure}

Figure \ref{figure5} shows the correlations between the statistical
parameters of $\rm RM$ for repeating FRBs. We see that the three parameters,
i.e. the average level, the dispersion range and the peak of
RM distribution, are positively correlated with each other.
However, it should be noted that FRB 20190520B seems to be an
outlier in these plots. This is mainly due to its large
$|\overline{\rm RM}|$ value and the DM abnormality possibly produced
by special conditions in the local or its host galaxy, as
discussed in Section (\ref{dispersion measure}).
\citet{127} argued that FRB 20190520B may differ from
other repeating FRB sources mainly in its local circum-burst
environment.

\begin{deluxetable}{ccccccccc}[htbp]
\tabletypesize{\scriptsize}
 \tablecaption{Detailed statistical characteristics of $\rm RM$ for repeating FRBs.
 This table contains the main statistical parameters about the normal distribution
 of RM for each FRB. $\mu$ and $\sigma$ represent the two parameters of normal
 distribution, characterizing the typical value and the dispersion range, respectively.
 $Average$ stands for the mean value, which is slightly different from $\mu$ here because
 $\mu$ is derived by a normal distribution fitting. $Peak$ represents the $\rm RM$ value
 that is recorded most frequently in observations.    }
\label{table3}
\tablehead{
    \colhead{$\rm FRB name$}   & &
    \colhead{$\mu$}  & &
    \colhead{$\sigma$}  & &
    \colhead{$\rm Average$}  & &
    \colhead{$\rm Peak$}  \\
    \colhead{}  & &
    \colhead{($\rm rad~m^{-2}$)}  & &
    \colhead{($\rm rad~m^{-2}$)}  & &
    \colhead{($\rm rad~m^{-2}$)}  & &
    \colhead{($\rm rad~m^{-2}$)}
}
\startdata
20121102A & & 69204.60  & & 12365.60  & & 78152.95  & & 69204.63  \\
20180301A & & 81.96  & & 1226.54  & & 81.96  & & 191.21  \\
20180916B & & -112.90  & & 20.50  & & -107.83  & & -112.91  \\
20190520B & & 9641.80  & & 10016.06  & & 3603.18  & & 9641.80  \\
20201124A & & -586.59  & & 105.72  & & -586.59  & & -601.85  \\
\enddata
\end{deluxetable}

\subsection{Bandwidth}
\label{bandwidth}

FRB emission is usually in a narrow frequency range in the radio
band. The observed bandwidth is an important parameter of FRBs,
which is also limited by the passband of telescopes. For this
reason, we need to analyze the observational data from each
telescope separately.

\begin{figure}[htbp]
\gridline{\fig{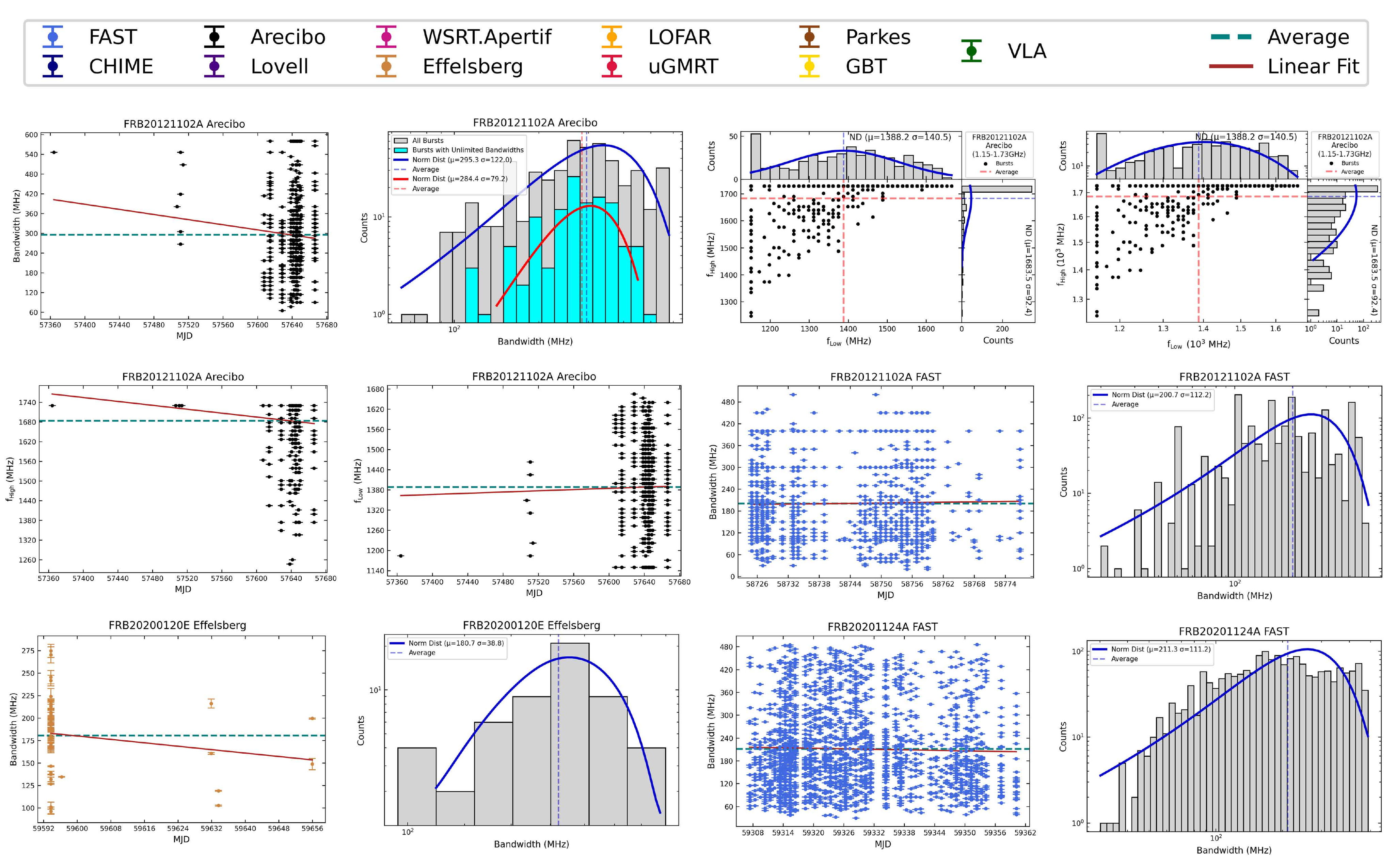}{1\textwidth}{}}
 \caption{The observed bandwidth of FRBs plotted versus time for the active repeating FRBs. The horizontal dashed
 line corresponds to the average of the bandwidth for each FRB. The solid line is
 our linear fit to the data points. For the FRBs detected by different telescopes,
 the distribution of the bandwidth is illustrated via histogram. $f_{\rm High}$ and $f_{\rm Low}$
 stand for the upper and lower frequency of the observed bandwidth, respectively.}
\label{figure6}
\end{figure}

Figure \ref{figure6} plots the bandwidth versus time (MJD) for the
most active repeating FRBs. The distribution of the bandwidth is
also illustrated via histogram. We see that the bandwidth does not
show any obvious time evolution trend and it is generally normally
distributed. For the Arecibo data of FRB 20121102A, it can be seen
that the upper and lower frequency of the observed bandwidth
($f_{\rm High}$ and $f_{\rm Low}$) stack up at the upper and lower
bounds of Arecibo's passband, respectively (see Figure
\ref{figure6}, Row 1, Column 3). It is due to the cutoff of the
limiting band of Arecibo. The intrinsic bandwidths of these
bursts should be beyond the passband.
This is more evident in Figure \ref{figure6} (Row 1, Column 4),
which is shown in logarithmic scale. Note
that the limiting passband of telescopes could also lead to the
distortion in the fluence data \citep{165}.

\begin{figure}[htbp]
\gridline{\fig{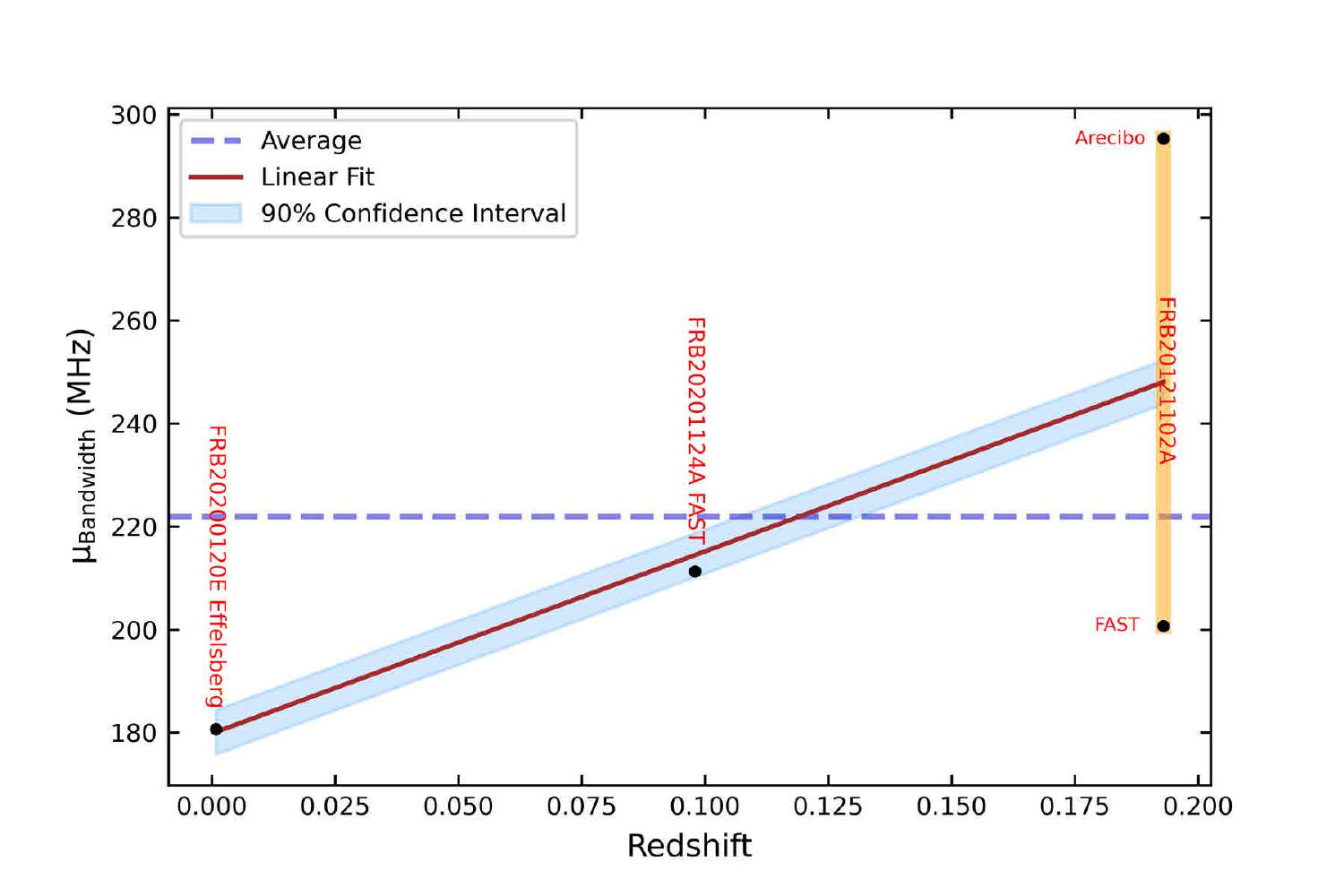}{0.5\textwidth}{}}
 \caption{Correlation between the redshift and the characteristic bandwidth for repeating FRBs.
 $\mu_{\rm Bandwidth}$ is the characteristic bandwidth derived from the normal distribution
 fitting, which could represent the average bandwidth.}
\label{figure7}
\end{figure}

Figure \ref{figure7} plots the characteristic bandwidth versus
redshift for repeating FRBs. Here the parameter $\mu_{\rm Bandwidth}$
can be regarded as the typical bandwidth of an FRB source. We see
that $\mu_{\rm Bandwidth}$ has a clear positive correlation with the
redshift ($Cor$ = 1.000, $P$ = 0.333).
Note that Arecibo (1.15-1.73 $\rm GHz$), FAST (1.05-1.45
$\rm GHz$) and Effelsberg (1.2-1.6 $\rm GHz$) have similar passbands.
Their data are used in Figure \ref{figure7} to minimize
the selection effect caused by instrumentation. Such a correlation
may be induced by the cosmological broadening of the bandwidth,
but it could also be due to cosmological evolution of the central
engines.

\begin{figure}[htbp]
\gridline{\fig{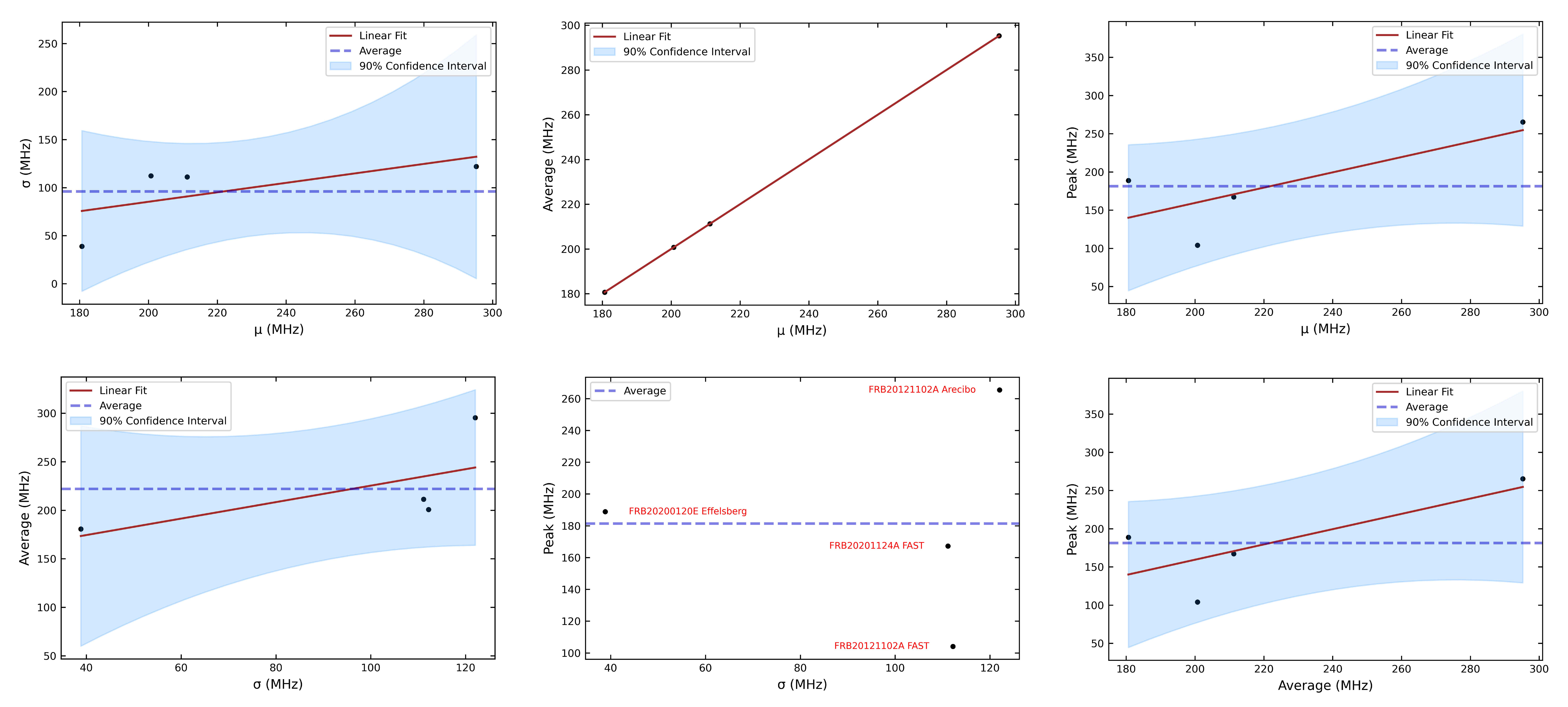}{1\textwidth}{}}
 \caption{Correlation between the statistical parameters of bandwidth for repeating FRBs.
 $\mu$ and $\sigma$ represent the two parameters derived from the normal distribution
 fitting of the bandwidth for each source, characterizing the typical bandwidth and the
 dispersion range, respectively. $Average$ stands for the mean value. $Peak$
 represents the value observed most frequently.}
\label{figure8}
\end{figure}

Detailed statistical parameters of the bandwidth are listed in
Table \ref{table4}. Figure \ref{figure8} shows the correlation
between these parameters for repeating FRBs. We see that the
parameters of bandwidth are positively correlated with each other
except for the $Peak$-$\sigma$ plot. Note that these correlations
are generally very weak (except for the $\mu$-$Average$ plot). A
larger sample size will help clarify the correlations in the
future.

\begin{deluxetable}{ccccccccc}[htbp]
\tabletypesize{\scriptsize}
 \tablecaption{Detailed characteristics of the bandwidth for repeating FRBs.
 $\mu$ and $\sigma$ represent the two parameters derived from the normal distribution fitting
 of the observed bandwidth for each source, characterizing the typical bandwidth and the
 dispersion range, respectively. $Average$ stands for the mean value. $Peak$
 represents the value observed most frequently. }
\label{table4}
\tablehead{
    \colhead{$\rm FRB$}   & &
    \colhead{$\mu$}  & &
    \colhead{$\sigma$}  & &
    \colhead{$\rm Average$}  & &
    \colhead{$\rm Peak$}  \\
    \colhead{}  & &
    \colhead{($\rm MHz$)}  & &
    \colhead{($\rm MHz$)}  & &
    \colhead{($\rm MHz$)}  & &
    \colhead{($\rm MHz$)}
}
\startdata
20121102A Arecibo & & 295.26  & & 122.04  & & 295.26  & & 265.44  \\
20121102A FAST & & 200.73  & & 112.19  & & 200.73  & & 104.11  \\
20200120E Effelsberg & & 180.65  & & 38.84  & & 180.65  & & 188.85  \\
20201124A FAST & & 211.26  & & 111.15  & & 211.26  & & 167.21  \\
\enddata
\end{deluxetable}

\subsection{Pulse Width}
\label{pulse width}

Pulse width ($w$), or the duration of a burst, is strongly influenced by the
plasma on the line of sight when an FRB propagates toward us. Multipath
propagation caused by scattering broadens the pulse and attenuates the
peak flux. It also leads to a delay in the arrival time. The broadened pulse
width of the FRB varies notably with the frequency, leading to a
frequency-dependent pulse trailing \citep{166,167,168} as
\begin{eqnarray}
\label{formula12}
w\propto {\rm DM}^2\cdot\nu^{-4}.
\end{eqnarray}
The measured width is also affected by the sensitivity and threshold of
telescopes, so we analyze the observational data of each telescope
separately.

\begin{figure}[htbp]
\gridline{\fig{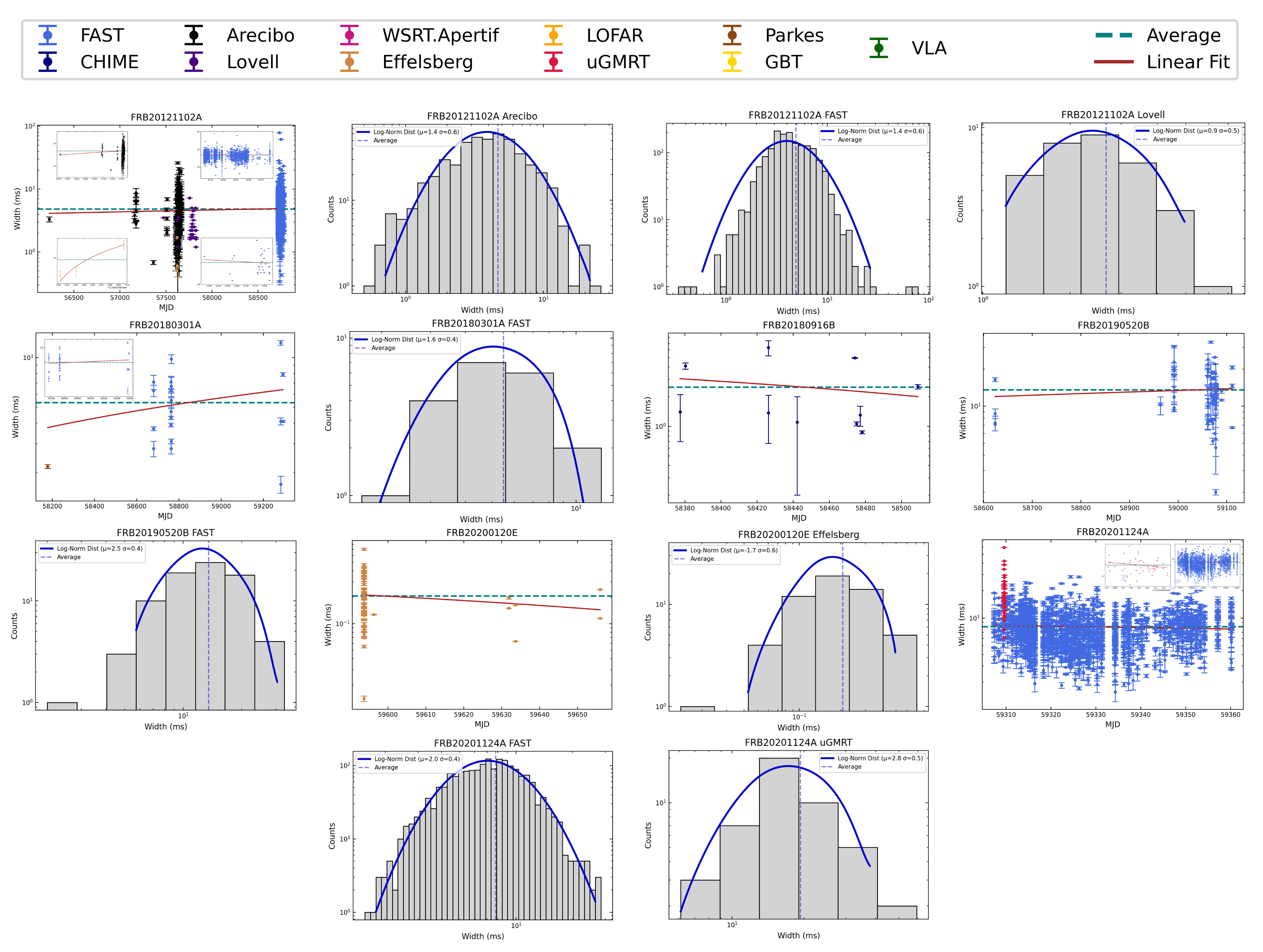}{1\textwidth}{}}
 \caption{Pulse width plotted versus time for repeating FRBs. The horizontal
 dashed line corresponds to the average of the pulse width for each source.
 The solid line is our linear fit to the data points. The distribution of
 the data points is also illustrated via histogram. The inset shows the
 observational data points from each telescope. }
\label{figure9}
\end{figure}

\begin{figure}[htbp]
\gridline{\fig{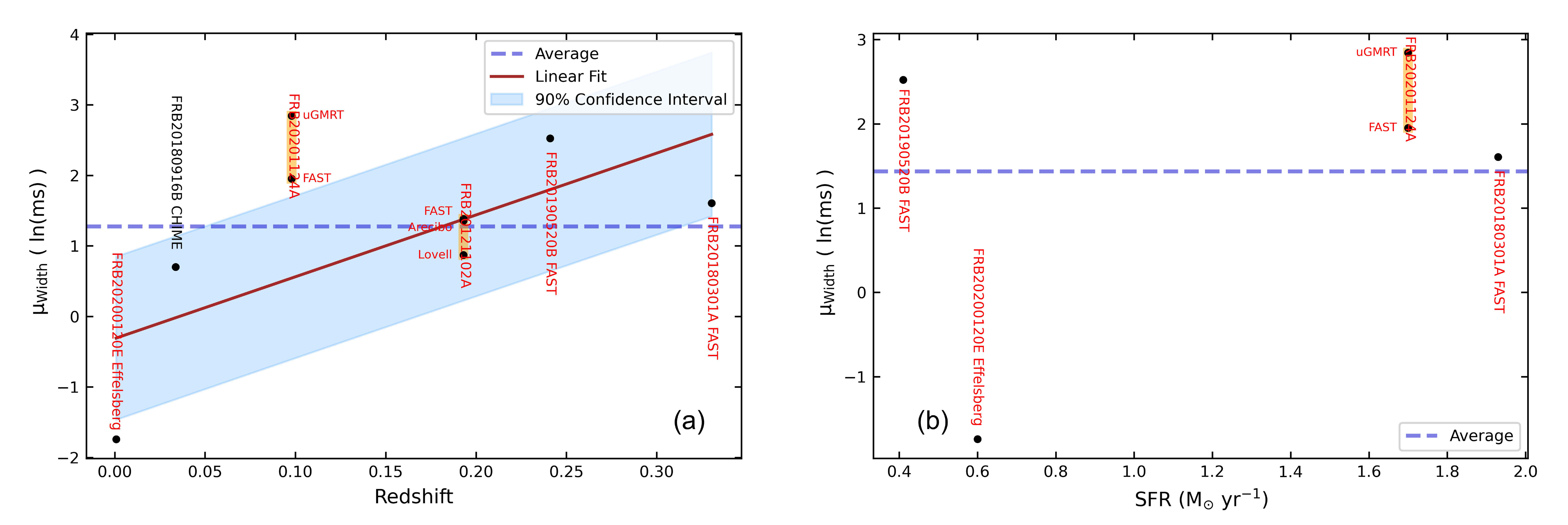}{1\textwidth}{}}
\caption{Correlation between the redshift, SFR and pulse width for repeating
FRBs. Panel (a): $\mu_{\rm Width}$ versus redshift. Panel (b): $\mu_{\rm Width}$
versus SFR. Note that $\mu_{\rm Width}$ is derived by using a lognormal
function to fit the observed pulse width distribution for each source,
and thus represents the typical pulse width. FRB 20180916B is not used
for statistical analysis due to the small amount of bursts with pulse width measurements}, but is
also shown in the figure for completeness.
\label{figure10}
\end{figure}

Figure \ref{figure9} plots the pulse width versus the MJD time for repeating FRBs.
We see that the width does not have an apparent variation over time. The
distribution of width can generally be well described by a lognormal function.
$\mu_{\rm Width}$ is derived in such a lognormal fit,
which represents the typical pulse width for the FRB source.
Detailed parameters derived are listed in Table \ref{table5}.
Figure \ref{figure10} illustrates the correlation between the redshift, SFR and
pulse width for repeating FRBs. There is a positive correlation between
$\mu_{\rm Width}$ and redshift ($Cor$ = 0.600, $P$ = 0.136). It may be
caused by a combination of the cosmological expansion and the propagation
effects. Cosmological expansion will broaden the width at high redshift, and
the scattering in a longer path will also broaden the width. However,
the width varies over several orders of magnitude and the currently
localized FRBs generally have a redshift of less than 0.5. So the contribution
of cosmological expansion should not be dominant. Consequently, cosmological
evolution of FRB sources may play an important role in the correlation, which
however still requires further investigation.
Note that the observational selection effect should also be considered.
Due to the limited sensitivity of our radio telescopes, fewer weak bursts with
small pulse width would be detected from distant sources.
FRB 20201124A observed by uGMRT seems to be an outlier in
the $\mu_{\rm Width}-z$ diagram, the reason of which still needs to be clarified.
Besides, there seems to be no correlation between $\mu_{\rm Width}$
and SFR, but the data points are too few to draw any firm conclusion. Future
observations will help to clarify it.

\begin{figure}[htbp]
\gridline{\fig{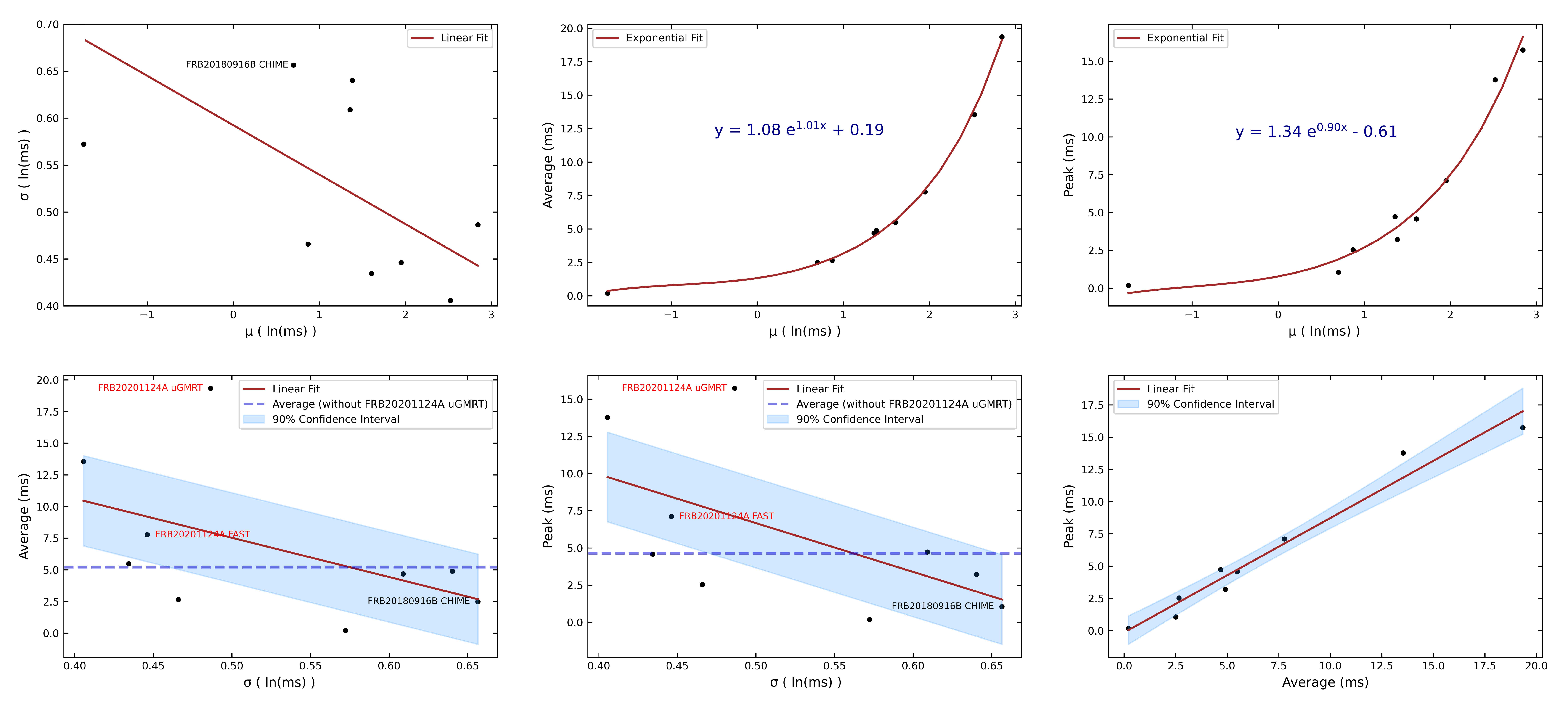}{1\textwidth}{}}
\caption{Correlation between various statistical parameters derived by
fitting the pulse width distribution with a lognormal function for repeating
FRBs. $\mu$ and $\sigma$ characterize the typical width and the dispersion range,
respectively. $Average$ stands for the mean value of pulse width. $Peak$
represents the width value observed most frequently. FRB 20180916B is not used
for statistical analysis due to the small amount of bursts with pulse width measurements}, but is
also shown in the figure for completeness.
\label{figure11}
\end{figure}

Figure \ref{figure11} illustrates the correlation between various statistical
parameters derived by fitting the pulse width distribution with a lognormal
function for each repeating source. The parameters generally correlate with
each other. Especially, the dispersion range of width is negatively correlated
with the typical width itself, the dispersion range is negatively correlated
with the peak, and the typical width is positively correlated with the peak.
The correlation between $\mu$ and $Average$/$Peak$ is a natural result,
since they all characterize the typical pulse width to some extent.
Note that the typical width and peak width of FRB 20201124A observed by uGMRT
are higher than the fitting results. This is primarily caused by the higher
detection threshold of fluence of uGMRT \citep{85}.

\begin{deluxetable}{ccccccccc}[htbp]
\tabletypesize{\scriptsize}
\tablecaption{Parameters derived by fitting the pulse width distribution with a
lognormal function for repeating FRBs. $\mu$ and $\sigma$ characterize the typical
width and the dispersion range, respectively. $Average$ stands for the mean
value of pulse width. $Peak$ represents the width value observed most frequently.}
\label{table5}
\tablehead{
    \colhead{$\rm FRB$}   & &
    \colhead{$\mu$}  & &
    \colhead{$\sigma$}  & &
    \colhead{$\rm Average$}  & &
    \colhead{$\rm Peak$}  \\
    \colhead{}  & &
    \colhead{($\rm ln(ms)$)}  & &
    \colhead{($\rm ln(ms)$)}  & &
    \colhead{($\rm ms$)}  & &
    \colhead{($\rm ms$)}
}
\startdata
20121102A Arecibo & & 1.36  & & 0.61  & & 4.68  & & 4.72  \\
20121102A FAST & & 1.38  & & 0.64  & & 4.90  & & 3.21  \\
20121102A Lovell & & 0.87  & & 0.47  & & 2.66  & & 2.53  \\
20180301A FAST & & 1.61  & & 0.43  & & 5.49  & & 4.57  \\
20180916B CHIME* & & 0.70  & & 0.66  & & 2.50  & & 1.06  \\
20190520B FAST & & 2.52  & & 0.41  & & 13.54  & & 13.77  \\
20200120E Effelsberg & & -1.74  & & 0.57  & & 0.21  & & 0.17  \\
20201124A FAST & & 1.95  & & 0.45  & & 7.77  & & 7.11  \\
20201124A uGMRT & & 2.84  & & 0.49  & & 19.34  & & 15.74  \\
\enddata
\tablecomments{
* With a small sample size.
}
\end{deluxetable}

\subsection{Waiting Time}
\label{waiting time}

Waiting time is the time interval between two consequent bursts.
It is closely related to the trigger mechanism of repeating FRBs,
and is also an important factor that should be considered when
modeling repeating FRBs. Note that the waiting time can be
extracted only when the source is continuously monitored, which
means the upper limit of the observed waiting time is subject to
the observation duration. The wide field of CHIME makes it
effective in measuring waiting times, since it has a relatively
long monitoring period on any particular source, i.e.,
\begin{eqnarray}
\label{formula13} t_{\rm observ}=86164 {\rm
s}\cdot\frac{\theta_{\rm E-W}}{360^\circ}\cdot\sec\delta,
\end{eqnarray}
where $\theta_{\rm E-W}$ is the width of the east-west field of view
of the telescope (in degrees), and $\delta$ is the declination of
the FRB source. For CHIME, $\theta_{\rm E-W} \sim 1.3^\circ -
2.5^\circ$ \citep{197}. According to Equation (\ref{formula13}), the
monitoring period of CHIME for FRBs near the north celestial pole
can even be limitless.

\begin{figure}[htbp]
\gridline{\fig{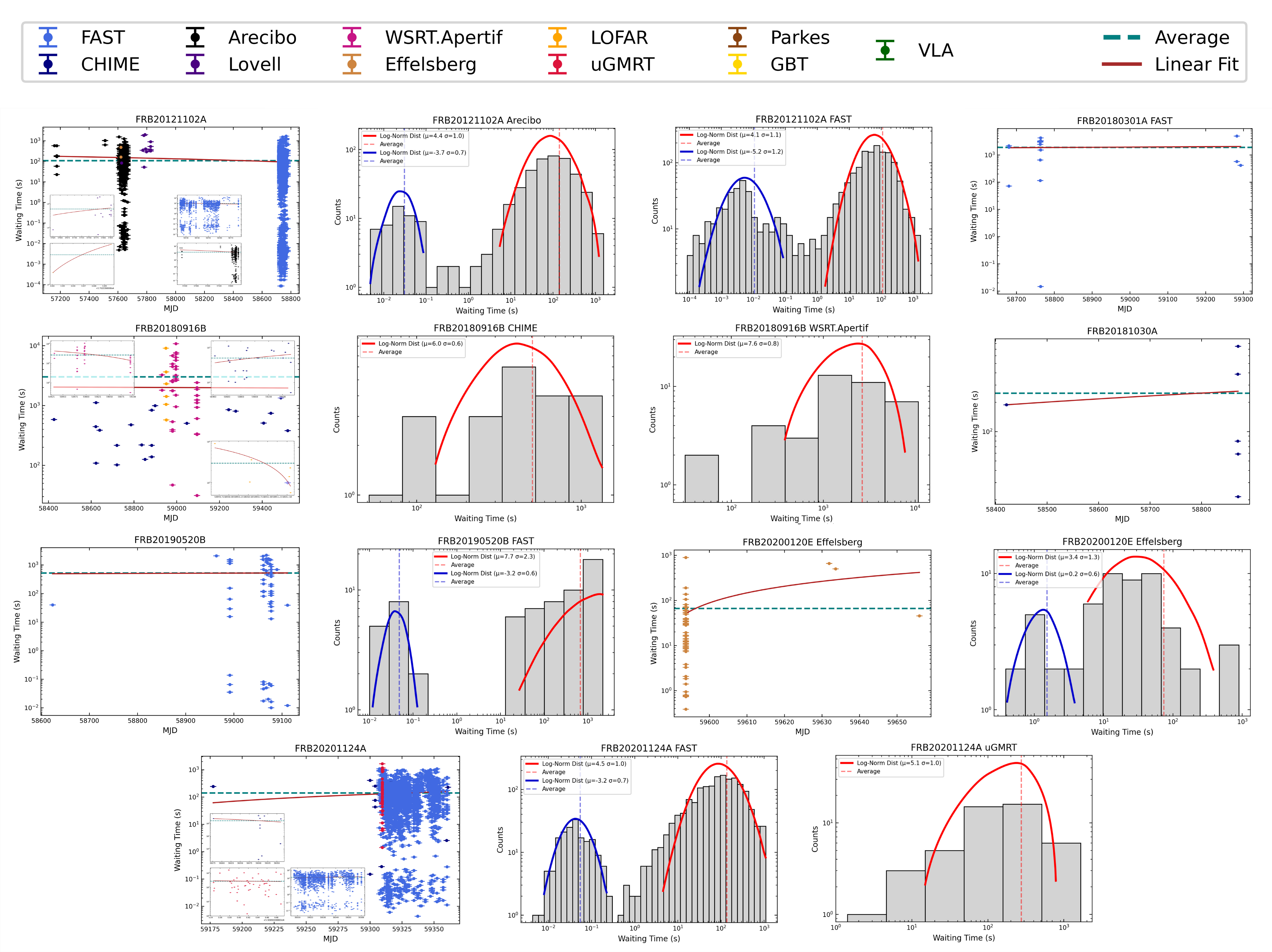}{1\textwidth}{}}
\caption{The waiting time plotted versus the MJD time for the most active repeating
FRBs. The horizontal dashed line corresponds to the average value of waiting time
for each source. The solid line is the best linear fit to the data points. The
distribution of the waiting time is also illustrated via histogram. The inset
shows the observational data points from each telescope. }
\label{figure12}
\end{figure}

Figure \ref{figure12} plots the waiting time versus the MJD time for the
most active repeating FRBs. We see that the waiting time does not have any
evolution trend with time. The distribution of the waiting time is also
illustrated via histogram. The sample size of the bursts detected by CHIME,
Apertif and uGMRT is generally small \citep{85,125}. For all other samples,
we could clearly see an interesting bimodal distribution in the histogram
\citet{140}. The two groups of bursts typically peak at $\sim 10$ ms and
$\sim 100$ s, respectively. The events separated by $\sim 10$ ms were argued
by some authors as one single burst with a multiple-pulse sub-structure
\citep{111,169,141,170}. Some of them may even be due to the strong
gravitational lensing effect \citep{172}. FRB pulses can also
be affected by the diffractive scintillation of IGM \citep{173} and ISM
turbulence \citep{174}. There are generally two kinds of scintillation,
diffraction and refraction \citep{198}. The timescale of refraction
usually ranges from days to months, and is thus completely
irrelevant to the waiting time discussed here. The timescale of
diffractive scintillation is much smaller, i.e. of the order of
minutes \citep{198}. We see that it still can hardly account for
the small waiting time of a few milliseconds. Anyway, whether
the cause of small waiting time is intrinsic or not is still controversial.
On the other hand, the waiting time of $\sim 100$ s could be closely
connected to the trigger mechanism of repeating FRBs \citep{171}.

\begin{figure}[htbp]
\gridline{\fig{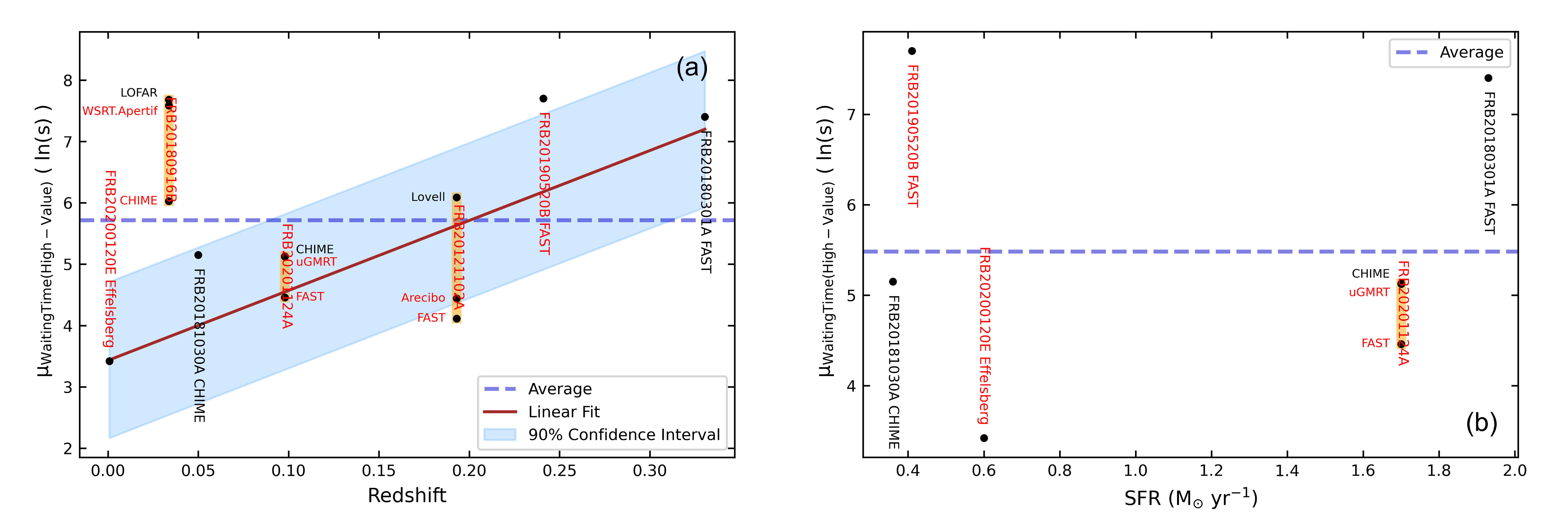}{1\textwidth}{}}
 \caption{Relation between the redshift, SFR and the typical waiting time for repeating
 FRB sources. Panel (a): $\mu_{\rm WaitingTime(High-Value)}$ versus the redshift. Panel (b):
 $\mu_{\rm WaitingTime(High-Value)}$ versus SFR. Here $\mu_{\rm WaitingTime(High-Value)}$
 refers to the typical waiting time of the right peak in the bimodal distribution,
 derived by using a lognormal function to fit the peak for each FRB. FRB 201800301A and FRB 20181030A are not used for statistical analysis due to the small amount of bursts that can be used for calculating the waiting time, but are
 shown in the figure for completeness. }
\label{figure13}
\end{figure}

The parameter $\mu_{\rm WaitingTime(High-Value)}$ is a measure of the
typical waiting time of the right component in the bimodal
distribution in Figure \ref{figure12}. It is derived by using a
lognormal function to fit the component for each source. The
detailed results of such fits are presented in Table \ref{table6}.
Figure \ref{figure13} illustrates the relation between the redshift,
SFR and the typical waiting time for repeating FRB sources. There
is an obvious linear correlation between
$\mu_{\rm WaitingTime(High-Value)}$ and redshift
($Cor$ = 0.600, $P$ = 0.233). Again, it may be
induced by the joint effects of cosmological expansion as well as
the cosmological source evolution,
but it may also be subject to the observational selection
effect since fewer weak bursts can be detected from distant sources
than from nearby sources due to the incompleteness of detection.
Note that FRB 20180916B is significantly higher than the fitting
curve, which may be caused by the relatively long duty time of
Apertif and CHIME observations \citep{116}. There is no correlation between
$\mu_{\rm WaitingTime(High-Value)}$ and SFR in Figure \ref{figure13}.
However, the sample size is still very limited in this plot and more
observations are needed to further clarify it in the future.

\begin{figure}[htbp]
\gridline{\fig{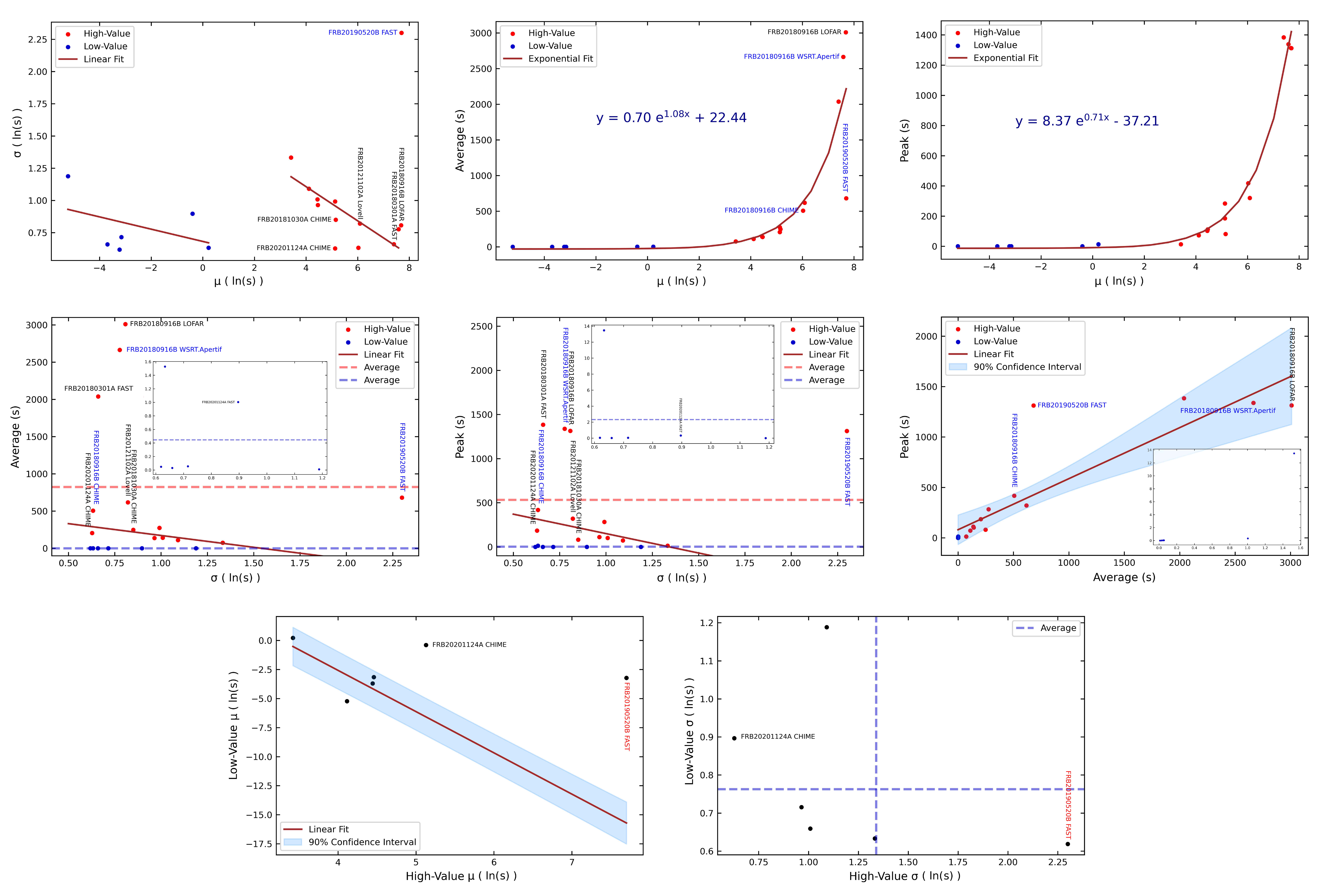}{1\textwidth}{}}
 \caption{Correlation between various parameters concerning the waiting
time for repeating FRBs. $\mu$ and $\sigma$ are the two parameters
derived by fitting the waiting time through a lognormal
distribution for each FRB, characterizing the typical waiting time
and the dispersion range, respectively. The red dots represent the
high-value waiting time while the blue dots are the low-value
waiting time. $Average$ stands for the mean value. $Peak$
represents the value observed most frequently. The FRBs annotated
in black are not used for parametric analysis due to the small
sample size, but are also shown here for completeness. }
\label{figure14}
\end{figure}

The correlation between various parameters concerning the waiting
time are shown in Figure \ref{figure14}. The dispersion range of
the waiting time is negatively correlated with the typical value,
and with the peak as well.
Such negative correlations are unusual and are in contradiction
with normal expectations from the observational selection effect as
explained below. 
First, we notice that the distribution of the high-value waiting 
time of repeaters detected by different telescopes generally peaks 
at $\sim 100$ s, which means that the intrinsic distribution of 
high-value waiting time is similar for different repeaters. 
Second, note that it is impossible for us to record a very large waiting 
time due to the limited observation length of our telescopes. When the 
observation length is larger, we will be able to detect some long waiting 
time events. It means both the mean waiting time and the dispersion range 
will be larger. On the other hand, when the observation length is smaller, 
both the measured mean time and the dispersion range will be smaller 
synchronously. As a result, the dispersion range is expected to 
increase together with the average waiting time. 
The negative correlations revealed in Figure \ref{figure14} thus are 
not induced by the above selection effect, but should be due to some 
intrinsic physical mechanisms associated with FRBs.
However, the exact mechanism is still unknown and need to be studied further
in the future. The typical value of waiting time is positively
correlated with the peak, which is quite reasonable since these two parameters
are both characteristic values of the waiting time. In Figure \ref{figure14} (Row 1, Column
1), the correlation between $\mu$ and $\sigma$ is different for the high-value
($Cor$ = -0.810, $P$ = 0.011) and low-value waiting
times ($Cor$ = -0.400, $P$ = 0.483), which indicates that
the causes of them are different. Similar phenomena can also be
seen in other two panels of (Row 2, Column 1) and  (Row 2, Column
2). It is worth noting that FRB 20190520B again is quite different
from other sources in Figure \ref{figure14}.

\begin{deluxetable}{ccccccccccccccccc}[htbp]
\tabletypesize{\scriptsize}
 \tablecaption{ Parameters derived by fitting the distribution of
 the waiting time with a lognormal function for repeating FRBs.
 $\mu$ and $\sigma$ characterize the typical waiting time and the dispersion range,
 respectively. $Average$ stands for the mean value of waiting time.
 $Peak$ represents the value of waiting time observed most frequently.
 }
 \label{table6}
 \tablehead{
    \colhead{$\rm FRB$}   & &
    \multicolumn{7}{c}{$\rm High-Value$}   & &
    \multicolumn{7}{c}{$\rm Low-Value$} \\
    \cmidrule{3-9} \cmidrule{11-17}
    \colhead{}  & &
    \colhead{$\mu$}  & &
    \colhead{$\sigma$}  & &
    \colhead{$\rm Average$}  & &
    \colhead{$\rm Peak$}  & &
    \colhead{$\mu$}  & &
    \colhead{$\sigma$}  & &
    \colhead{$\rm Average$}  & &
    \colhead{$\rm Peak$} \\
    \colhead{}  & &
    \colhead{($\rm ln(s)$)}  & &
    \colhead{($\rm ln(s)$)}  & &
    \colhead{($\rm s$)}  & &
    \colhead{($\rm s$)}  & &
    \colhead{($\rm ln(s)$)}  & &
    \colhead{($\rm ln(s)$)}  & &
    \colhead{($\rm s$)}  & &
    \colhead{($\rm s$)}
}

\startdata
20121102A Arecibo & & 4.44  & & 1.01  & & 141.54  & & 101.89  & & -3.69  & & 0.66  & & 0.03  & & 0.02  \\
20121102A FAST & & 4.11  & & 1.09  & & 111.02  & & 72.20  & & -5.22  & & 1.19  & & 0.01  & & 0.00  \\
20121102A Lovell* & & 6.09  & & 0.82  & & 617.50  & & 320.28  & & -     & & -     & & -     & & - \\
20180301A FAST* & & 7.40  & & 0.66  & & 2037.87  & & 1384.06  & & -     & & -     & & -     & & - \\
20180916B CHIME & & 6.03  & & 0.63  & & 506.29  & & 417.98  & & -  & & -  & & -  & & -  \\
20180916B LOFAR* & & 7.68  & & 0.81  & & 3009.12  & & 1313.66  & & -     & & -     & & -     & & - \\
20180916B WSRT.Apertif & & 7.59  & & 0.78  & & 2664.54  & & 1338.21  & & -  & & -  & & -  & & -  \\
20181030A CHIME* & & 5.15  & & 0.85  & & 247.60  & & 81.49  & & -  & & -  & & -  & & -  \\
20190520B FAST & & 7.70  & & 2.30  & & 681.24  & & 1312.10  & & -3.22  & & 0.62  & & 0.05  & & 0.05  \\
20200120E Effelsberg & & 3.42  & & 1.33  & & 74.35  & & 37.59  & & 0.22  & & 0.63  & & 1.53  & & 9.20  \\
20201124A CHIME* & & 5.13  & & 0.63  & & 205.46  & & 185.05  & & -0.40  & & 0.90  & & 1.00  & & 0.33  \\
20201124A FAST & & 4.46  & & 0.96  & & 137.67  & & 111.52  & & -3.16  & & 0.72  & & 0.06  & & 0.04  \\
20201124A uGMRT & & 5.12  & & 0.99  & & 274.93  & & 283.16  & & -     & & -     & & -     & & - \\
\enddata
\tablecomments{
* The sample size is small.
}
\end{deluxetable}

\subsection{Peak Flux}
\label{peak flux}

The peak flux, which represents the intensity of a burst,
is a basic parameter for FRBs. Usually the fluence is
defined as the integration of flux over time, and then the
boxcar equivalent width is derived from the fluence and the
peak flux.

\begin{figure}[htbp]
\gridline{\fig{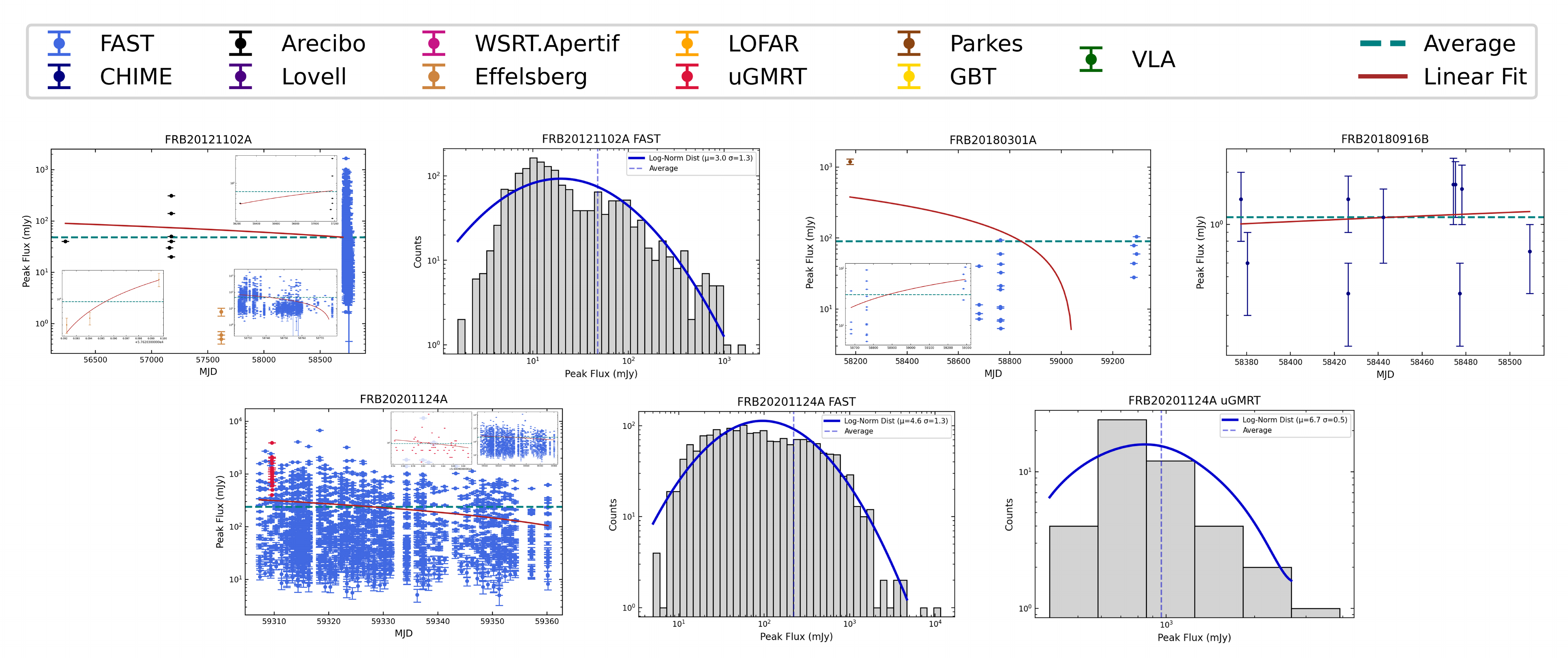}{1\textwidth}{}} \caption{The peak
flux plotted versus the MJD time for repeating FRBs. The inset
shows the observational data points of different telescopes
separately. The horizontal dashed line corresponds to the average
of peak flux for each source. The solid line is the best linear
fit to the data points. The distributions of the peak flux
observed by different telescopes are also illustrated via
histogram for each source.  } \label{figure15}
\end{figure}

The peak flux is plotted versus the MJD time for repeating FRBs in
Figure \ref{figure15}. The distribution of the peak flux is also
illustrated via histogram for each source. It could be seen that
the peak flux has a weak decreasing trend on a long-time
scale, especially for FRBs 20121102A ($Cor$ = -0.170,
$P$ = 2.233 $\times 10^{-25}$) and 20201124A ($Cor$ = -0.101, $P$ = 4.270 $\times 10^{-11}$),
which has been previously noted by \citet{111} on the bimodal
burst energy distribution of FRB 20121102A. In other words, strong
bursts generally appear in an early stage, and late bursts are
weaker in principle. Interestingly, for FRB 20121102A, Figure
\ref{figure15} (Row 1, Column 2) shows that the number of weak
bursts is slightly larger than the fitting curve, which is
different from that of FRB 20201124A (Row 2, Column 2, and Row 2,
column 3). In the future, when more FRBs are detected from these
two sources, the histogram should be re-examined again to see
whether it is still different for these two sources.

\begin{figure}[htbp]
\gridline{\fig{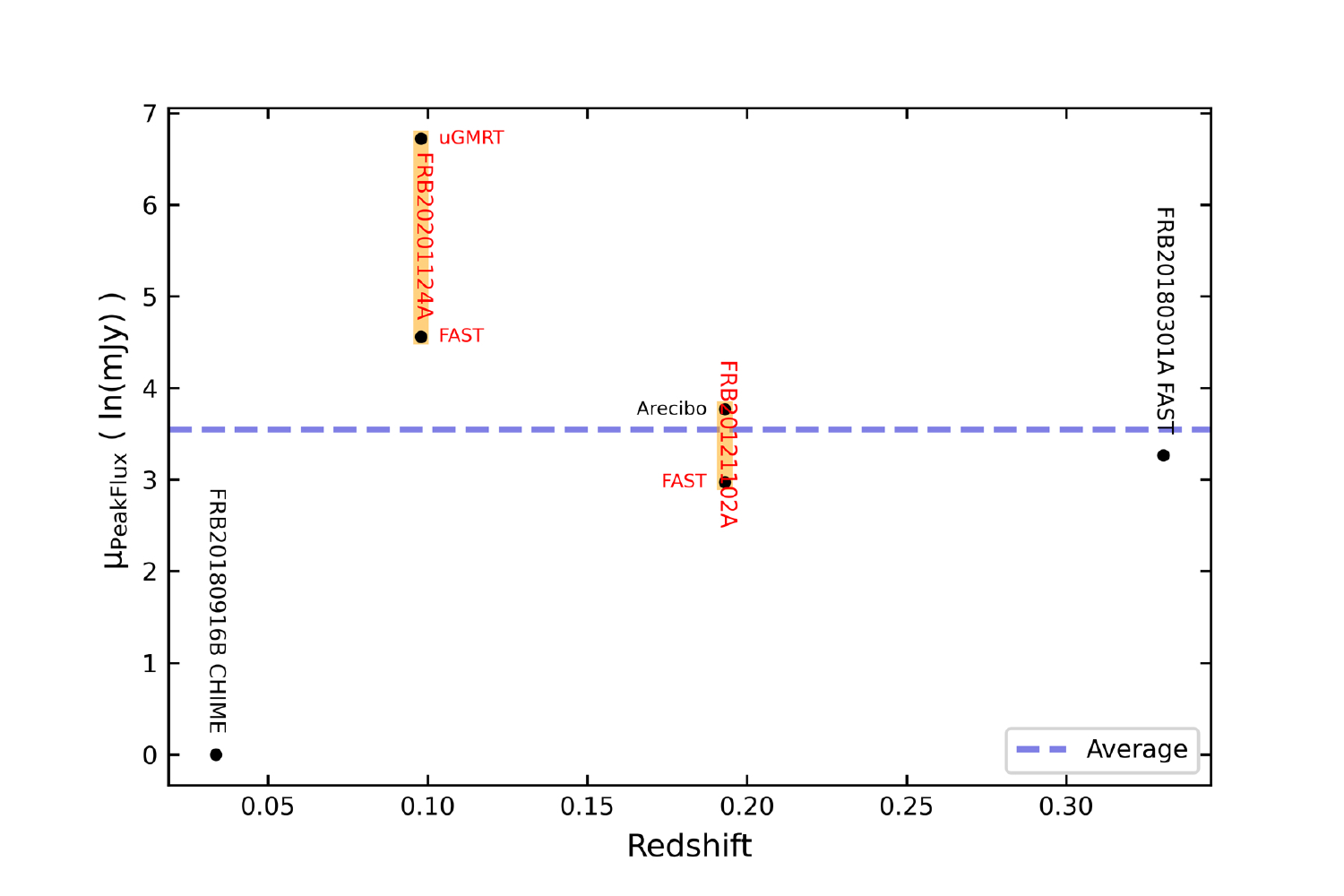}{0.5\textwidth}{}}
 \caption{ $\mu_{\rm PeakFlux}$ plotted versus the redshift
 for repeating FRBs. Here, $\mu_{\rm PeakFlux}$ is derived by fitting the
 distribution of peak flux with a lognormal function, which is a measure of
 the typical peak flux for the source. FRBs 20180916B and 20180301A are not
 used for statistical analysis due to the small amount of bursts with peak flux measurement, but are
 also shown in the figure for completeness.}
 \label{figure16}
\end{figure}

\begin{figure}[htbp]
\gridline{\fig{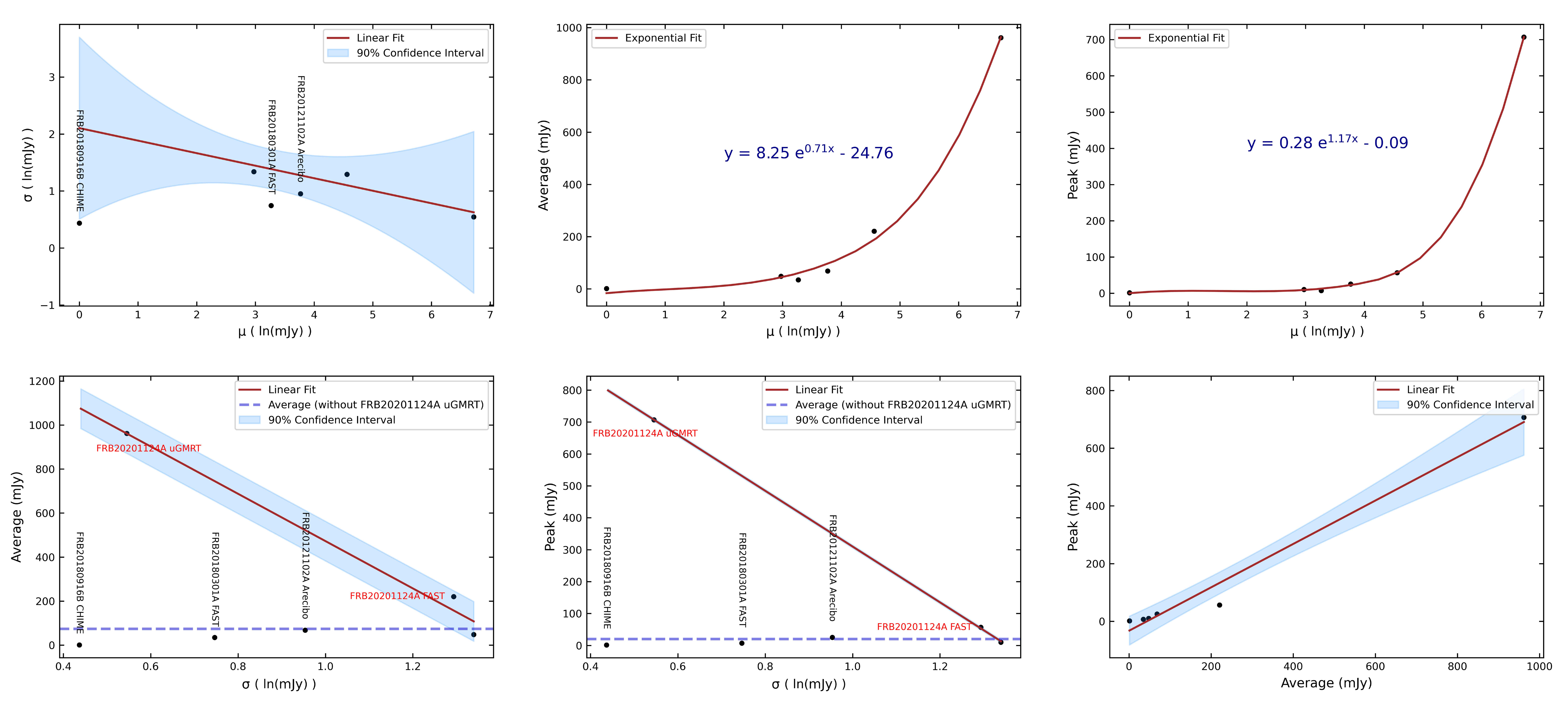}{1\textwidth}{}}
 \caption{Correlation between various statistical parameters derived by fitting the
peak flux distribution with a lognormal function for repeating
FRBs. $\mu$ and $\sigma$ characterize the typical peak flux and
the dispersion range, respectively. $Average$ stands for the mean
value of peak flux. $Peak$ represents the value observed most
frequently. The FRBs annotated in black are not used for
statistical analysis due to the small amount of bursts with peak flux measurement, but are
also shown in the figure for completeness. }
 \label{figure17}
\end{figure}

In order to compare the characteristics of peak flux in different
repeating sources, we still use a lognormal function to fit its
distribution. The derived parameters are presented in Table
\ref{table7}. Here, $\mu_{\rm PeakFlux}$ is a measure of the typical
peak flux. $\mu_{\rm PeakFlux}$ is plotted versus the redshift for
repeating FRBs in Figure \ref{figure16}. It is still difficult to
determine whether there is a correlation between $\mu_{\rm PeakFlux}$
and redshift or not, due to the very limited data points in the
figure. Correlations between various statistical parameters
concerning the peak flux are shown in Figure \ref{figure17}. We
see that the dispersion range of the peak flux is negatively
correlated with the average peak flux
($Cor$ = -1.000, $P$ = 0.333). It can be caused
by the observational selection effect of telescope sensitivity.
For a more sensitive telescope, much more weaker bursts can be
detected, resulting in a smaller average peak
flux and a larger dispersion range.

\begin{deluxetable}{ccccccccc}[htbp]
\tabletypesize{\scriptsize} \tablecaption{ Parameters derived by
fitting the peak flux distribution with a lognormal function for
repeating FRBs. $\mu$ and $\sigma$ characterize the typical peak
flux and the dispersion range, respectively. $Average$ stands for
the mean value of peak flux. $Peak$ represents the value observed
most frequently. }
\label{table7}
\tablehead{
    \colhead{$\rm FRB$}   & &
    \colhead{$\mu$}  & &
    \colhead{$\sigma$}  & &
    \colhead{$\rm Average$}  & &
    \colhead{$\rm Peak$}  \\
    \colhead{}  & &
    \colhead{($\rm ln(mJy)$)}  & &
    \colhead{($\rm ln(mJy)$)}  & &
    \colhead{($\rm mJy$)}  & &
    \colhead{($\rm mJy$)}
}
\startdata
20121102A Arecibo* & & 3.77  & & 0.95  & & 68.18  & & 25.13  \\
20121102A FAST & & 2.97  & & 1.34  & & 47.92  & & 10.12  \\
20180301A FAST* & & 3.27  & & 0.75  & & 34.70  & & 7.14  \\
20180916B CHIME* & & 0.00003  & & 0.44  & & 1.10  & & 1.51  \\
20201124A FAST & & 4.56  & & 1.29  & & 220.64  & & 56.81  \\
20201124A uGMRT & & 6.72  & & 0.55  & & 961.70  & & 706.82  \\
\enddata
\tablecomments{
* Sample size is small.
}
\end{deluxetable}

\subsection{Fluence}
\label{fluence}

For an FRB source, the fluence represents the energetics of a
burst. As mentioned earlier in Section (\ref{bandwidth}), the
measured fluence are affected by the passband, the detection
threshold, and the instrument sensitivity of telescopes. The
energy release of an FRB can be calculated through their fluence
as
\begin{eqnarray}
\label{formula14}
 E=\int_{\nu}{\frac{4\pi}{1+z}D_{\rm L}^2F_{\rm \nu}} d \nu,
\end{eqnarray}
Where $D_{\rm L}$ is the luminosity distance and $F_{\rm \nu}$ is the fluence
at frequency $\nu$. \citet{175} suggested to use the center
frequency of telescope passband to calculate the burst energy.
Using this method, \citet{111} argued that the burst energy of FRB
20121102A shows a bimodal distribution.

\begin{figure}[htbp]
\gridline{\fig{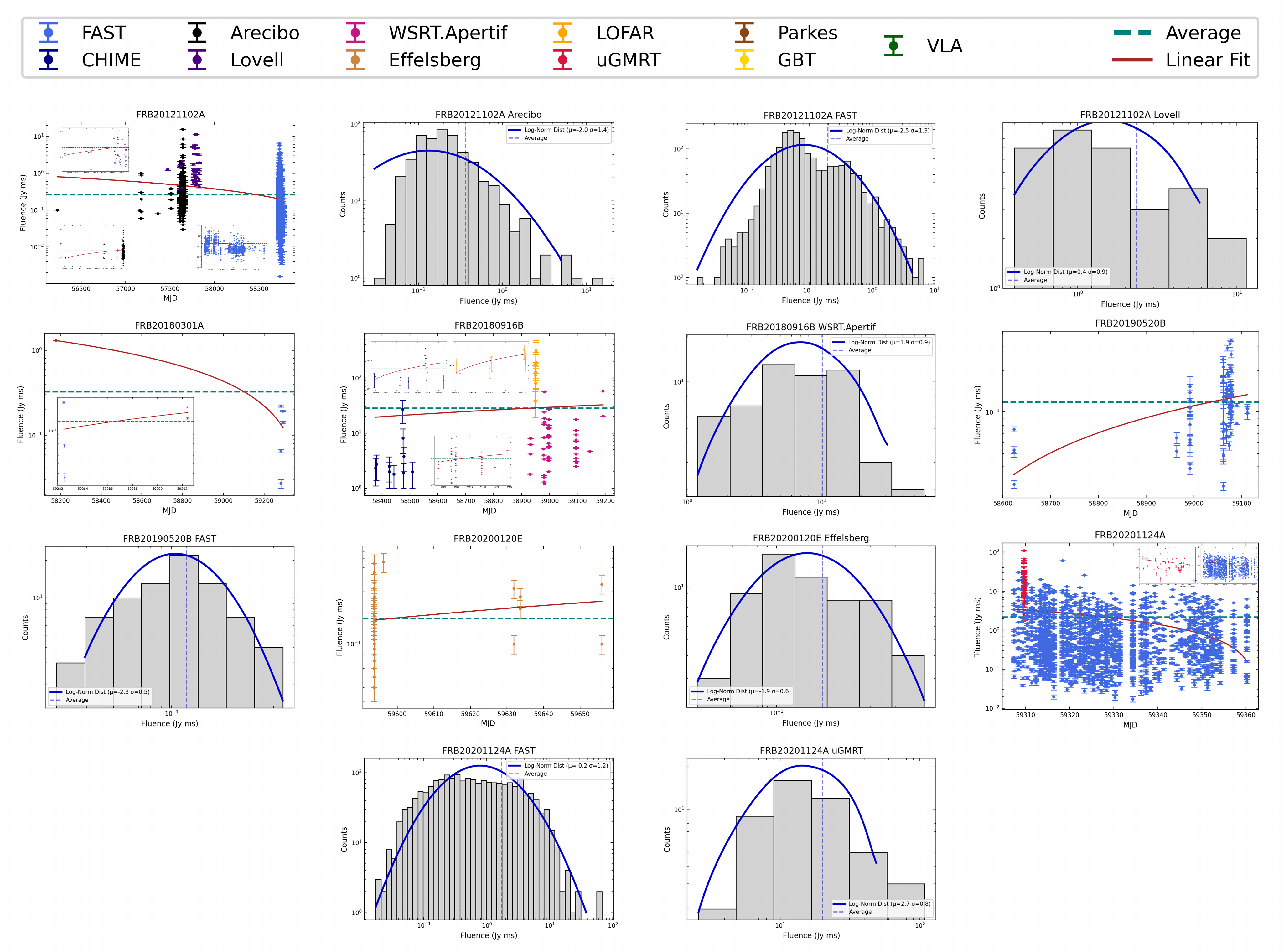}{1\textwidth}{}}
 \caption{ The fluence plotted versus time for repeating FRBs. The horizontal
 dashed line corresponds to the average fluence of each FRB. The solid line
 is the best linear fit to the data points. The inset shows the observational
 data points from different telescopes separately. For each source, the
 distribution of the fluence is also illustrated via histogram according to
 the detection telescope. }
\label{figure18}
\end{figure}

Figure \ref{figure18} plots the fluence versus the MJD time for
repeating FRB sources. We see that the fluence generally does not
have any obvious evolution on a long-time scale, which is slightly
different to the behavior of peak flux. Comparing Figure
\ref{figure18} with Figure \ref{figure15}, it could be seen that
the amplitude of fluence variation over time is actually very
small. Meanwhile, the distribution of the fluence is well fitted
by a lognormal function. Like the peak flux in Section (\ref{peak
flux}), for FRB 20121102A, the number of low fluence bursts is
slightly higher than the fitting curve for both the Arecibo
observations (Figure \ref{figure18}: Row 1, Column 2) and the FAST
observations (Figure \ref{figure18}: Row 1, Column 3). Note that the deadline of the observational data used in this
study is June 26, 2022. The recent bursts from FRB 20201124A detected
by FAST \citep{179} and the bursts from FRB 20121102A detected by
Arecibo \citep{195} are not included in our analysis.

\begin{figure}[htbp]
\gridline{\fig{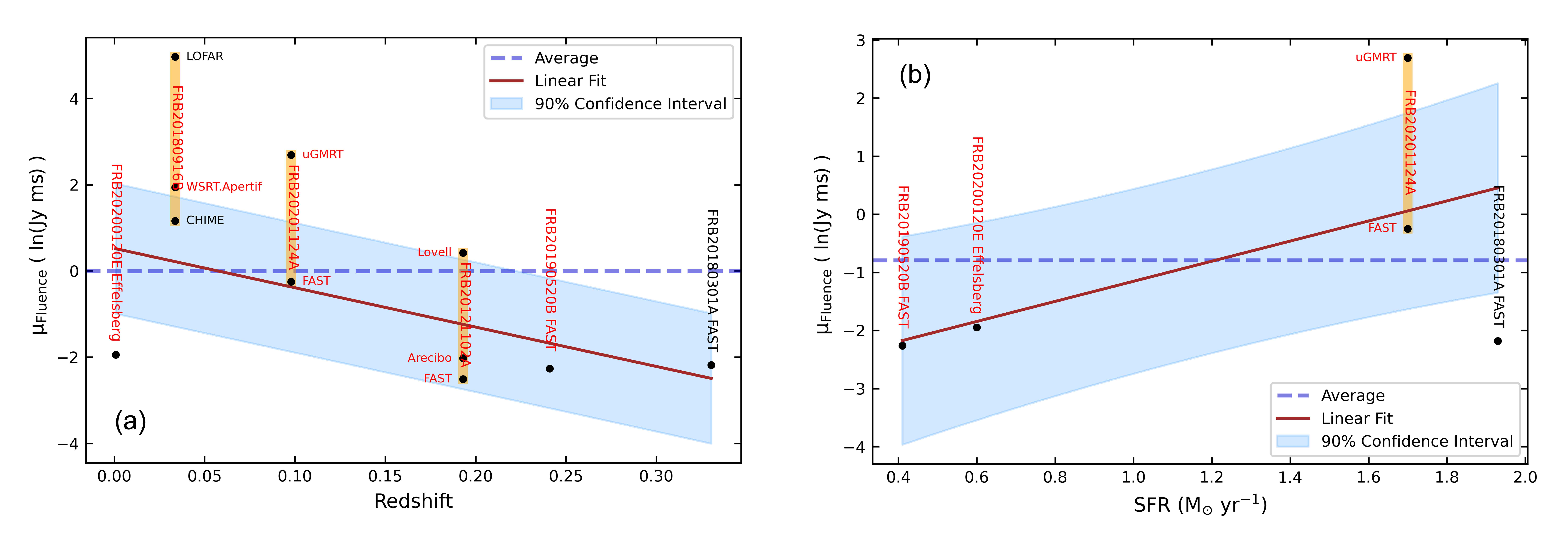}{1\textwidth}{}}
 \caption{Correlation between the redshift, SFR and the typical
fluence ($\mu_{\rm Fluence}$) for repeating FRB sources. Panel (a):
$\mu_{\rm Fluence}$ versus redshift. Panel (b): $\mu_{\rm Fluence}$ versus
SFR. $\mu_{\rm Fluence}$ is derived by fitting the distribution of
fluence with a lognormal function for each FRB, which represents
the typical fluence. FRB 20180301A is not used for statistical
analysis due to the small sample size, but is also shown in the
figure for completeness.}
 \label{figure19}
\end{figure}

A lognormal function is used to fit the distribution of fluence.
The derived parameters are presented in Table \ref{table8}. Here,
$\mu_{\rm Fluence}$ is a measure of the typical fluence. It is plotted
versus the redshift and SFR for repeating FRBs in Figure
\ref{figure19}. We see that $\mu_{\rm Fluence}$ is negatively
correlated with the redshift ($Cor$ = -0.582, $P$ = 0.081).
It indicates that the intrinsic energies released by the
sources are distributed in a relatively narrow range, which leads to the fact
that the characteristic fluence is lower for a distant source.
$\mu_{\rm Fluence}$ is positively correlated with
SFR ($Cor$ = 0.913, $P$ = 0.071). Such a positive correlation is likely due to
the evolution of source features at high distance scales.
As discussed in Section (\ref{dispersion measure}), Figure \ref{figure2}(b)
indicates that younger magnetars which are more active due to the action of
their strong magnetic fields are more likely found to be associated with a
higher SFR. The possible link between the magnetic field strength
and the mean FRB fluence still needs further investigation in the future.
If confirmed, it would shed new light on the trigger mechanism and
the emission process of FRBs.
Note that the data point corresponding to uGMRT detection of FRB
20201124A seems to be an outlier in both panels of Figure
\ref{figure19}. Its value of $\mu_{\rm Fluence}$ is large, which may
be due to the relatively higher detection threshold of uGMRT \citep{85}.

\begin{figure}[htbp]
\gridline{\fig{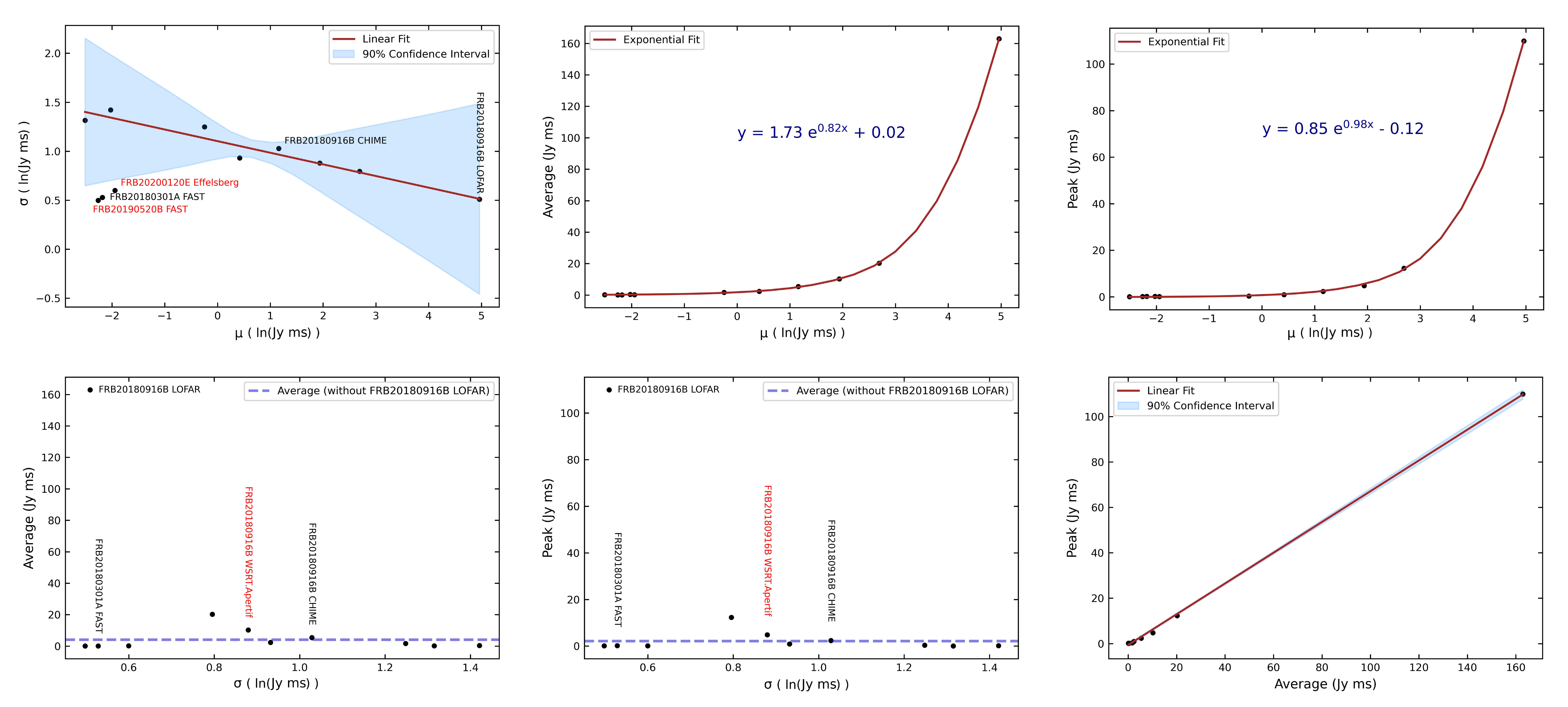}{1\textwidth}{}}
 \caption{ Correlation between various statistical parameters derived by
fitting the fluence distribution with a lognormal function for
repeating FRBs. $\mu$ and $\sigma$ characterize the typical
fluence and the dispersion range, respectively. $Average$ stands
for the mean value of fluence. $Peak$ represents the value
observed most frequently. The FRBs annotated in black are not used
for statistical analysis due to the small amount of bursts with fluence measurement, but
are also shown in the figure for completeness. }
 \label{figure20}
\end{figure}

Relations between various statistical parameters concerning the
fluence are shown in Figure \ref{figure20}. The average fluence is
positively correlated with the peak. The dispersion range of fluence
is negatively correlated with the typical fluence
($Cor$ = -0.905, $P$ = 0.003) (Figure \ref{figure20}: Row 1,
Column 1), but note that FRB 20190520B is still an obvious outlier in the
plot. Such a negative correlation may be due to the observational
selection effect. In fact, for a telescope with a relatively higher
detection threshold, it would be impossible for it to detect low fluence
bursts. As a result, the dispersion range of observed fluence will be
suppressed. In other words, it will give a higher mean fluence, together
with a narrower dispersion range.

\begin{deluxetable}{ccccccccc}[htbp]
\tabletypesize{\scriptsize}
 \tablecaption{ Parameters derived by
fitting the fluence distribution with a lognormal function for
repeating FRBs. $\mu$ and $\sigma$ characterize the typical peak
flux and the dispersion range, respectively. $Average$ stands for
the mean value of peak flux. $Peak$ represents the value observed
most frequently. }
 \label{table8}
 \tablehead{
    \colhead{$\rm FRB$}   & &
    \colhead{$\mu$}  & &
    \colhead{$\sigma$}  & &
    \colhead{$\rm Average$}  & &
    \colhead{$\rm Peak$}  \\
    \colhead{}  & &
    \colhead{($\rm ln(Jy\ ms)$)}  & &
    \colhead{($\rm ln(Jy\ ms)$)}  & &
    \colhead{($\rm Jy\ ms$)}  & &
    \colhead{($\rm Jy\ ms$)}
}
\startdata
20121102A Arecibo & & -2.03  & & 1.42  & & 0.36  & & 0.19  \\
20121102A FAST & & -2.51  & & 1.32  & & 0.19  & & 0.05  \\
20121102A Lovell & & 0.42  & & 0.93  & & 2.35  & & 0.92  \\
20180301A FAST* & & -2.18  & & 0.53  & & 0.13  & & 0.18  \\
20180916B CHIME* & & 1.16  & & 1.03  & & 5.41  & & 2.36  \\
20180916B LOFAR* & & 4.96  & & 0.51  & & 162.89  & & 109.93  \\
20180916B WSRT.Apertif & & 1.94  & & 0.88  & & 10.21  & & 4.80  \\
20190520B FAST & & -2.26  & & 0.50  & & 0.12  & & 0.11  \\
20200120E Effelsberg & & -1.94  & & 0.60  & & 0.17  & & 0.10  \\
20201124A FAST & & -0.25  & & 1.25  & & 1.70  & & 0.34  \\
20201124A uGMRT & & 2.69  & & 0.80  & & 20.21  & & 12.28  \\
\enddata
\tablecomments{
* Sample size is small.
}
\end{deluxetable}


\section{Double-Parameter Analysis}
\label{sec4:double-parameter analysis}

In this section, we analyze the correlation between various parameter pairs
among $\rm DM$, $\rm RM$, bandwidth, pulse width, waiting time, peak flux
and fluence, hoping to put further constraints on the
triggering and emission mechanism of repeating FRBs.

\subsection{Pulse Width, Peak Flux and Fluence}
\label{pulse width, peak flux and fluence}

The pulse width, peak flux and fluence are three basic parameters of FRBs,
which are strongly connected with the energetics of the burst.
The fluence is approximately a product of the width and the peak flux.
In Figure \ref{appxfigure1}, FRBs are plotted on the peak flux - pulse width
plane. We see that there is no obvious correlation between these two parameters,
which means that the peak flux and the width are two independent parameters.
An interesting feature is that, for each FRB source, the peak flux typically
varies in a wide range (spanning for more than two magnitudes), but the width
only varies in a much narrower range (spanning for less than one magnitude).
This is especially clear for the two most active sources of FRBs 20121102A
and 20201124A. It could also be due to the observational selection
effect that weak bursts with small pulse width are more difficult to be detected.
In Figure \ref{appxfigure2}, FRBs are plotted on the width - fluence plane.
There is a weak positive correlation between the width and fluence for each
source (values of $Cor$ and $P$ are marked in the figure).
It is mainly due to the fact that the fluence is generated by
multiplying the width with the peak flux.
\citet{85} reported the positive correlation between the width
and fluence for FRB 20201124A initially in their work. Note that \citet{176}
suggested an alternative explanation for the weak correlation between pulse
width and fluence for FRB 20121102A, namely the bursts may be divided into
two subtypes according to their brightness temperature and may be produced
by different physical processes. FRBs are plotted on the peak
flux - fluence plane in Figure \ref{appxfigure3}. We could see a clear
positive correlation between the peak flux and the fluence
(values of $Cor$ and $P$ are marked in the figure). Again, the
reason is that the fluence is a product of the peak flux and the width, with
the width varying only in a narrow range.

\begin{figure}[htbp]
\gridline{\fig{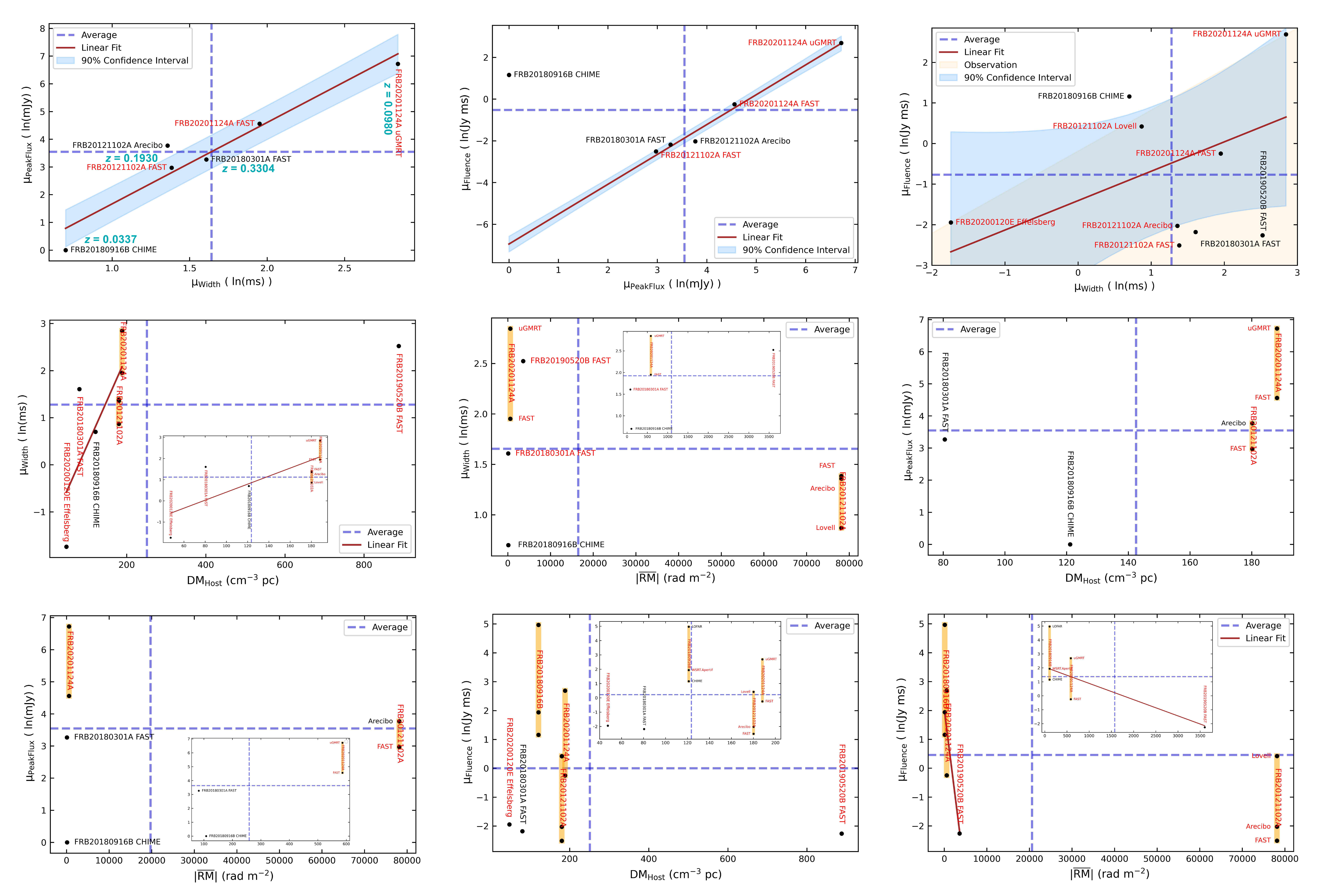}{1\textwidth}{}}
 \caption{ Correlation between the typical pulse width ($\mu_{\rm Width}$),
 peak flux ($\mu_{\rm PeakFlux}$) and fluence ($\mu_{\rm Fluence}$) for repeating
 FRB sources. These parameters are also plotted versus ${\rm DM}_{\rm Host}$ and
 $|\overline{\rm RM}|$. Here, ${\rm DM}_{\rm Host}$ represents the DM component
 contributed by the host galaxy and the local circum-burst environment.
 The FRBs annotated in black are not used for the double-parameter analysis
 due to the small sample size, but are also shown in the figure for completeness. }
\label{figure21}
\end{figure}

Correlations between the typical pulse width ($\mu_{\rm Width}$), peak flux
($\mu_{\rm PeakFlux}$) and fluence ($\mu_{\rm Fluence}$) are plotted in
Figure \ref{figure21} (Row 1) for repeating FRB sources. These parameters
are also plotted versus ${\rm DM}_{\rm Host}$ and $|\overline{\rm RM}|$.
There is a positive correlation between the typical peak flux and pulse
width ($Cor$ = 1.000, $P$ = 0.333)
(Figure \ref{figure21}: Row 1, Column 1).
Note that the scattering effect caused by electrons during the
propagation of radio waves would make the pulse width larger and the
peak flux smaller, which then should lead to a negative correlation between
the peak flux and pulse width. Therefore, the positive correlation shown
in Figure \ref{figure21} should not be due to the propagation effect. It
might be connected with the trigger mechanism. For example, repeating FRBs
could be produced by multiple collisions of asteroids with a neutron
star \citep{20}.
According to the calculations of \citet{19}, when the asteroid mass is larger,
it would produce a stronger FRB with both a high peak flux and a large pulse
width. Thus a positive correlation between the peak flux and the pulse width is naturally expected.
$\mu_{\rm PeakFlux}$ and $\mu_{\rm Fluence}$ are tightly
correlated ($Cor$ = 1.000, $P$ = 0.333) as shown in
Figure \ref{figure21} (Row 1, Column 2), while the correlation between
$\mu_{\rm Width}$ and $\mu_{\rm Fluence}$ is very weak ($Cor$ = 0.200, $P$ = 0.719)
(Figure \ref{figure21}: Row 1, Column 3). Again it is mainly due to
the fact that $\mu_{\rm PeakFlux}$ varies in a wide range, but $\mu_{\rm Width}$ is
distributed in a much narrower range. As a result, the influence of
peak flux on fluence is more significant than that of pulse width. Additionally,
$\mu_{\rm Width}$ is affected by cosmological expansion and propagation effects
(see Section \ref{pulse width}), which further interferes the correlation
between $\mu_{\rm Width}$ and $\mu_{\rm Fluence}$.
Note that the linear correlation between $\mu_{\rm Width}$, $\mu_{\rm PeakFlux}$ and
$\mu_{\rm Fluence}$ actually corresponds to a power-law function among the
average of width, peak flux and fluence.

Figure \ref{figure21} (Row 2, Column 1) shows that $\mu_{\rm Width}$
is positively correlated with ${\rm DM}_{\rm Host}$ for most FRB
sources ($Cor$ = 0.582, $P$ = 0.081).
This is not strange, since a larger $\rm DM$ tends to broaden the
pulse width. However, note that FRB 20190520B markedly deviates
from this general trend, the reason of which is still unknown.

\subsection{Bandwidth, Pulse Width and Waiting Time}
\label{bandwidth, pulse width and waiting time}

\begin{figure}[htbp]
\gridline{\fig{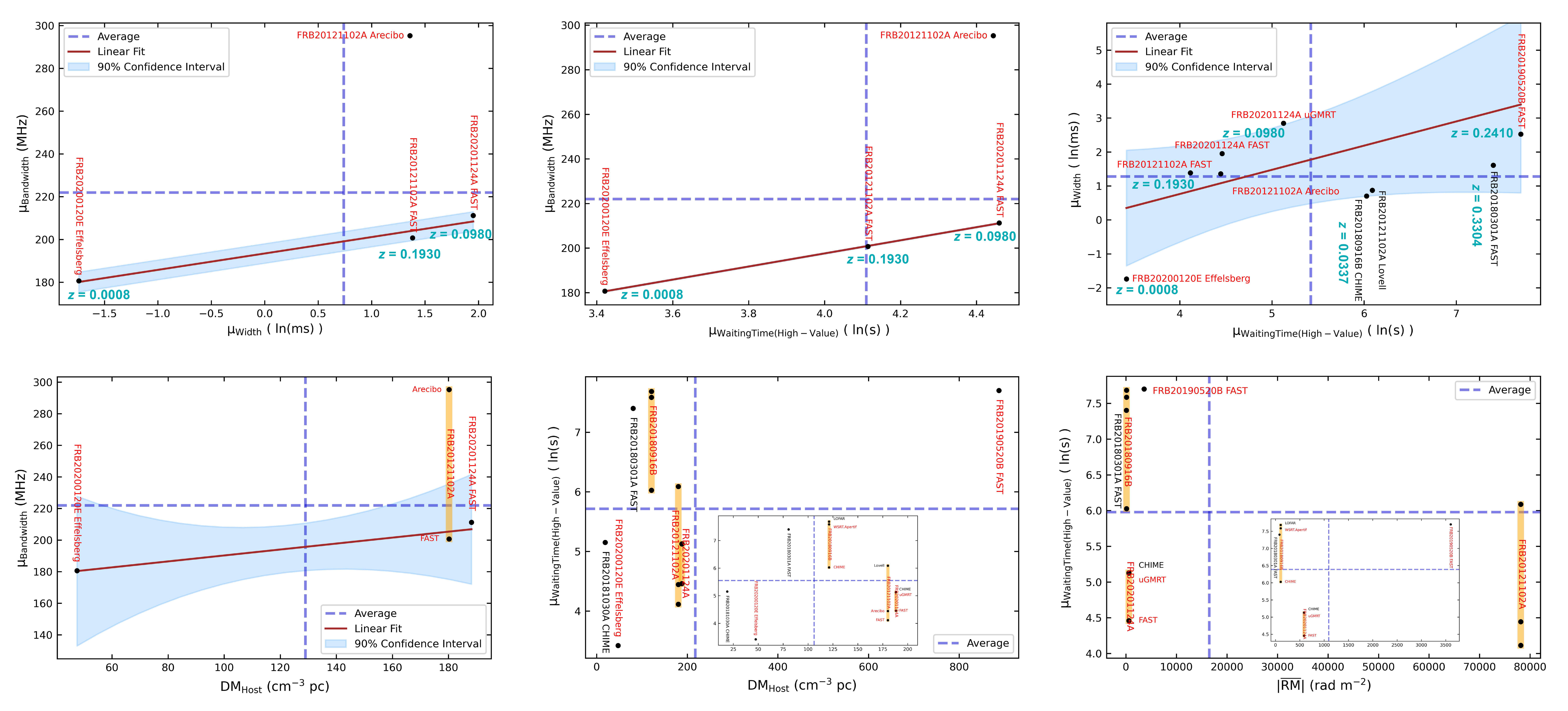}{1\textwidth}{}}
 \caption{ Correlation between the typical bandwidth ($\mu_{\rm Bandwidth}$),
 pulse width ($\mu_{\rm Width}$) and waiting time ($\mu_{\rm WaitingTime(High-Value)}$)
 for repeating FRB sources. These parameters are also plotted versus ${\rm DM}_{\rm Host}$
 and $|\overline{\rm RM}|$. Here, ${\rm DM}_{\rm Host}$ represents the DM component
 contributed by the host galaxy and the local circum-burst environment.
 The FRBs annotated in black are not used for the double-parameter analysis
 due to the small sample size, but are also shown in the figure for completeness.
 }
\label{figure22}
\end{figure}

The bandwidth, pulse width and waiting time are all influenced by cosmological
expansion, and might also be related to potential redshift-dependent evolution
of the central engine (see Sections \ref{bandwidth}, \ref{pulse width},
\ref{waiting time}). Correlations between the typical bandwidth ($\mu_{\rm Bandwidth}$),
pulse width ($\mu_{\rm Width}$) and waiting time ($\mu_{\rm WaitingTime(High-Value)}$)
are illustrated in Figure \ref{figure22} for repeating FRB sources. These
parameters are also plotted versus ${\rm DM}_{\rm Host}$ and $|\overline{\rm RM}|$.
From Row 1 of this figure, we see that $\mu_{\rm Bandwidth}$, $\mu_{\rm Width}$ and
$\mu_{\rm WaitingTime(High-Value)}$ are positively correlated with each other.
Note that the $\mu_{\rm Bandwidth}$ - $\mu_{\rm Width}$ correlation ($Cor$ = 1.000, $P$ = 0.333)
(Figure \ref{figure22}: Row 1, Column 1) and the $\mu_{\rm Bandwidth}$ -
$\mu_{\rm WaitingTime(High-Value)}$ correlation ($Cor$ = 1.000, $P$ = 0.333) (Figure \ref{figure22}: Row 1,
Column 2) are both very tight, except that the Arecibo data point of
FRB 20121102A seems to be an obvious outlier in the two panels.
The deviation of FRB 20121102A (Arecibo observations) may be due to the
relatively wider passband of Arecibo, which is 1.15 --- 1.73 $\rm GHz$).
As a comparison, the passband of Effelsberg is 1.2 --- 1.6 $\rm GHz$
and the band of FAST is 1.05 --- 1.45 $\rm GHz$ (see Section \ref{bandwidth}).
However, the $\mu_{\rm Width}$ - $\mu_{\rm WaitingTime(High-Value)}$ correlation
is very dispersive ($Cor$ = 0.733, $P$ = 0.056) as compared with the above two correlations. It is not
clear whether such a dispersion is due to the increased sample size or not.
In the future, more observational data points are needed to further clarify
this issue. In Figure \ref{figure22} (Row 2, Column 1), there is a weak
correlation between $\mu_{\rm Bandwidth}$ and ${\rm DM}_{\rm Host}$
($Cor$ = 1.000, $P$ = 0.333). In other panels of
Figure \ref{figure22}, no clear correlation is observed.

\subsection{Bandwidth, Waiting Time and Peak Flux, Fluence}
\label{bandwidth, waiting time and peak flux, fluence}

\begin{figure}[htbp]
\gridline{\fig{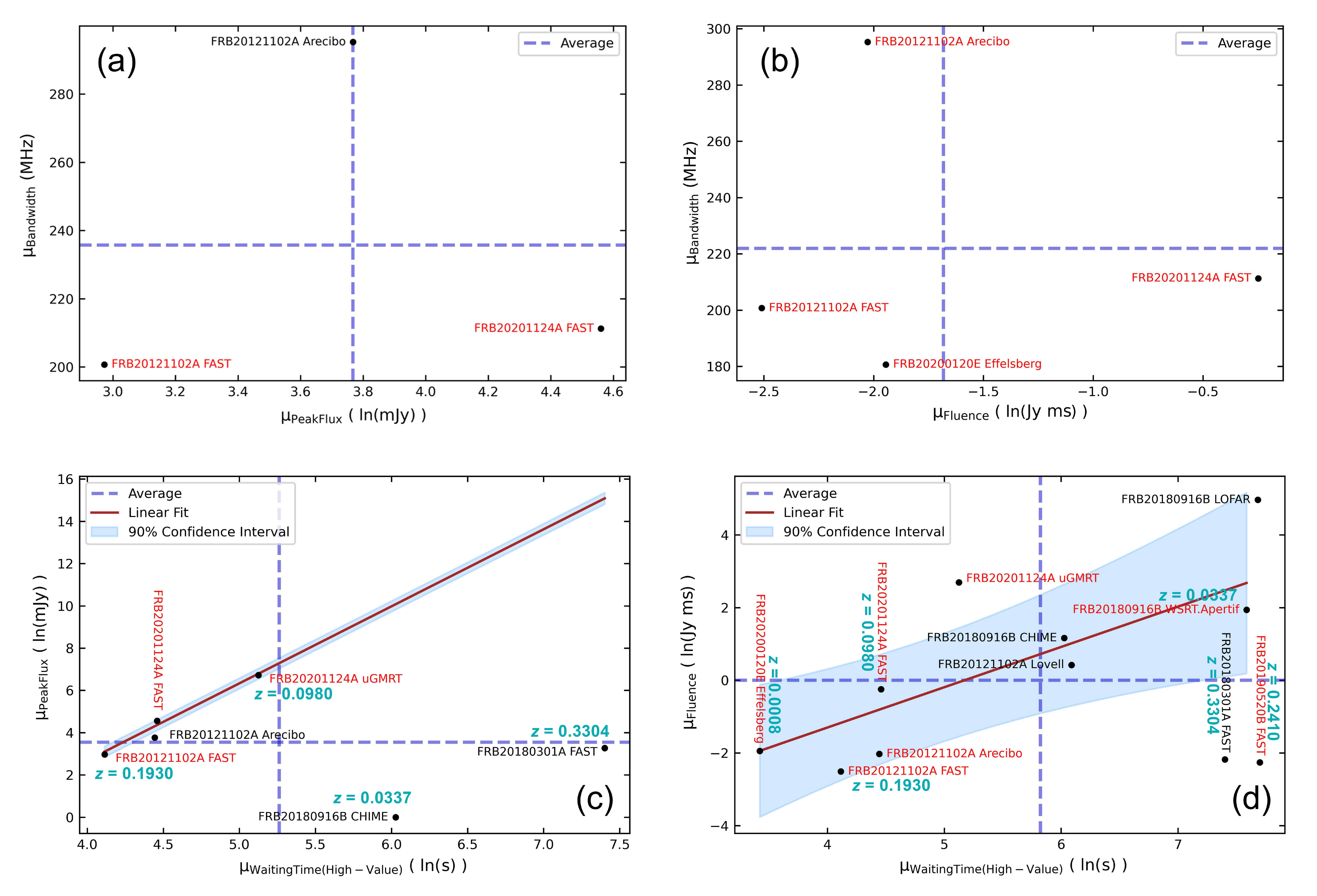}{1\textwidth}{}}
 \caption{ Correlation between the typical bandwidth ($\mu_{\rm Bandwidth}$),
 waiting time ($\mu_{\rm WaitingTime(High-Value)}$), peak flux ($\mu_{\rm PeakFlux}$),
 and fluence ($\mu_{\rm Fluence}$) for repeating FRB sources.
 The FRBs annotated in black are not used for the double-parameter analysis
 due to the small sample size, but are also shown in the figure for completeness.
 }
\label{figure23}
\end{figure}

\begin{figure}[htbp]
\gridline{\fig{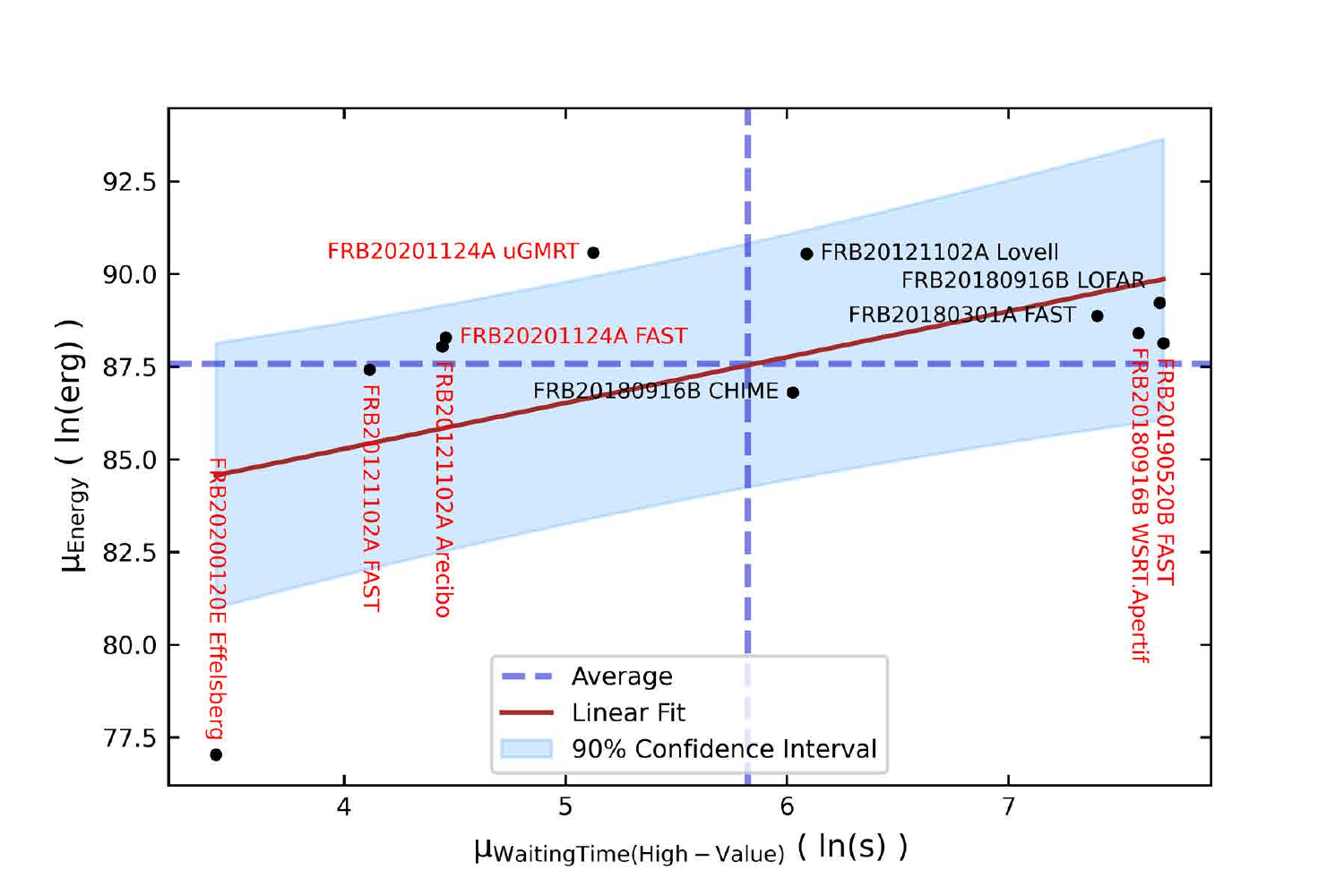}{0.7\textwidth}{}}
 \caption{ Correlation between the typical energy ($\mu_{\rm
 Energy}$) and waiting time ($\mu_{\rm WaitingTime(High-Value)}$) for
 repeating FRB sources.
 The FRBs annotated in black are not used for the double-parameter analysis
 due to the small sample size of their bursts, but are shown here for
 completeness.
 }
\label{figure24}
\end{figure}

Correlations between the typical bandwidth ($\mu_{\rm Bandwidth}$),
waiting time ($\mu_{\rm WaitingTime(High-Value)}$), and peak flux ($\mu_{PeakFlux}$), fluence ($\mu_{\rm Fluence}$) are illustrated in Figure \ref{figure23}
for repeating FRB sources. In Panels (a) and (b), since the observational
data points are too few, no firm conclusion could be drawn on the
correlation between the corresponding parameter pairs. In Panel (c),
there seems to be a positive correlation between
$\mu_{\rm WaitingTime(High-Value)}$ and $\mu_{\rm PeakFlux}$ ($Cor$ = 1.000, $P$ = 0.333), especially
when the data point corresponding to FAST observations of FRB 20180301A is
expelled due to its small sample size. However, be aware that the available
data points are still very limited in this panel. A positive
correlation could also be interestingly observed between
$\mu_{\rm WaitingTime(High-Value)}$ and $\mu_{\rm Fluence}$
($Cor$ = 0.600, $P$ = 0.136) in Panel (d),
and FRB 20190520B is still an outlier. It is
noteworthy that $\mu_{\rm WaitingTime(High-Value)}$ is positively
correlated with redshift in Section (\ref{waiting time}) and
$\mu_{\rm Fluence}$ is negatively correlated with redshift in
Section (\ref{fluence}). However, there is a positive correlation
between $\mu_{\rm WaitingTime(High-Value)}$ and $\mu_{\rm Fluence}$. It means
that for different repeaters, when the high-value waiting time is larger, the
energetics of bursts is also larger. Such a positive correlation is not
dominated by the distance, but should be connected with the triggering mechanism
of repeating FRBs. There is also a positive correlation
between $\mu_{\rm Energy}$ and $\mu_{\rm WaitingTime(High-Value)}$
($Cor$ = 0.619, $P$ = 0.069) as shown in Figure \ref{figure24}.
Here $\mu_{\rm Energy}$ is calculated from $\mu_{\rm Fluence}$ by using Equation
(\ref{formula14}), which characterizes the typical burst energy of
the FRB sources. Such a correlation indicates that when the waiting time is
longer, the burst energy will also be relatively larger. It could place
interesting constraints on the trigger mechanism of FRBs.
The typical burst energies are distributed in
a relatively narrow range (also see Section (\ref{fluence})). It is interesting
to note that the typical energy of FRB 20190520B does not differ significantly
from other repeaters, which strongly indicates that its central engine
should also be similar to others'. The special features of FRB 20190520B in
DM, RMSE of DM, DM$_{\rm host}$ and RM may be due to the different conditions
of its local environment \citep{127}.

\subsection{Dispersion Measure and Rotation Measure}
\label{dispersion measure and rotation measure}

$\rm DM$ and $\rm RM$ can reflect the physical conditions of the
surrounding media around the source region. For repeating FRBs
with a measured redshift, we can estimate ${\rm DM}_{\rm Host}$ based on
the Macquart ${\rm DM}_{\rm IGM}-z$ relation and the MW free electron
models such as NE2001 or YMW16 (see Section \ref{dispersion
measure}). Note that ${\rm DM}_{\rm Host}$ refers to the $\rm DM$ contributed
by the host galaxy and the local circum-burst environment of the
FRB source.

\begin{figure}[htbp]
\gridline{\fig{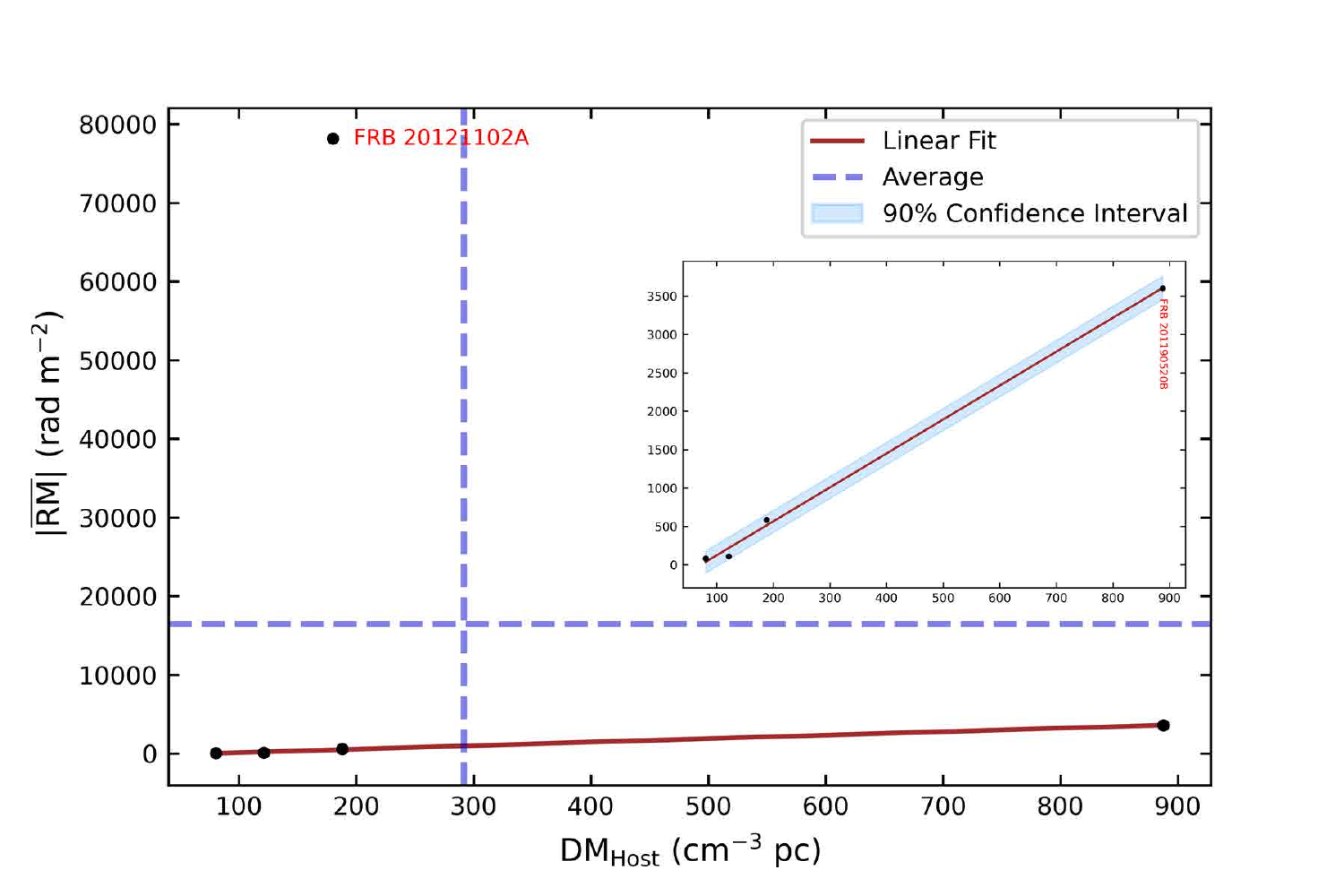}{0.5\textwidth}{}}
 \caption{ Correlation between $|\overline{\rm RM}|$ and ${\rm DM}_{\rm Host}$ for
 repeating FRBs. ${\rm DM}_{\rm Host}$ represents the DM component
 contributed by the host galaxy and the local circum-burst environment.
}
\label{figure25}
\end{figure}

Correlation between $|\overline{\rm RM}|$ and ${\rm DM}_{\rm Host}$ are
plotted in Figure \ref{figure25} for repeating FRBs. It is
striking that these two parameters are tightly
correlated ($Cor$ = 1.000, $P$ = 0.083) and can
be well fitted by a linear function as
\begin{eqnarray}
\label{formula15} \left(\frac{|\overline{\rm RM}|}{\rm rad\ m^{-2}} \right)=4.420\left(\frac{{\rm DM}_{\rm Host}}{\rm cm^{-3}\ pc} \right)-316.898.
\end{eqnarray}
From Equation (\ref{formula11}), we know that $\overline{\rm RM}$
should be proportional to ${\rm DM}_{\rm Host}$. The correlation in
Figure \ref{figure25} further shows that all these repeaters follow the
same linear function, which strongly indicates that the mean magnetic field in their
local environment should not vary too much with respect to each other.
However, FRB 20121102A deviates from the best fit line
markedly. Note that it is also an outlier in
Figure \ref{figure21} (Row 3, Column 3). Particularly,
comparing with other repeaters, FRB 20121102A is the first
repeater discovered by observers and has the longest observation
history. Its evolution over time thus can be better assessed.
FAST recently detected a series of bursts from FRB
20121102A again, with the $\rm DM$ being found to be significantly
smaller \citep{180}, which makes the deviation even more
significant. It seems that this repeater may have a much stronger
magnetic field. Such a conclusion is supported by some
recent researches that attempt to directly constrain the
strength of the magnetic field in the local environment of
repeaters (See Table 3 of \citet{205}).

The linear correlation between $|\overline{\rm RM}|$ and ${\rm DM}_{\rm Host}$
can potentially help us estimate the distance of those FRBs
whose redshift is not measured. For this purpose, we first need to
measure the $|\overline{\rm RM}|$. Then the $|\overline{\rm RM}|$ -
${\rm DM}_{\rm Host}$ can be applied to estimate the ${\rm DM}_{\rm Host}$.
Finally, we can calculate ${\rm DM}_{\rm IGM}$ and use it to estimate the
distance. However, this approach is mainly limited to repeating
FRBs, because measuring $|\overline{\rm RM}|$ requires observations
over long-time scales.
Specifically, the RM of repeating FRBs varies over time
so the determination of $\overline{\rm RM}$, the statistical mean value
of $\rm RM$, requires multiple measurements spanned
in a long period.


In our study, we have also explored possible correlations between
many other parameter pairs, but fail to find any clear correlation
among them. More double-parameter distributions of individual FRBs where no
correlation is observed are illustrated in the Appendix Figures of
\ref{appxfigure4} --- \ref{appxfigure27}.


\section{Conclusions and Discussion}
\label{sec5:conclusions and discussion}

In this study, a sample of 16 FRB repeaters is collected, with
at least five bursts being detected from each source. Among them,
four repeaters, i.e. FRBs 20121102A, 20180916B, 20190520B and
20201124A, are the most active sources. More than 5,000 bursts are
included, which are observed by 11 different telescopes. Data of
many key parameters are collected from the literature, including
the arrival time (MJD), pulse width, dispersion measure, Faraday
rotation measure, bandwidth, waiting time, peak flux and
fluence. A thorough statistical investigation is performed based
on the available data and the behaviors of different repeaters are
compared. The DM of FRB 20121102A is found to increase
continuously on a long timescale. For most repeating sources, $\rm DM$
typically varies in a very narrow range of $\pm 3$ cm$^{-3}$ pc.
However, FRB 20190520B is found to have a large variation of $\pm
12$ cm$^{-3}$ pc. The Macquart relation \citep{148}, i.e. the
tight positive correlation between ${\rm DM}_{\rm IGM}$ and $z$, is
confirmed with this expanded sample. Furthermore, $\rm DM_{\rm Host}$ is also
found to be correlated with SFR. The
absolute mean value of $\rm RM$ seems to correlate negatively with
SFR, but the observational data points are still too sparse and
more observations are needed to further test this correlation in the future.
$\rm RM$ evolves with time in a much more chaos behavior in different
repeaters, which may be due to orbital motion of the central
engine that resides in a binary system. The pulse width
generally does not evolve with time. The waiting time shows a
bimodal distribution in most repeaters. The absolute mean $\rm RM$ is also
found to be linearly dependent on $\rm DM_{\rm Host}$ (but note that FRB
20121102A is an obvious outlier), which possibly hints the
existence of a universal magnetic field in different sources.
It may also provide a useful method to estimate the redshift
for those repeating FRBs whose distance is unavailable.

Contrary to the increasing tendency of $\rm DM$ evolution in FRB
20121102A, the $\rm DM$ of FRB 20190520B seems to be decreasing over a
long time, and the decreasing tendency is rather strong. Such a
decreasing, together with the relatively wide variation range of
$\rm DM$ mentioned just above, indicates that it is quite special as
compared with other repeaters. In fact, FRB 20190520B deviates
from other sources in many plots in Figure \ref{figure2}. It also
has an extremely large $\rm DM$, which is $\sim 1208$ cm$^{-3}$ pc,
thus deviating from the Macquart relation seriously (see
Figure \ref{figure2}). Such a large $\rm DM$ may be mainly contributed by
$\rm DM_{\rm Host}$. The intrinsic reasons that lead to these
special features may be different local conditions. Note that the number of
bursts detected from FRB 20190520B is still not too large, i.e. about 150
in total. More observations on this repeater in the future will be
helpful for understanding its nature.

Only two repeaters have well measured $\rm RM$ values in a long
period, i.e. FRBs 20201124A and 20180916B. The $\rm RM$ of FRB
20201124A evolves in a very special pattern over a time range of
$\sim 54$ days (Figure \ref{figure2}), with an amplitude of $\sim
540$ rad m$^{-2}$. On the contrary, the $\rm RM$ evolution is quite
simple for FRB 20180916B, which only monotonously increases by a
small amplitude of $\sim 50$ rad m$^{-2}$ over a long period of
$\sim 1000$ days. It clearly indicates that there might be more
than one kind of triggering mechanisms even for repeating FRBs.

For the repeaters in our sample, the typical bandwidth is found to
be closely related to the typical pulse width and the typical
waiting time. However, the correlation between the typical
pulse width and the typical waiting time is not very tight, which
indicates that the cosmological time dilation effect and the observational
selection effect due to limited telescope sensitivity are not
significant for the sample. The correlation between the typical
bandwidth and the typical pulse width/typical waiting
time is then likely due to the evolution of the source characteristics
at different redshifts. It is also found that the typical
fluence is negatively correlated with redshift but is positively
correlated with SFR, which are likely due to the evolution of source
features at large distances that are related to the emission process.
Specifically, younger magnetars which are more active in generating
bursting activities are more like to be associated with a higher SFR.
They will also lead to a more turbulent local environment and a larger variation
in the DM. This is supported by the positive correlation found
between DM RMSE and SFR.
Besides, there is no correlation between the typical waiting time and
SFR. Since the waiting time is closely related to triggering mechanism,
the absence of correlation between the waiting time and SFR probably
indicates that FRBs might arise from external mechanisms rather than internal ones.

Interestingly, both the typical peak flux and the typical
fluence are positively correlated with the typical waiting time.
The typical energy is also positively correlated with the typical
waiting time, where FRB 20190520B is no longer an outlier anymore.
The implications of such correlations on the triggering mechanism
of repeating FRBs still need to be investigated.
Additionally, there is also a positive correlation between the
typical peak flux and the typical pulse width, which is
interestingly consistent with the NS-asteroid collision model
of repeating FRBs.


\section*{Acknowledgements}
We thank the anonymous referee for useful comments and suggestions
that lead to an overall improvement of the presentation.
This study is supported
by the National Key R\&D Program of China (2021YFA0718500),
by the National Natural Science Foundation of China (Grant Nos. 12041306, 12233002),
by National SKA Program of China No. 2020SKA0120300,
and by the Youth Innovations and Talents Project of Shandong
Provincial Colleges and Universities (Grant No. 201909118).

We acknowledge use of the CHIME/FRB Public Database, provided at https://www.chime-frb.ca/ by the CHIME/FRB Collaboration.


\bibliography{references}{}
\bibliographystyle{aasjournal}


\appendix

In addition to the above analyses, we have also explored a few
other double-parameter distributions of repeating FRB sources. But
generally, no clear correlation is observed. Here, we present
these additional double-parameter plots as an Appendix.

\graphicspath{{./}{appxfigures/}}

\begin{figure}[ht!]
\gridline{\fig{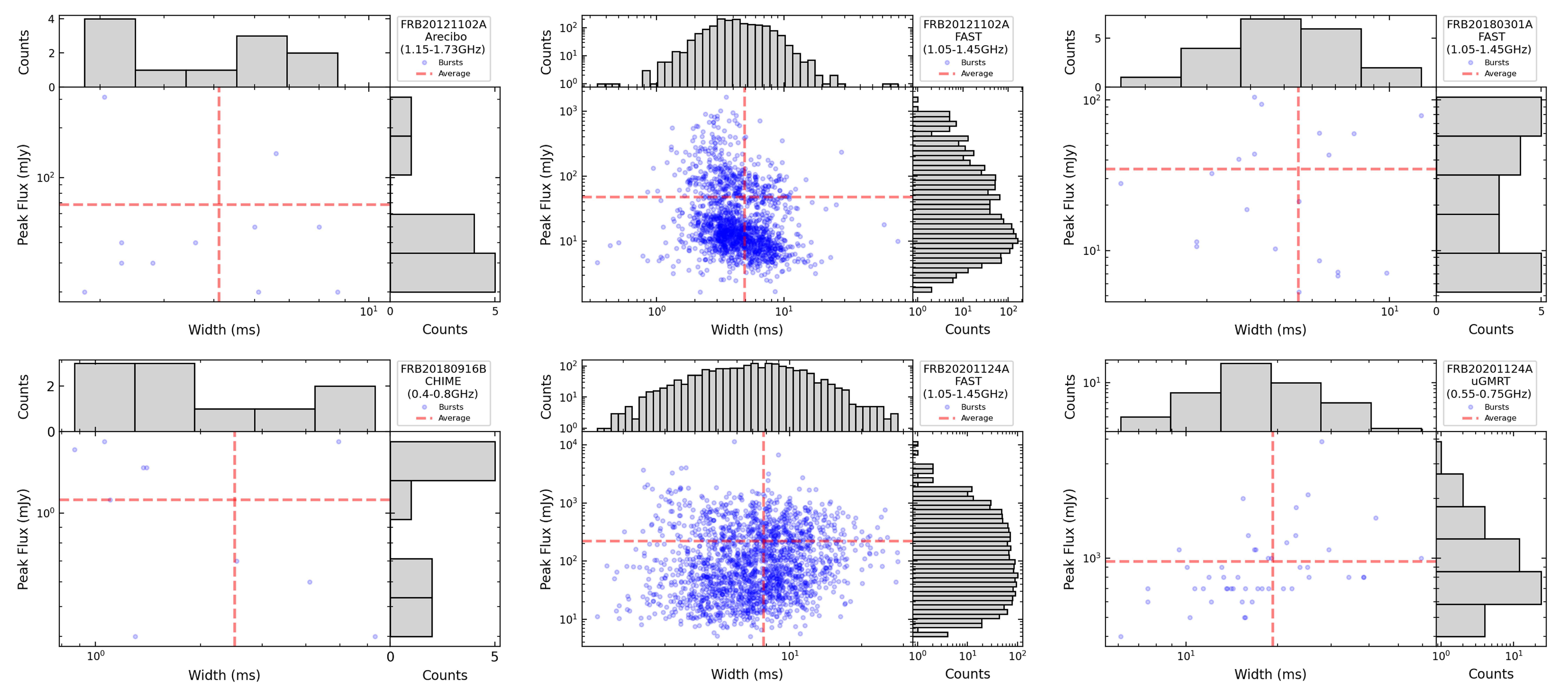}{0.75\textwidth}{}}
 \caption{ The distribution of FRBs on the peak flux - pulse width plane.
 Bursts from each source are plotted according to the detection telescope
 separately. The horizontal and vertical dashed lines mark the average
 values of the two parameters, respectively. }
\label{appxfigure1}
\end{figure}

\begin{figure}[ht!]
\gridline{\fig{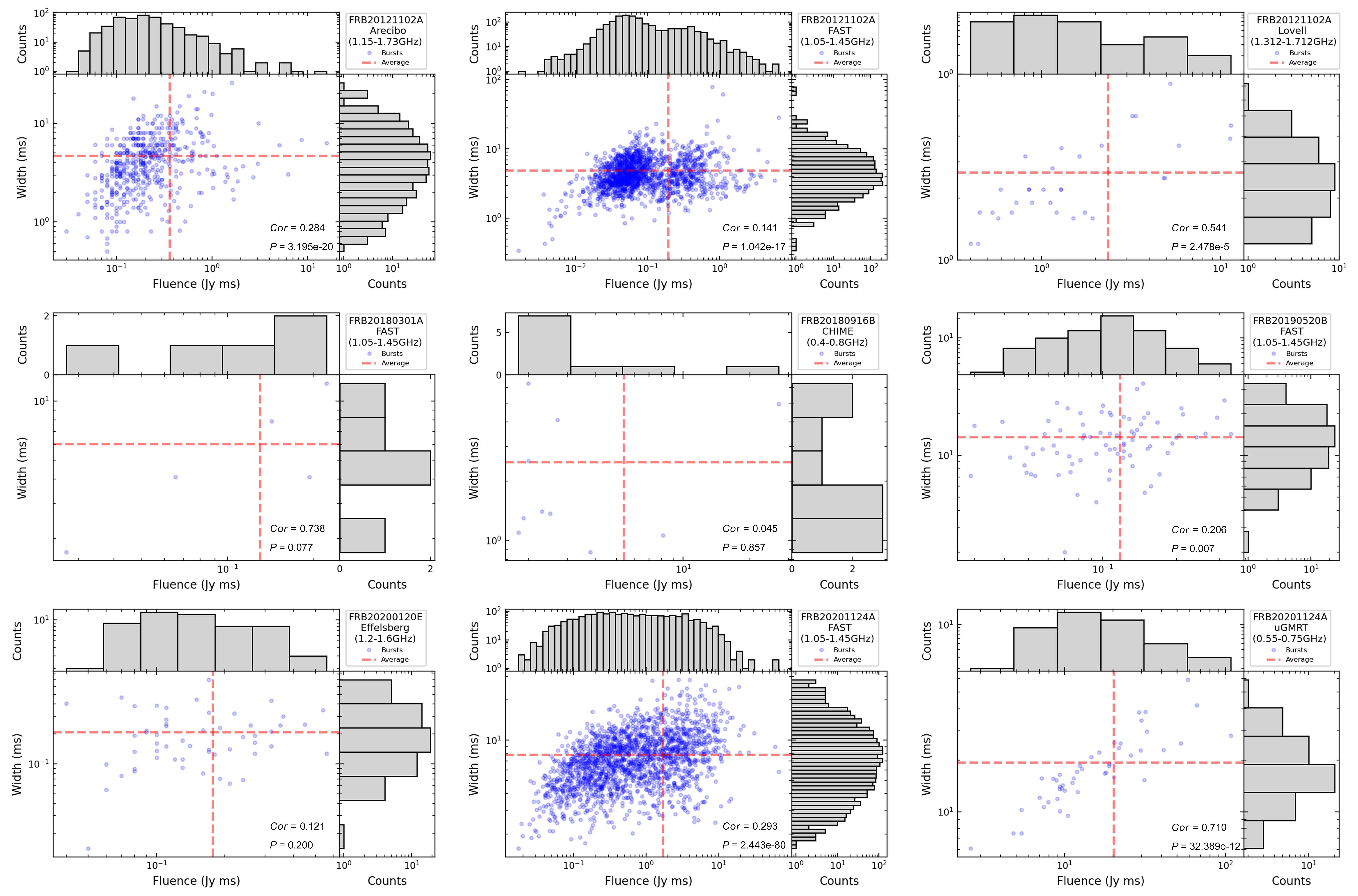}{0.75\textwidth}{}}
 \caption{ The distribution of FRBs on the width - fluence plane.
 Bursts from each source are plotted according to the detection telescope
 separately. The horizontal and vertical dashed lines mark the average
 values of the two parameters, respectively. }
\label{appxfigure2}
\end{figure}

\begin{figure}[ht!]
\gridline{\fig{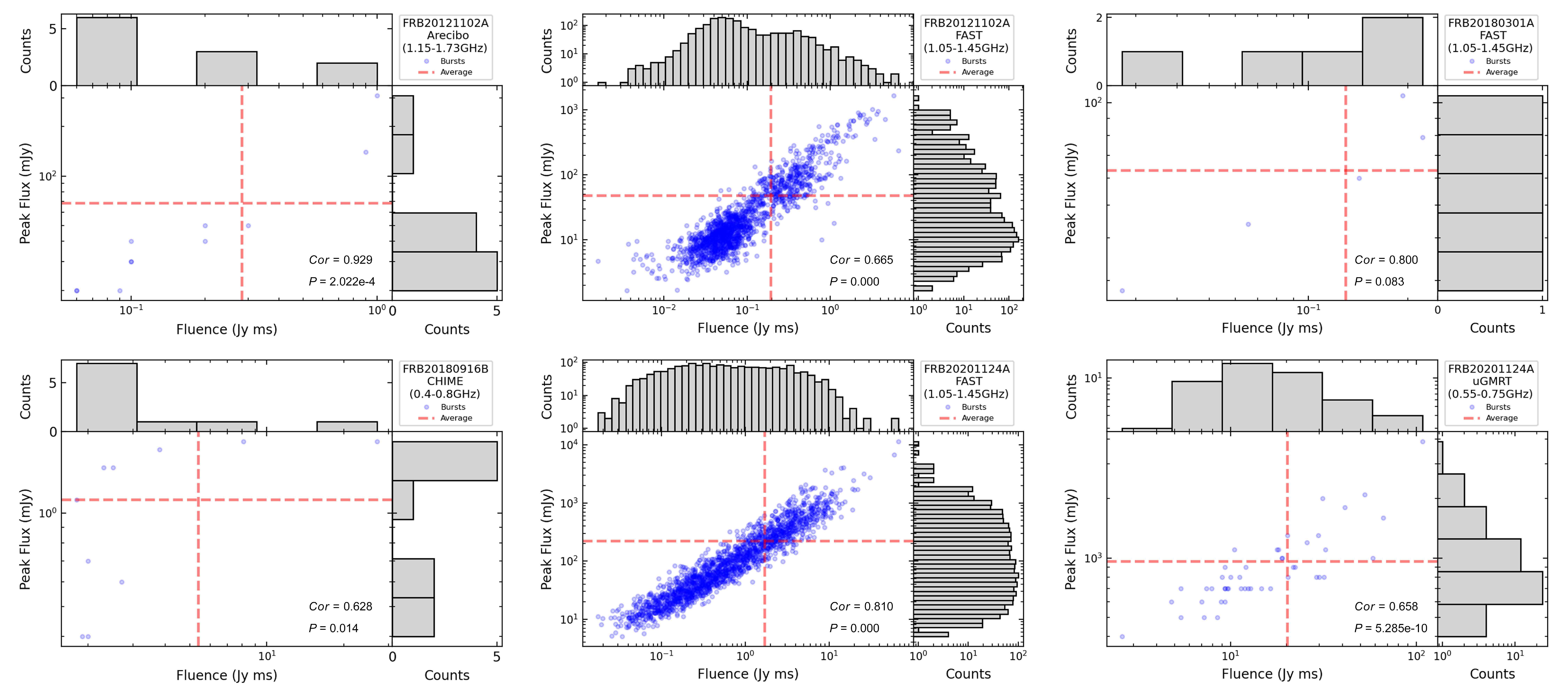}{0.75\textwidth}{}}
 \caption{ The distribution of FRBs on the peak flux - fluence  plane.
 Bursts from each source are plotted according to the detection telescope
 separately. The horizontal and vertical dashed lines mark the average
 values of the two parameters, respectively.   }
\label{appxfigure3}
\end{figure}

\begin{figure}[ht!]
\gridline{\fig{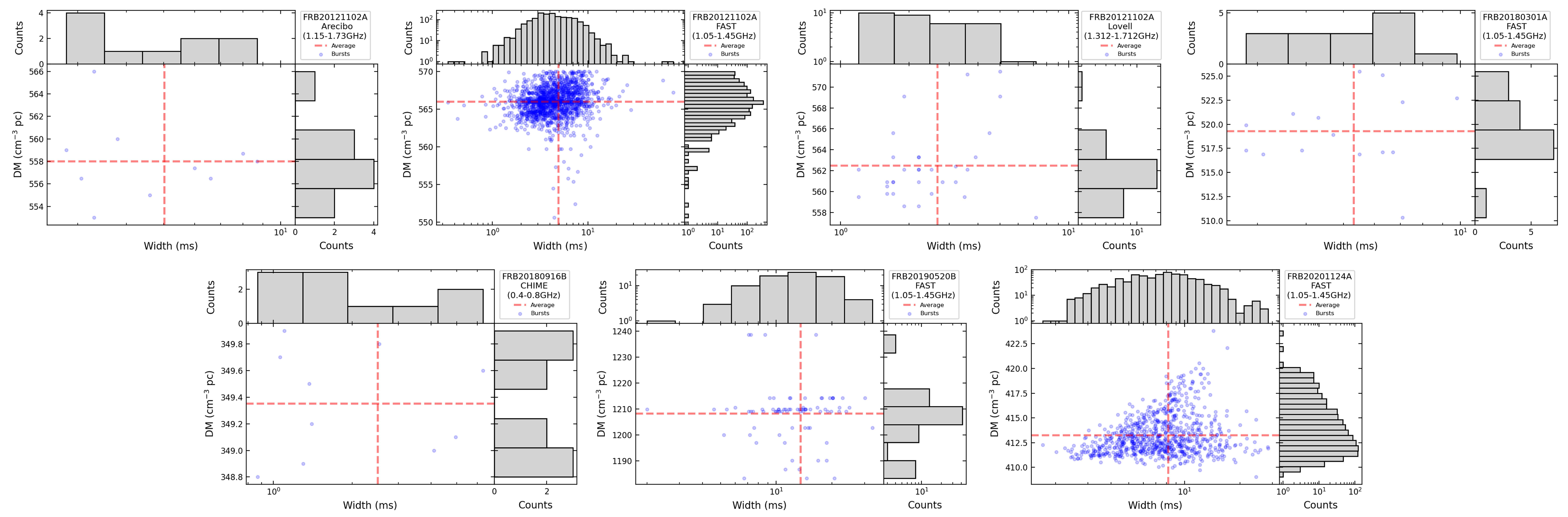}{1\textwidth}{}}
 \caption{ The distribution of FRBs on the $\rm DM$ - pulse width plane.
 Bursts from each source are plotted according to the detection telescope
 separately. The horizontal and vertical dashed lines mark the average
 values of the two parameters, respectively.
}
\label{appxfigure4}
\end{figure}

\begin{figure}[ht!]
\gridline{\fig{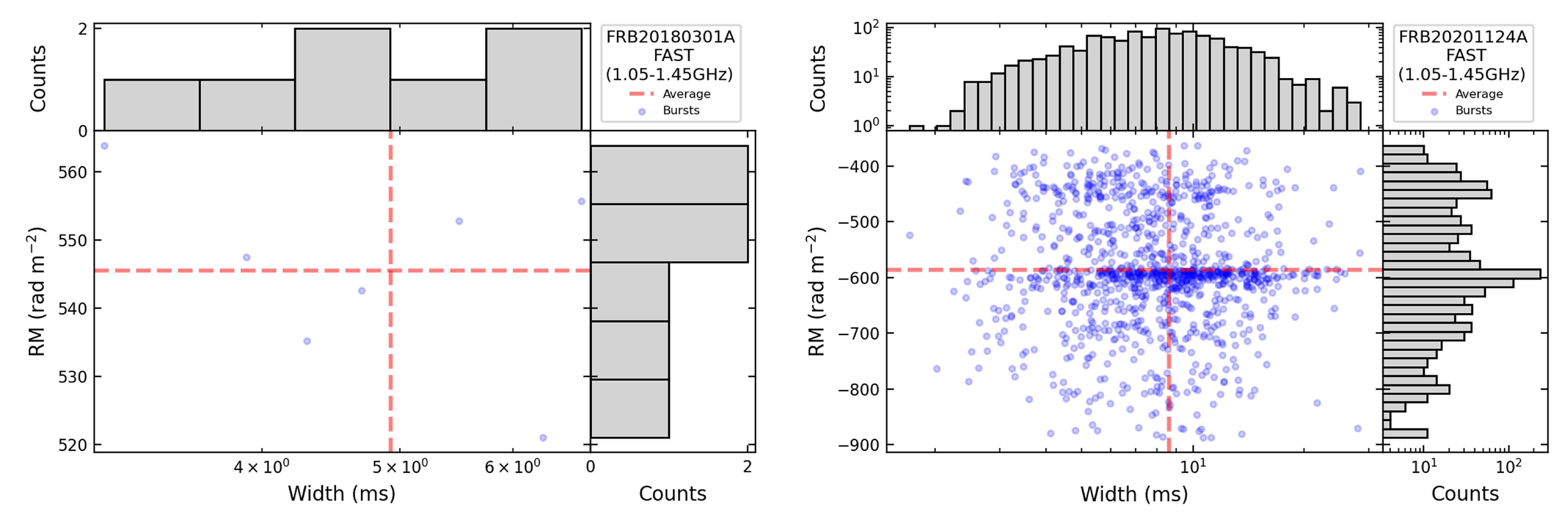}{0.5\textwidth}{}}
 \caption{ The distribution of observed bursts on the $\rm RM$ - pulse width plane,
 for FRB 20180301A and FRB 20201124A. All the bursts are detected by FAST.
 The horizontal and vertical dashed lines mark the average
 values of the two parameters, respectively.
}
\label{appxfigure5}
\end{figure}

\begin{figure}[ht!]
\gridline{\fig{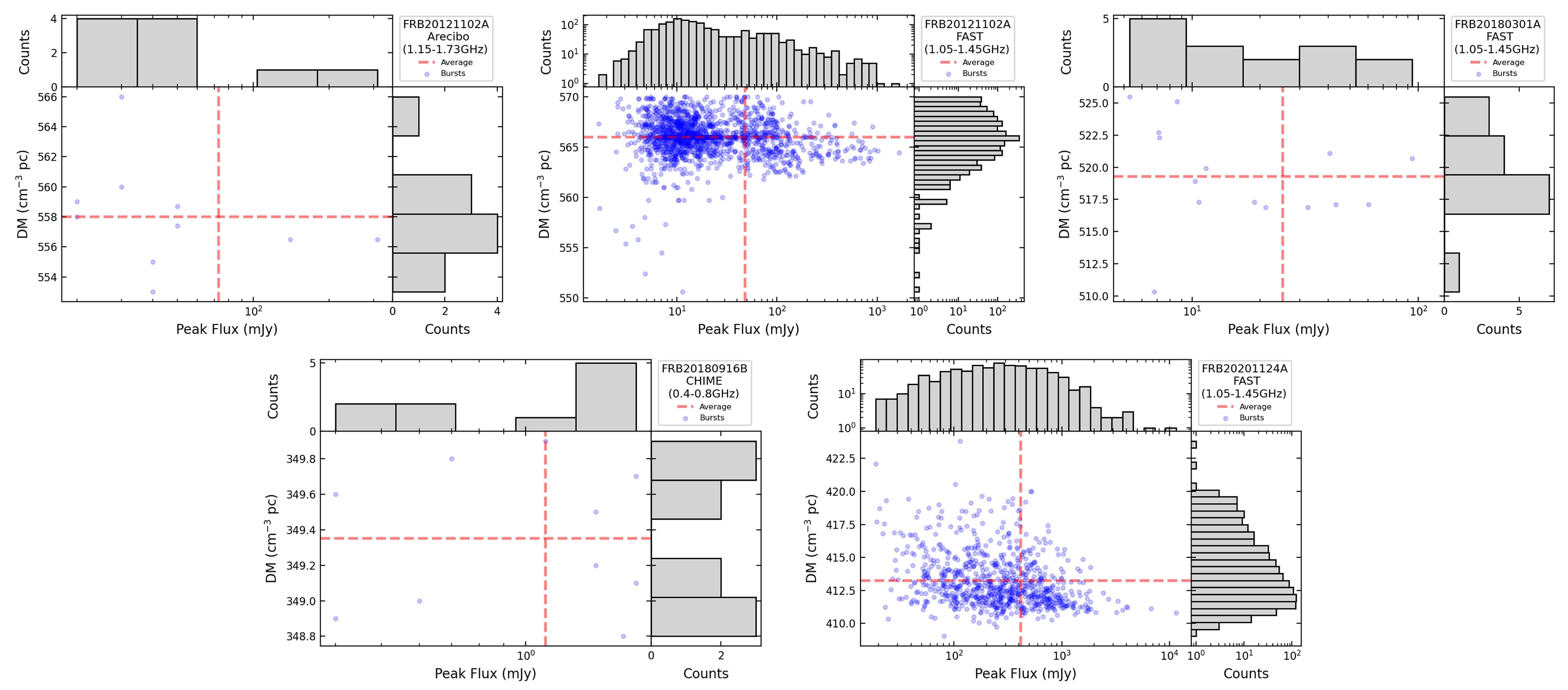}{0.75\textwidth}{}}
 \caption{ The distribution of FRBs on the $\rm DM$ - peak flux plane.
 Bursts from each source are plotted according to the detection telescope
 separately. The horizontal and vertical dashed lines mark the average
 values of the two parameters, respectively.
}
\label{appxfigure6}
\end{figure}

\begin{figure}[ht!]
\gridline{\fig{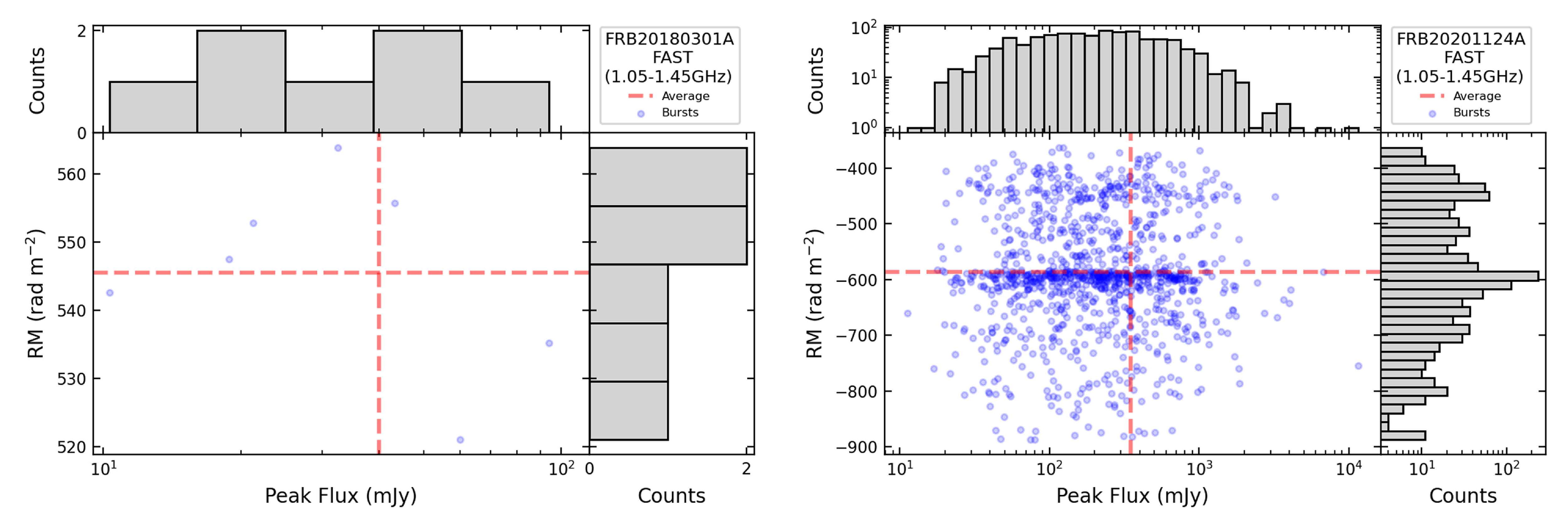}{0.5\textwidth}{}}
\caption{ The distribution of observed bursts on the $\rm RM$ - peak flux plane,
 for FRB 20180301A and FRB 20201124A. All the bursts are detected by FAST.
 The horizontal and vertical dashed lines mark the average
 values of the two parameters, respectively.
}
\label{appxfigure7}
\end{figure}

\begin{figure}[ht!]
\gridline{\fig{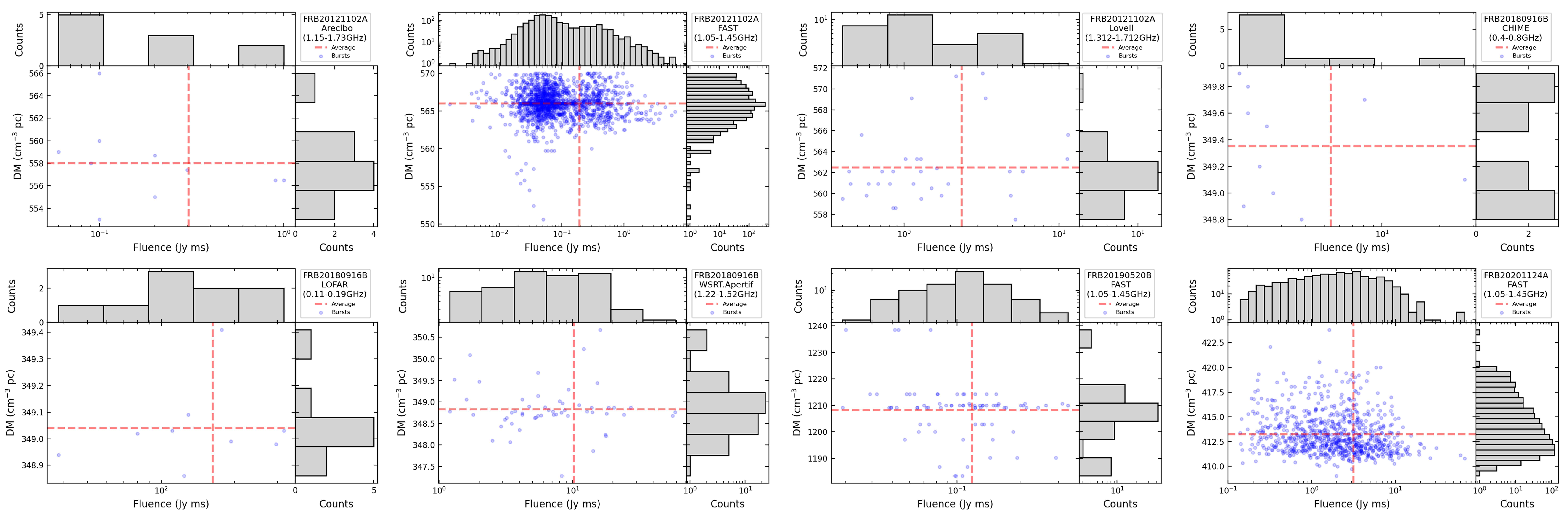}{1\textwidth}{}}
 \caption{ The distribution of FRBs on the $\rm DM$ - fluence plane.
 Bursts from each source are plotted according to the detection telescope
 separately. The horizontal and vertical dashed lines mark the average
 values of the two parameters, respectively.
}
\label{appxfigure8}
\end{figure}

\begin{figure}[ht!]
\gridline{\fig{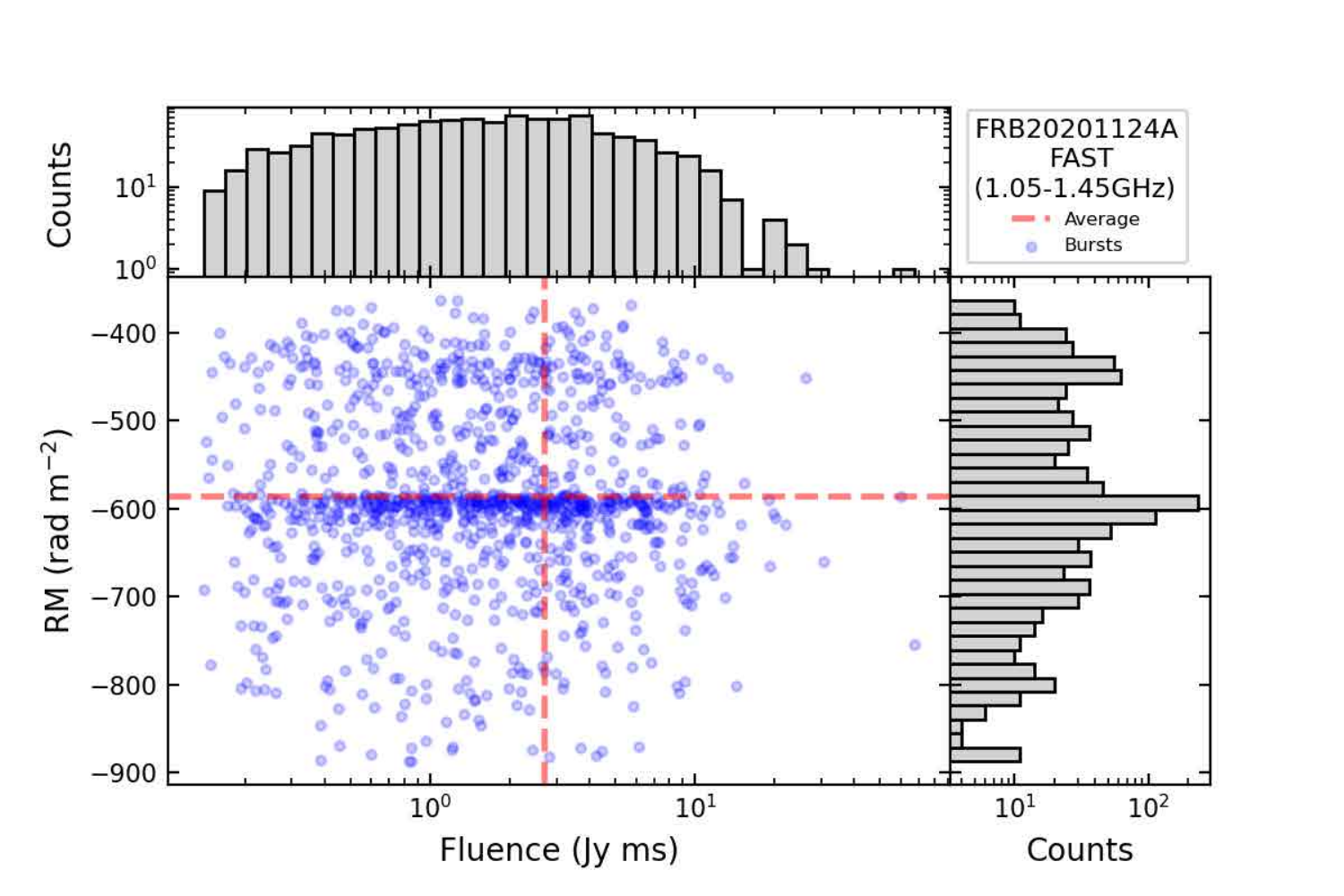}{0.35\textwidth}{}}
 \caption{ The distribution of observed bursts on the $\rm RM$ - fluence plane
 for FRB 20201124A. All the bursts are detected by FAST.
 The horizontal and vertical dashed lines mark the average
 values of the two parameters, respectively.
}
\label{appxfigure9}
\end{figure}

\begin{figure}[ht!]
\gridline{\fig{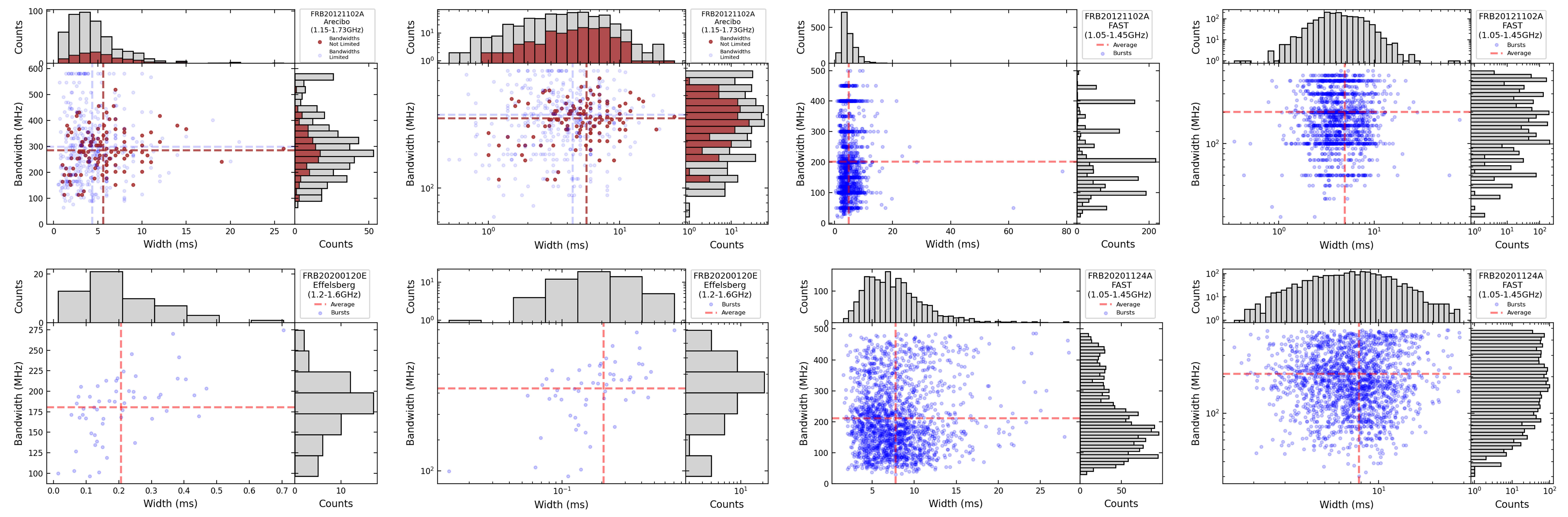}{1\textwidth}{}}
 \caption{ The distribution of FRBs on the bandwith - pulse width plane.
 Bursts from each source are plotted according to the detection telescope
 separately. The horizontal and vertical dashed lines mark the average
 values of the two parameters, respectively.
}
\label{appxfigure10}
\end{figure}

\begin{figure}[ht!]
\gridline{\fig{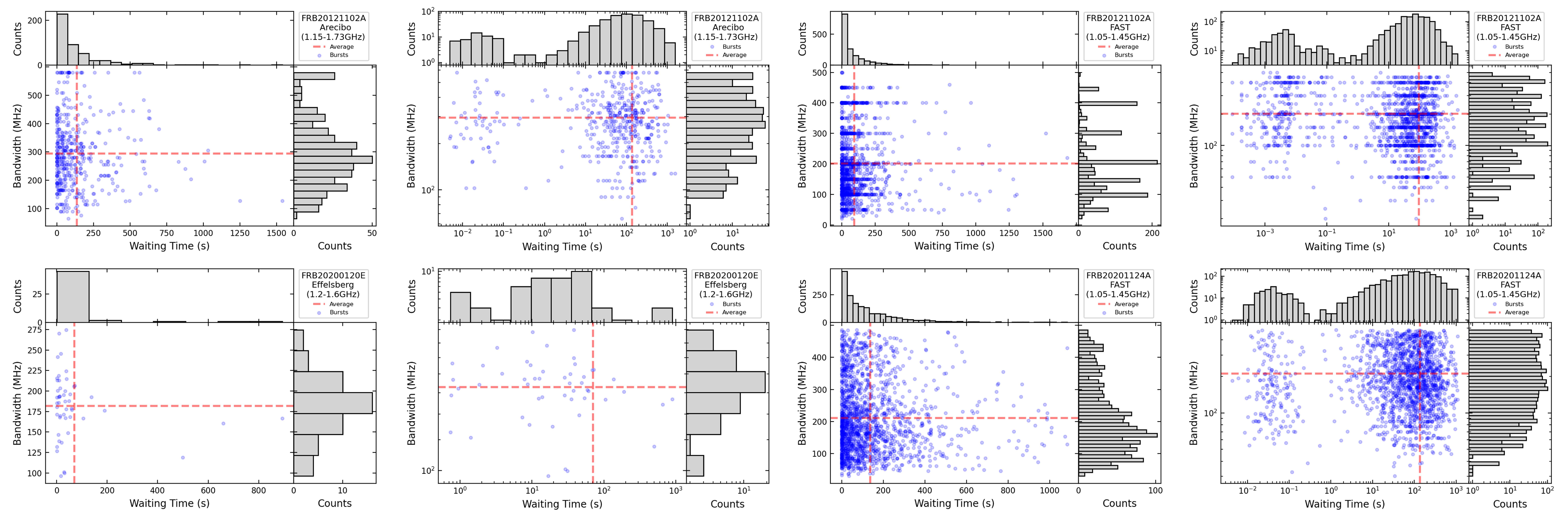}{1\textwidth}{}}
 \caption{ The distribution of FRBs on the bandwith - waiting time plane.
 Bursts from each source are plotted according to the detection telescope
 separately. The horizontal and vertical dashed lines mark the average
 values of the two parameters, respectively.
}
\label{appxfigure11}
\end{figure}

\begin{figure}[ht!]
\gridline{\fig{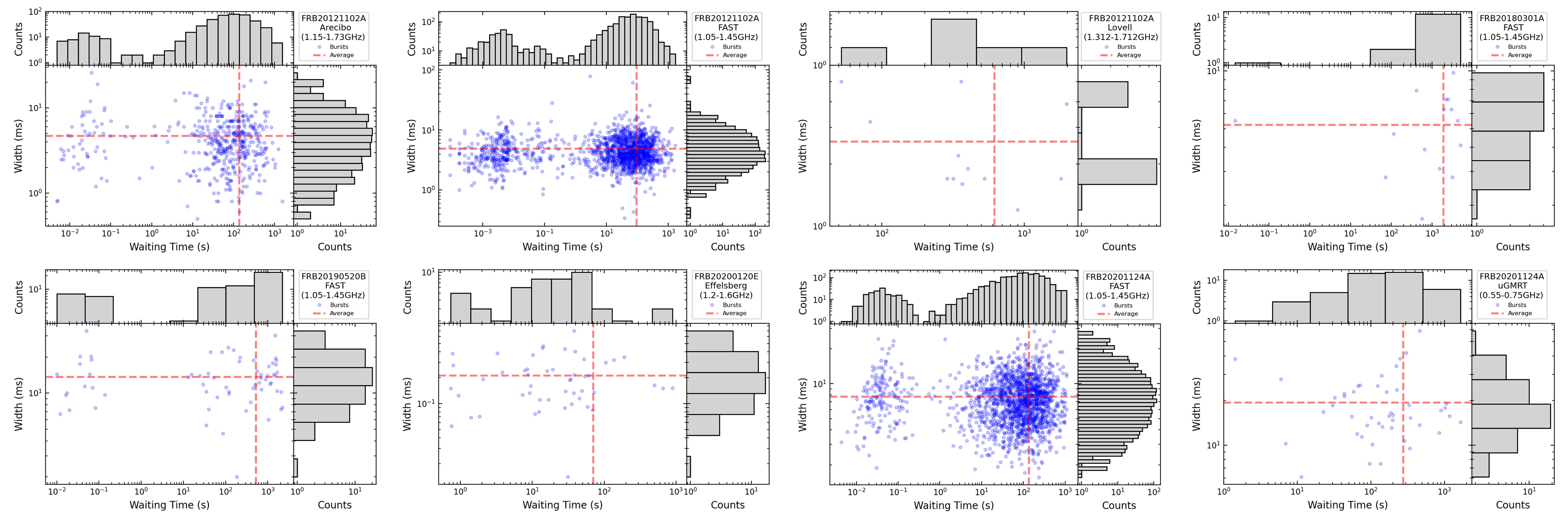}{1\textwidth}{}}
 \caption{ The distribution of FRBs on the pulse width - waiting time plane.
 Bursts from each source are plotted according to the detection telescope
 separately. The horizontal and vertical dashed lines mark the average
 values of the two parameters, respectively.
}
\label{appxfigure12}
\end{figure}

\begin{figure}[ht!]
\gridline{\fig{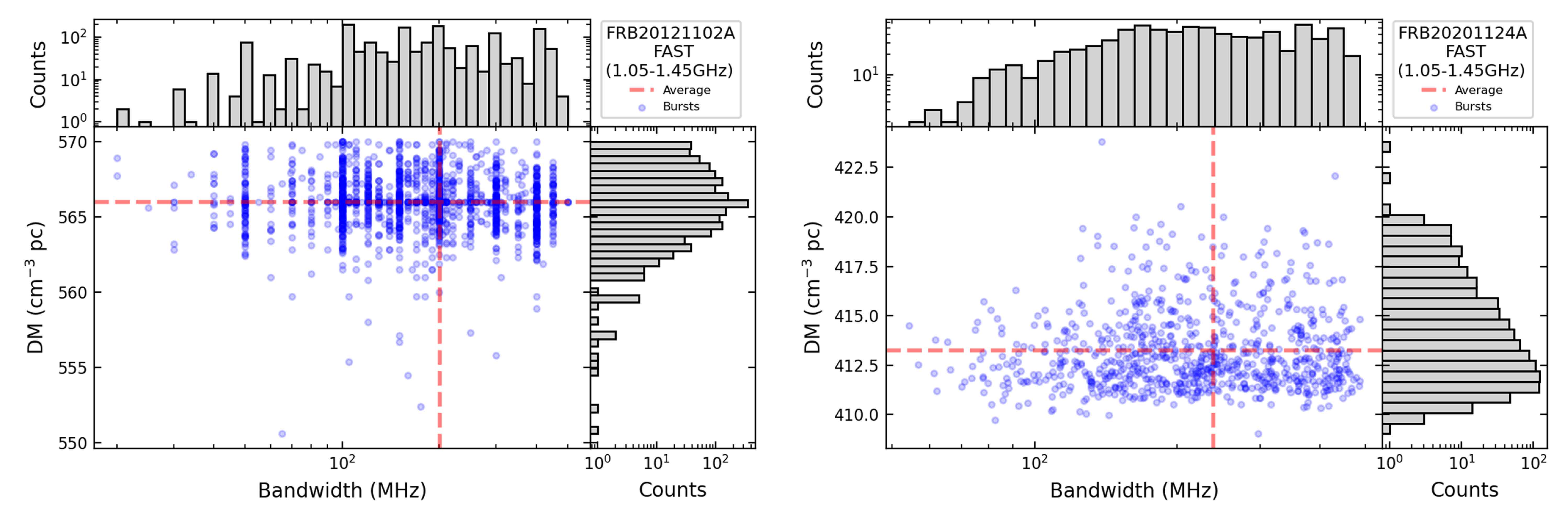}{0.5\textwidth}{}}
\caption{ The distribution of observed bursts on the $\rm DM$ - bandwidth plane,
 for FRB 20121102A and FRB 20201124A. All the bursts are detected by FAST.
 The horizontal and vertical dashed lines mark the average
 values of the two parameters, respectively.
}
\label{appxfigure13}
\end{figure}

\begin{figure}[ht!]
\gridline{\fig{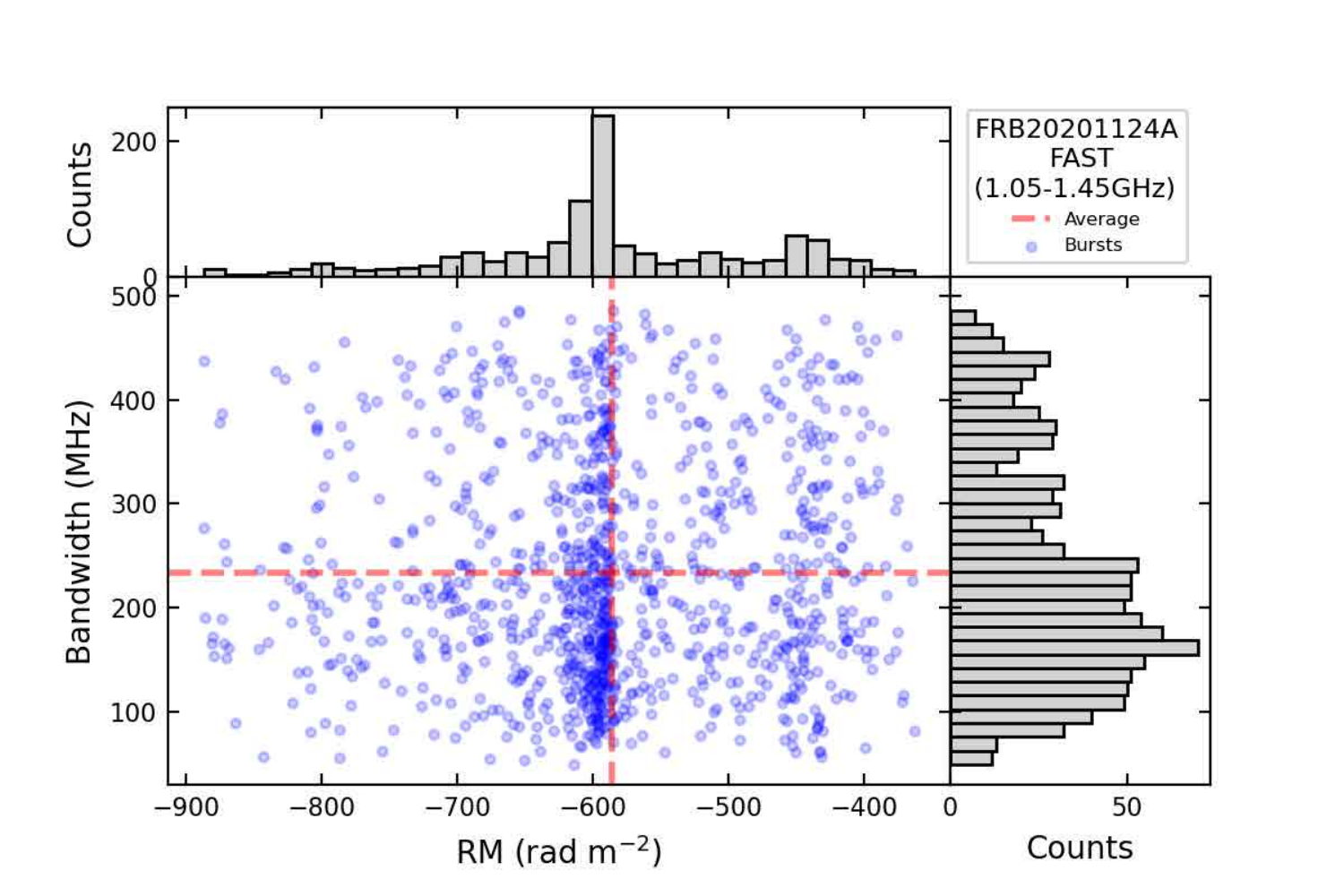}{0.35\textwidth}{}}
\caption{ The distribution of observed bursts on the $\rm RM$ - bandwith plane
 for FRB 20201124A. All the bursts are detected by FAST.
 The horizontal and vertical dashed lines mark the average
 values of the two parameters, respectively.
}
\label{appxfigure14}
\end{figure}

\begin{figure}[ht!]
\gridline{\fig{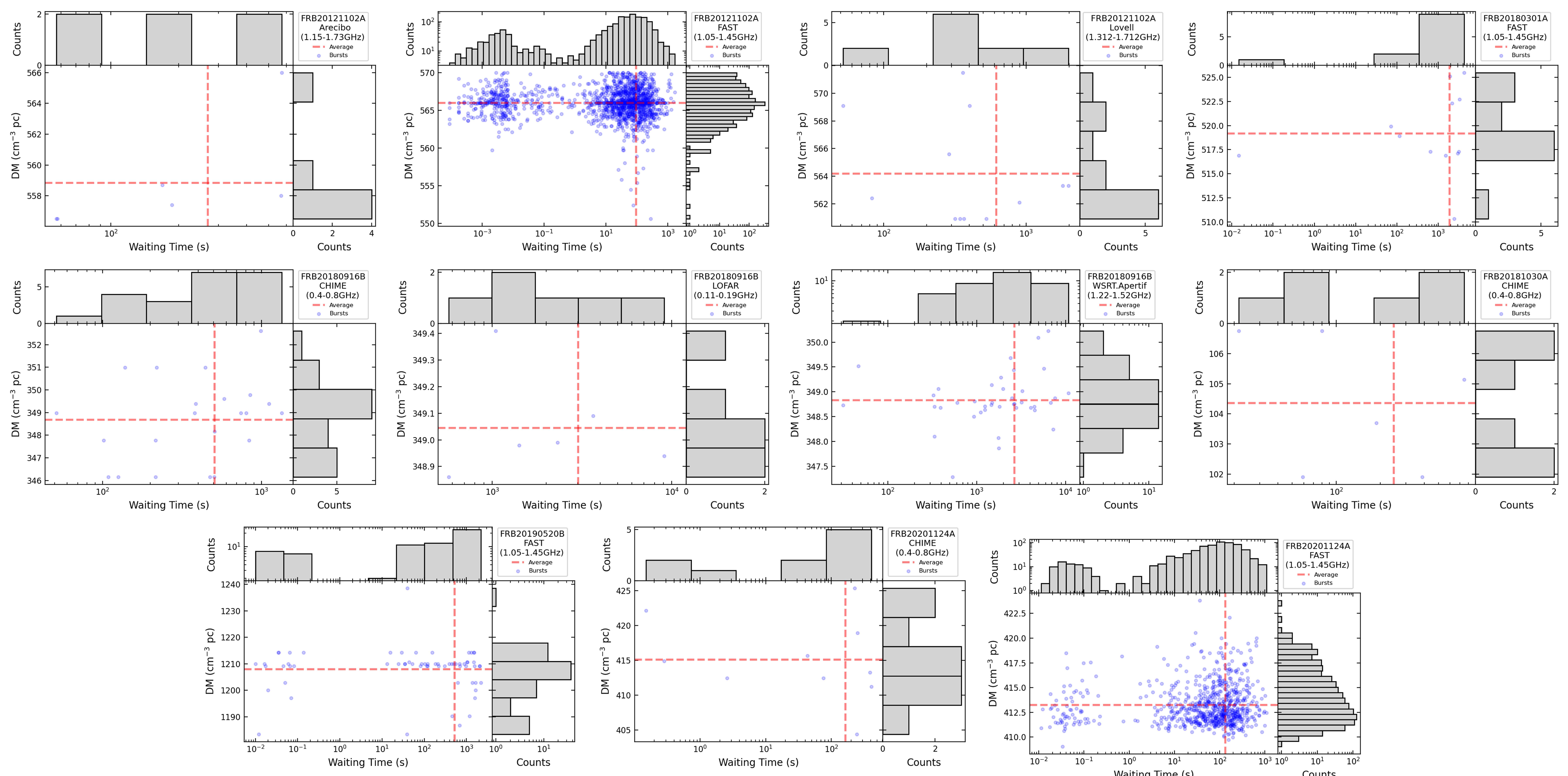}{1\textwidth}{}}
 \caption{ The distribution of FRBs on the $\rm DM$ - waiting time plane.
 Bursts from each source are plotted according to the detection telescope
 separately. The horizontal and vertical dashed lines mark the average
 values of the two parameters, respectively.
 }
\label{appxfigure15}
\end{figure}

\begin{figure}[ht!]
\gridline{\fig{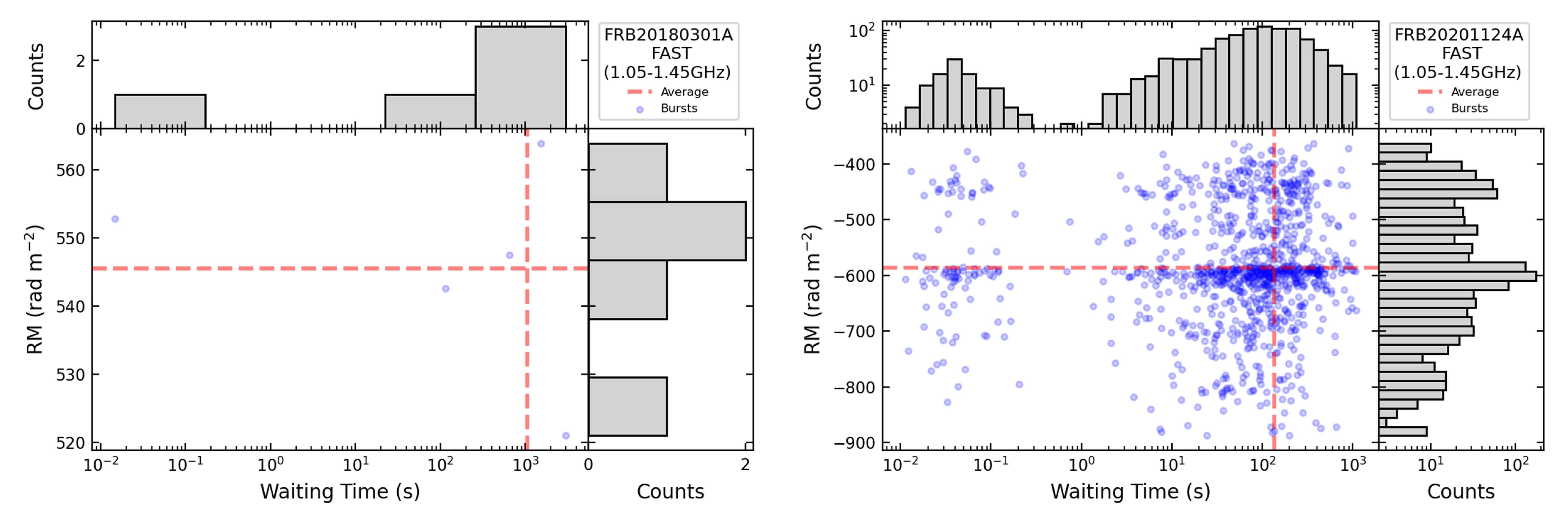}{0.5\textwidth}{}}
 \caption{ The distribution of observed bursts on the $\rm RM$ - waiting time plane,
 for FRB 20180301A and FRB 20201124A. All the bursts are detected by FAST.
 The horizontal and vertical dashed lines mark the average
 values of the two parameters, respectively.
}
 \label{appxfigure16}
\end{figure}

\begin{figure}[ht!]
\gridline{\fig{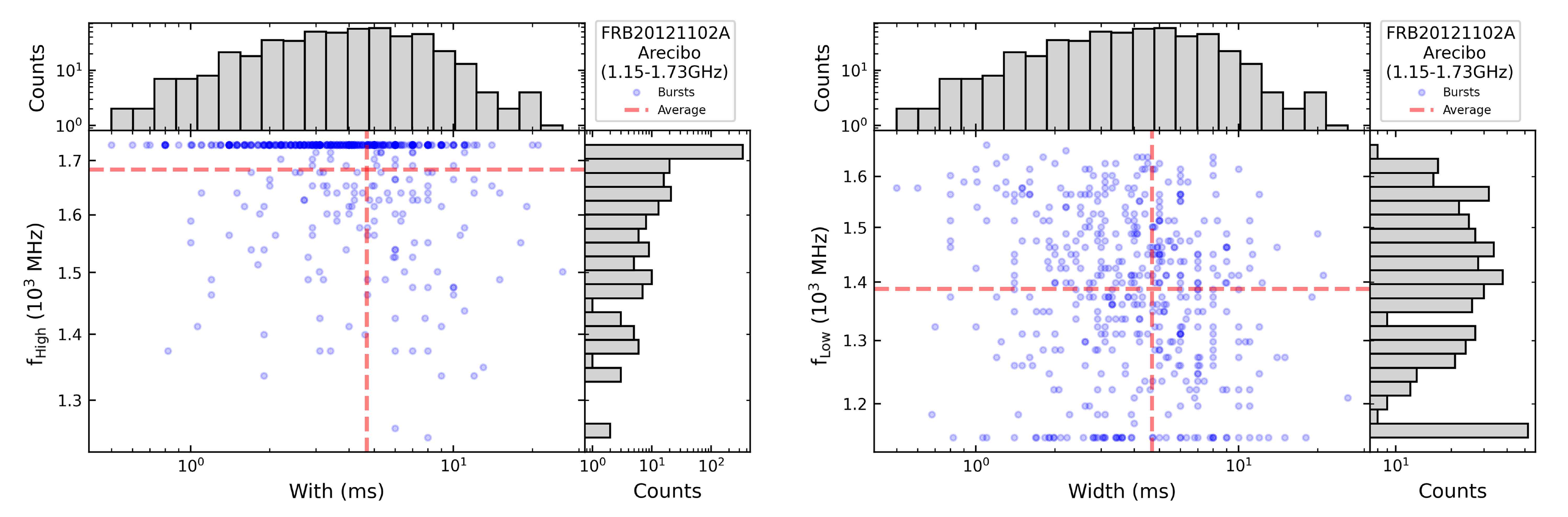}{0.5\textwidth}{}}
 \caption{ The distribution of observed bursts on the bandwidth bound - pulse width plane
 for FRB 20121102A. All the bursts are detected by Arecibo. $f_{\rm High}$ and $f_{\rm Low}$
 stand for the upper and lower bounds of observed bandwidth, respectively.
 The horizontal and vertical dashed lines mark the average
 values of the two parameters.
 }
\label{appxfigure17}
\end{figure}

\begin{figure}[ht!]
\gridline{\fig{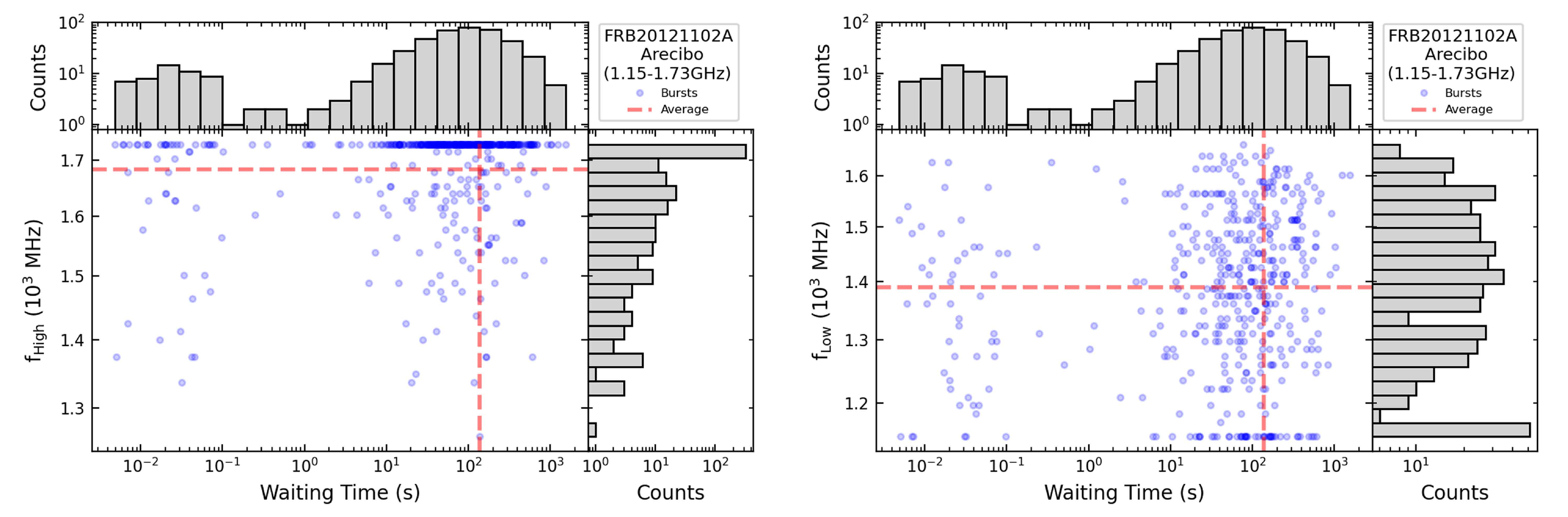}{0.5\textwidth}{}}
 \caption{ The distribution of observed bursts on the bandwidth bound - waiting time plane
 for FRB 20121102A. All the bursts are detected by Arecibo. $f_{\rm High}$ and $f_{\rm Low}$
 stand for the upper and lower bounds of observed bandwidth, respectively.
 The horizontal and vertical dashed lines mark the average
 values of the two parameters.
 }
 \label{appxfigure18}
\end{figure}

\begin{figure}[ht!]
\gridline{\fig{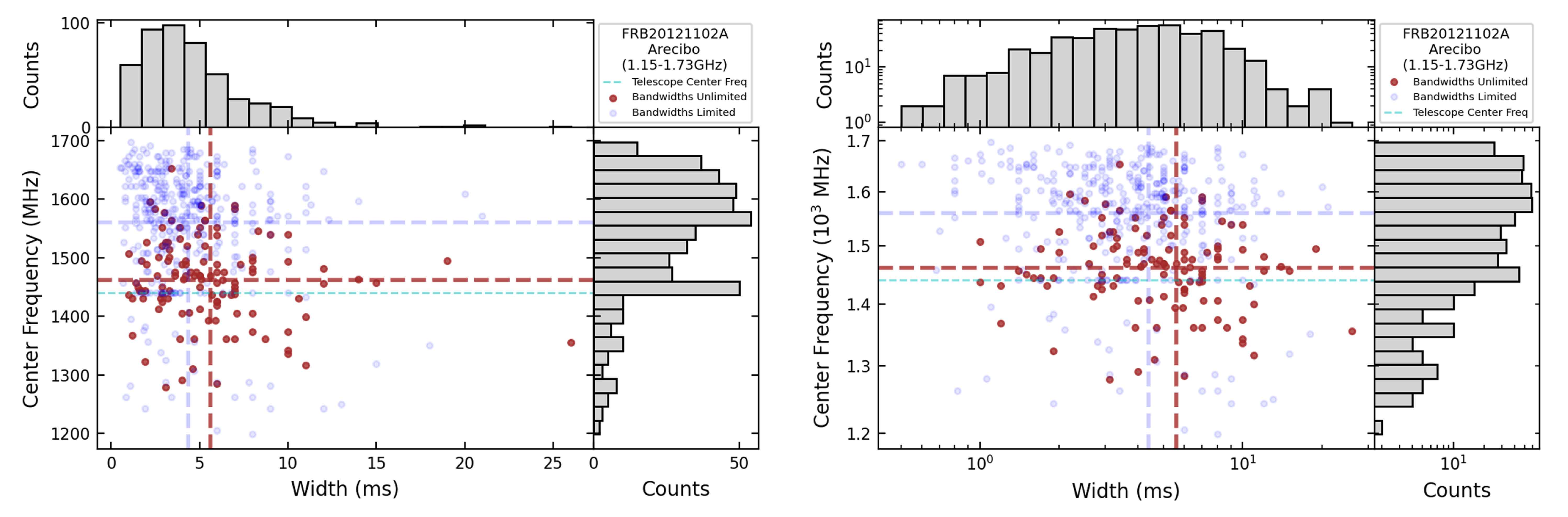}{0.5\textwidth}{}}
 \caption{ The observed center frequency plotted versus pulse width
 for FRB 20121102A. All the bursts are detected by Arecibo.
 The horizontal and vertical dashed lines mark the average
 values of the two parameters, respectively.
 }
\label{appxfigure19}
\end{figure}

\begin{figure}[ht!]
\gridline{\fig{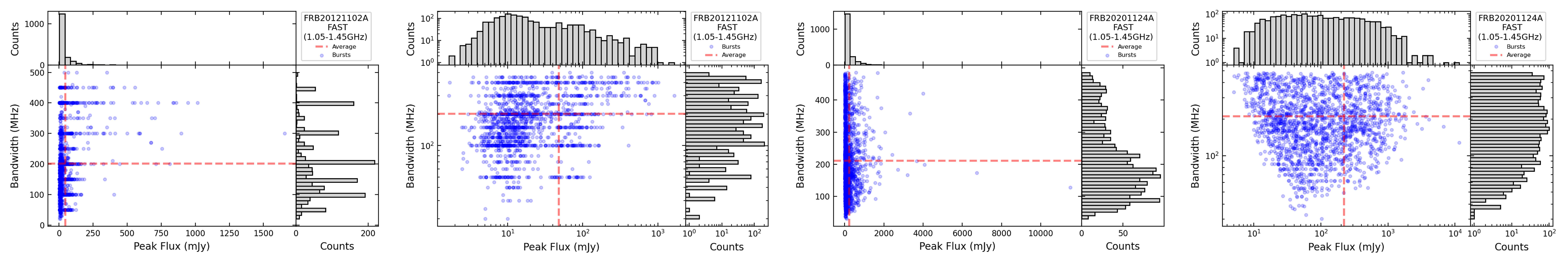}{1\textwidth}{}}
 \caption{ The distribution of FRBs on the bandwidth - peak flux plane.
 Bursts from each source are plotted according to the detection telescope
 separately. The horizontal and vertical dashed lines mark the average
 values of the two parameters, respectively.
 }
 \label{appxfigure20}
\end{figure}

\begin{figure}[ht!]
\gridline{\fig{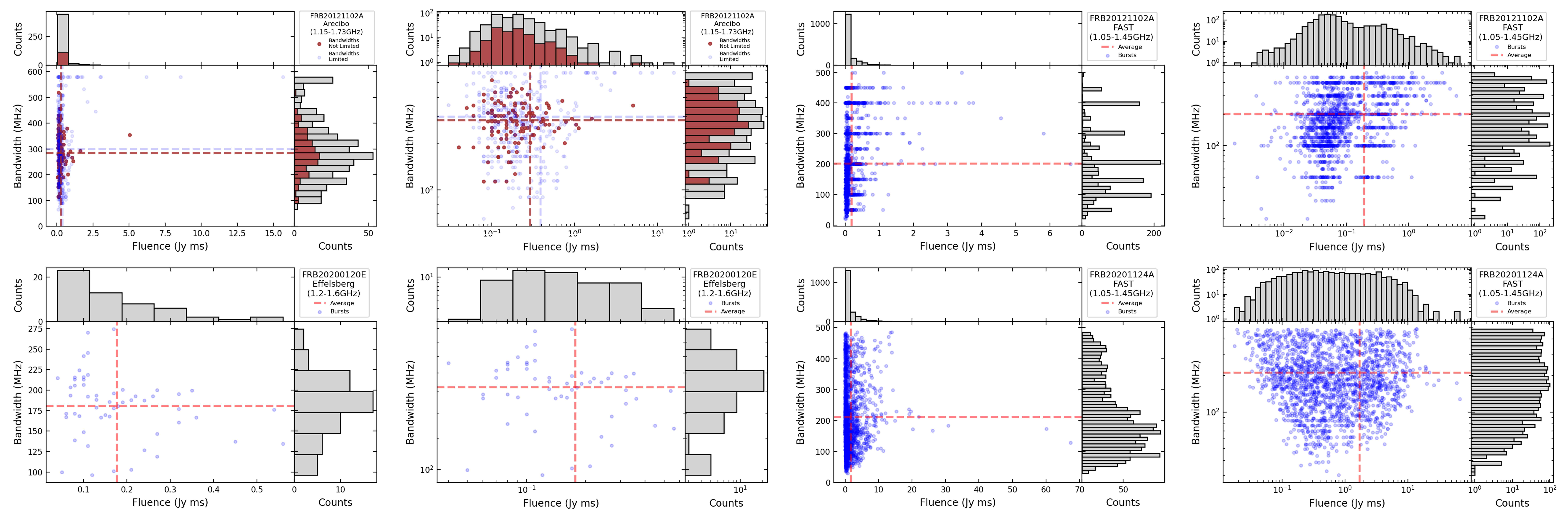}{1\textwidth}{}}
 \caption{ The distribution of FRBs on the bandwidth - fluence plane.
 Bursts from each source are plotted according to the detection telescope
 separately. The horizontal and vertical dashed lines mark the average
 values of the two parameters, respectively.
 }
 \label{appxfigure21}
\end{figure}

\begin{figure}[ht!]
\gridline{\fig{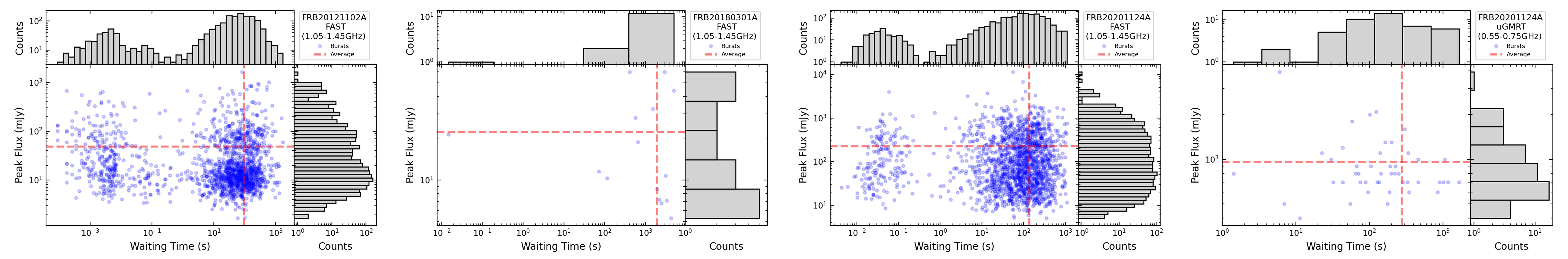}{1\textwidth}{}}
 \caption{ The distribution of FRBs on the peak flux - waiting time plane.
 Bursts from each source are plotted according to the detection telescope
 separately. The horizontal and vertical dashed lines mark the average
 values of the two parameters, respectively.
 }
 \label{appxfigure22}
\end{figure}

\begin{figure}[ht!]
\gridline{\fig{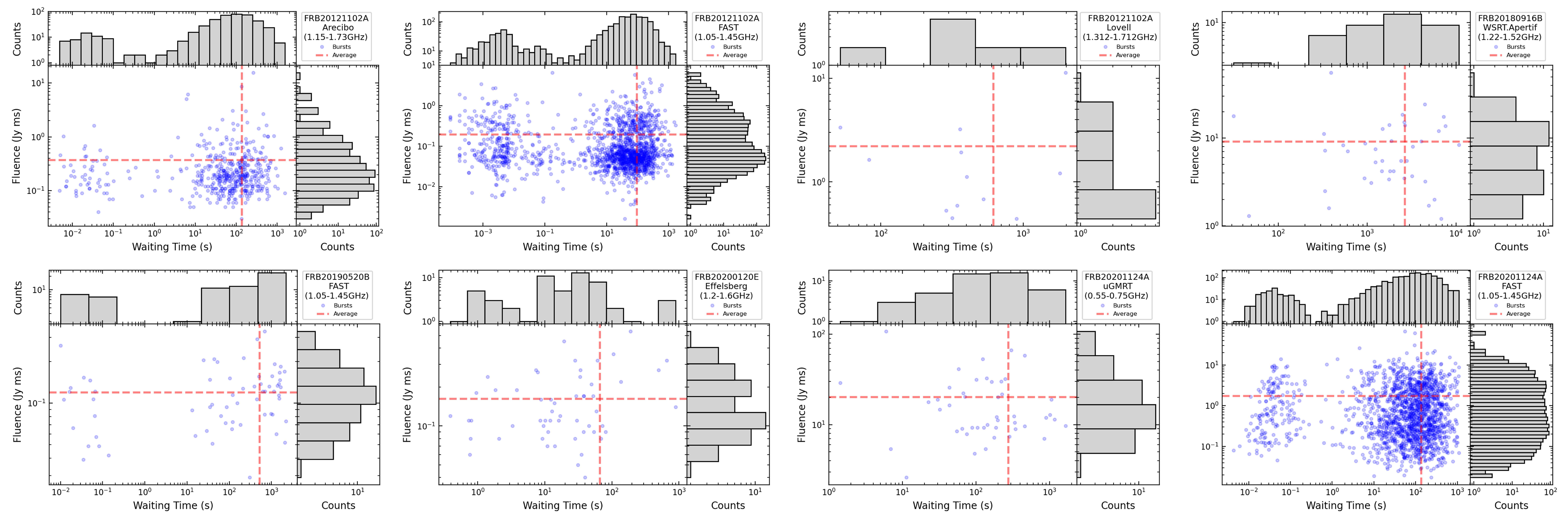}{1\textwidth}{}}
 \caption{ The distribution of FRBs on the fluence - waiting time plane.
 Bursts from each source are plotted according to the detection telescope
 separately. The horizontal and vertical dashed lines mark the average
 values of the two parameters, respectively.
 }
 \label{appxfigure23}
\end{figure}

\begin{figure}[ht!]
\gridline{\fig{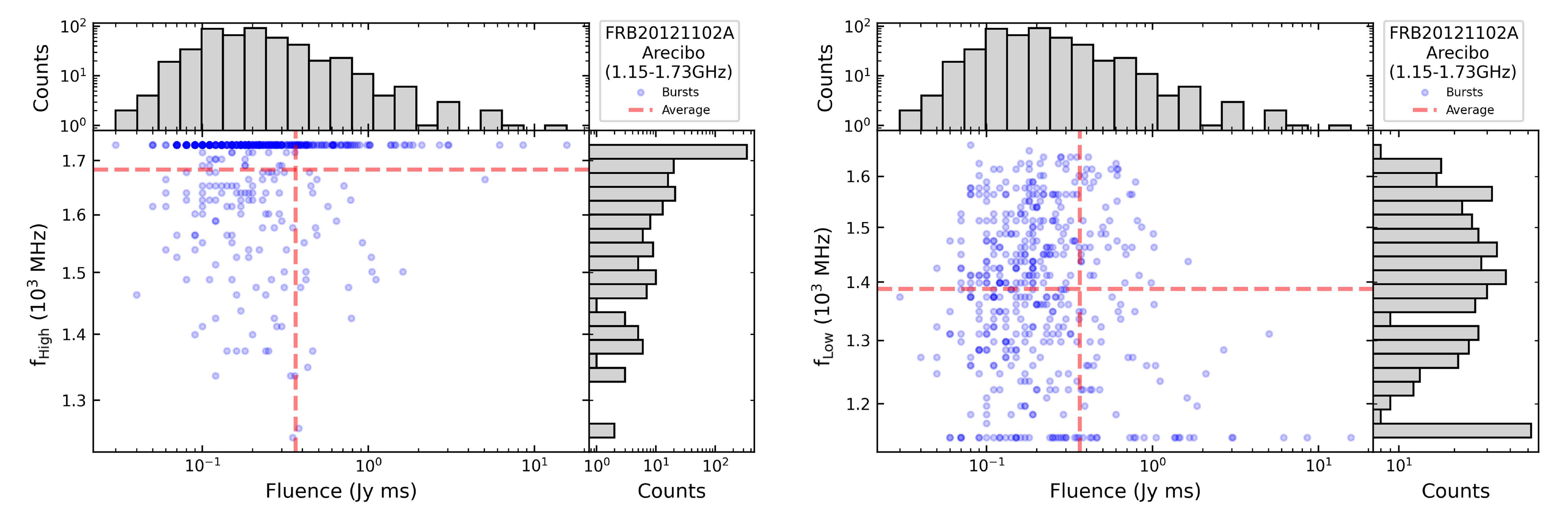}{0.5\textwidth}{}}
 \caption{ The distribution of observed bursts on the bandwidth bound - fluence plane
 for FRB 20121102A. All the bursts are detected by Arecibo. $f_{\rm High}$ and $f_{\rm Low}$
 stand for the upper and lower bounds of observed bandwidth, respectively.
 The horizontal and vertical dashed lines mark the average
 values of the two parameters.
 }
 \label{appxfigure24}
\end{figure}

\begin{figure}[ht!]
\gridline{\fig{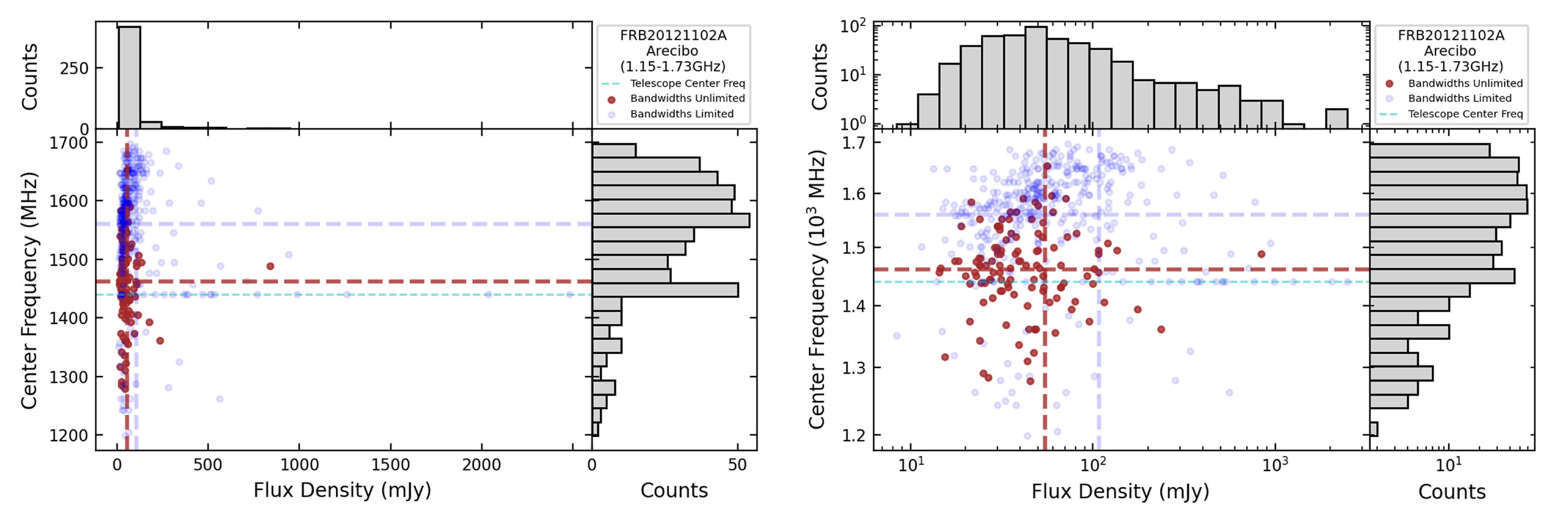}{0.5\textwidth}{}}
 \caption{ The observed center frequency plotted versus flux density
 for FRB 20121102A. All the bursts are detected by Arecibo.
 The horizontal and vertical dashed lines mark the average
 values of the two parameters, respectively.
 }
 \label{appxfigure25}
\end{figure}

\begin{figure}[ht!]
\gridline{\fig{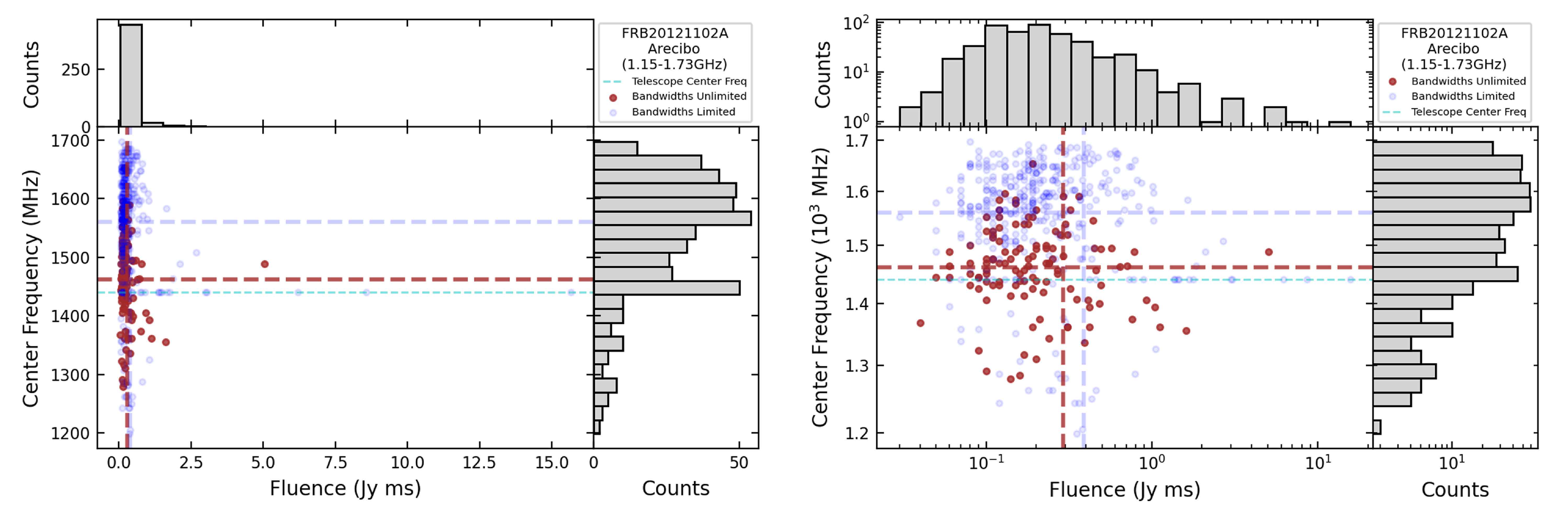}{0.5\textwidth}{}}
 \caption{ The observed center frequency plotted versus fluence
 for FRB 20121102A. All the bursts are detected by Arecibo.
 The horizontal and vertical dashed lines mark the average
 values of the two parameters, respectively.
 }
 \label{appxfigure26}
\end{figure}

\begin{figure}[ht!]
\gridline{\fig{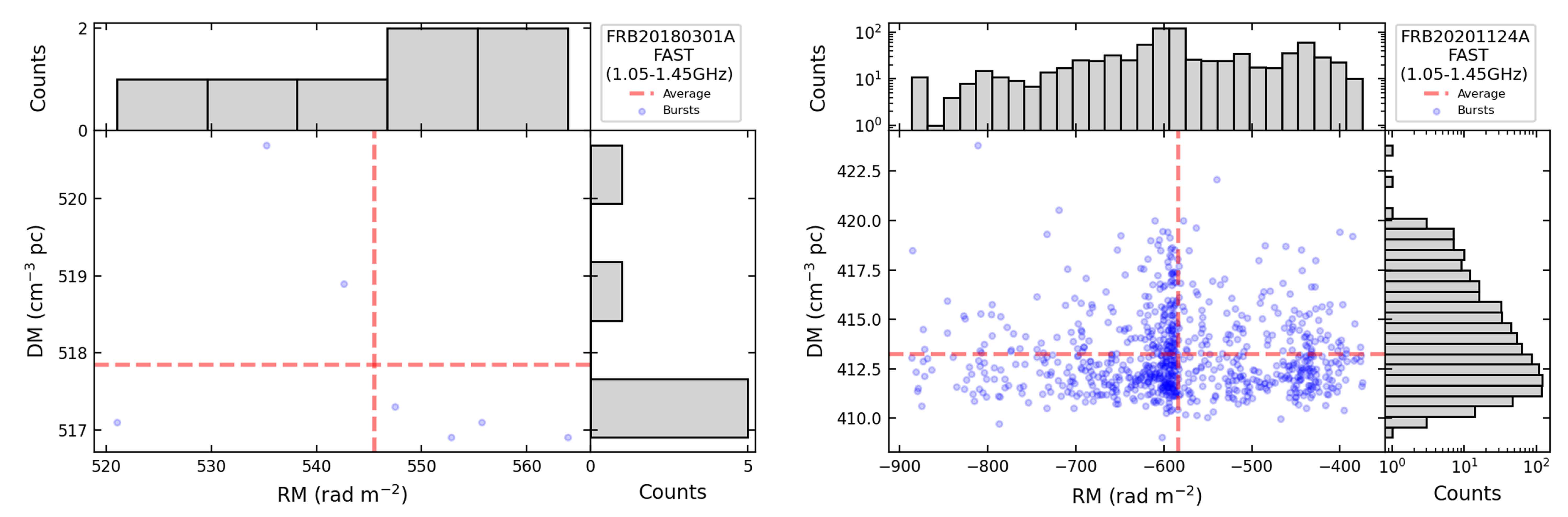}{0.5\textwidth}{}}
 \caption{ The distribution of observed bursts on the $\rm DM$ - $\rm RM$ plane
 for FRB 20180301A and FRB 20201124A. All the bursts are detected by FAST.
 The horizontal and vertical dashed lines mark the average
 values of the two parameters, respectively.
 }
 \label{appxfigure27}
\end{figure}

\end{CJK*}
\end{document}